%% file: main.tex
\documentclass[fleqn,usenatbib,useAMS]{mnras}

\usepackage{graphicx}
\usepackage{amsmath}
\usepackage{amssymb}    
\usepackage{multicol}      
\usepackage{bm}
\usepackage{float}
\usepackage{chngcntr}
\usepackage{soul}
\usepackage{pdflscape}    
\usepackage{caption}
\usepackage{subcaption}
\usepackage{enumitem}
\usepackage{array}
\usepackage[toc,page]{appendix}
\usepackage{xcolor}
\usepackage{threeparttable}
\definecolor{darkgreen}{RGB}{0,200,0}

\usepackage{xspace}
\newcommand{\kms}[1]{\text{\,kms$^{-1}$}\if\relax\detokenize{#1}\relax\else\xspace\fi#1}
\newcommand{\oiii}[1]{\text{[\ion{O}{III}]$\lambda$5007}\if\relax\detokenize{#1}\relax\else\xspace\fi#1}
\newcommand{\nii}[1]{\text{[\ion{N}{II}]$\lambda$6583}\if\relax\detokenize{#1}\relax\else\xspace\fi#1}

\newcommand{\ha}[1]{\text{H$\alpha$\if\relax\detokenize{#1}\relax\else\xspace\fi#1}}
\newcommand{\hb}[1]{\text{H$\beta$\if\relax\detokenize{#1}\relax\else\xspace\fi#1}}

\makeatletter

\renewcommand*\@fnsymbol[1]{%
  \ifcase#1\or
    ^\star\or
    ^\ddagger\or
    ^\S\or
    \P\or
    \|\or
    **\or
    \ddagger\ddagger
  \else\@ctrerr\fi
}

\makeatother

\title[Gas inflows to the nucleus of nearby galaxies]{Composite Bulges -- V. Detecting signatures of gas inflows in IFU data: The MUSE view of ionised gas kinematics in nearby galaxies}
\author[Kolcu et al.]{Tutku Kolcu$^{1,2}$\thanks{Contact e-mail: \href{mailto:Tutku.Kolcu@nottingham.ac.uk}{Tutku.Kolcu@nottingham.ac.uk}}, 
Witold Maciejewski$^{1}$\thanks{Contact e-mail: \href{mailto:w.maciejewski@ljmu.ac.uk}{w.maciejewski@ljmu.ac.uk}},
Peter Erwin$^{3,4,\dagger}$, 
Dimitri A. Gadotti$^{5}$, 
Francesca Fragkoudi$^{6}$, 
\newauthor
Paula R. T. Coelho$^{7}$, 
Victor P. Debattista$^{8}$, 
Adriana de Lorenzo-Cáceres$^{9,10}$, 
Camila de Sá-Freitas$^{11}$, 
\newauthor
Patricia Sánchez-Blázquez$^{12}$\\
$^{1}$ Astrophysics Research Institute, Liverpool John Moores University, IC2 Liverpool Science Park, 146 Brownlow Hill, L3 5RF, UK\\
$^{2}$ School of Physics and Astronomy, University of Nottingham, University Park, Nottingham NG7 2RD, UK\\
$^{3}$ Max-Planck-Institut für extraterrestrische Physik, Giessenbachstrasse, 85748 Garching, Germany\\
$^{4}$ Universitäts-Sternwarte München, Scheinerstrasse 1, D-81679 München, Germany\\
$^{\dagger}$Deceased\\
$^{5}$ Centre for Extragalactic Astronomy, Department of Physics, Durham University, South Road, Durham DH1 3LE, UK\\
$^{6}$ Institute for Computational Cosmology, Department of Physics, Durham University, DH1 3LE, UK\\
$^{7}$ Universidade de São Paulo, Instituto de Astronomia, Geofísica e Ciências Atmosféricas, Rua do Matão 1226, 05508-090, São Paulo-SP, Brazil\\
$^{8}$ Jeremiah Horrocks Institute, University of Lancashire, Preston PR1 2HE, UK\\
$^{9}$ Universidad de La Laguna, Departamento de Astrofísica. Avda. Astrofísico Francisco Sánchez SNR, E-38206 San Cristóbal de La Laguna, Tenerife, Spain\\
$^{10}$ Instituto de Astrofísica de Canarias. C/ Vía Láctea SNR, E-38205 San Cristóbal de La Laguna, Tenerife, Spain\\
$^{11}$ European Southern Observatory, Alonso de Córdova 3107, Vitacura, Región Metropolitana, Chile\\
$^{12}$ Departamento de Física Teórica, Universidad Autónoma de Madrid, E-28049 Cantoblanco, Spain
}

\date{Accepted XXX. Received YYY; in original form ZZZ}

\begin{document}
\maketitle

\begin{abstract}

Using VLT/MUSE data, we study the ionised-gas kinematics in a mass- and volume-limited ($M_* \geq 10^{10} M_\odot$, $D \leq 20$\,Mpc) sample of 21 nearby galaxies to identify signatures of extended shocks within their inner kiloparsec, which appear as coherent velocity jumps in kinematic maps. By removing angular momentum, shocks in gas cause inflows, which can trigger nuclear star formation and fuel AGN activity. To identify the signatures of extended shocks, we examine residual velocity fields after subtracting a modelled rotating disc, and we study velocity difference between various gas tracers. Combining our kinematic analysis with BPT ionisation diagnostic maps and dust morphology, we find that \textcolor{black}{12 of 21 galaxies ($\sim$57\%)} show extended shock signatures with velocity jumps consistent with models of bar-driven shocks. This is likely a lower limit, as three additional galaxies \textcolor{black}{($\sim$14\%)} exhibit shocks along bars, potentially reaching the nucleus but obscured by AGN outflows. We trace shock signatures inwards close to the resolution limit, which suggests that shocks may be the prevailing mechanism of inflow in the central kpc of galaxies. The only two unbarred galaxies in our sample are also the only systems with unperturbed kinematics and no shocks, strongly linking the perturbed gas dynamics in centres of galaxies to the presence of bars. All galaxies with inner bars show LINER- or Seyfert-like nuclear emission, whereas galaxies without inner bars exhibit all emission types, indicating that regardless of gas supply, inner bars suppress star formation in galactic nuclei.

\end{abstract}

\section{Introduction}

Gas inflows in galaxies can significantly affect their evolution by enhancing nuclear star formation \citep{Prieto_19} and feeding the active galactic nucleus (AGN) \citep{Davies_14,Audibert_19}, which consequently triggers AGN outflows \citep{Prieto_05,Davies_14,Audibert_19} that in turn may cause quenching by suppressing star formation \citep{Man_19, Donnari_21}. 
Numerous observational and theoretical studies have investigated the mechanisms driving gas inflows in galaxies \citep{Sanders_1976,Emsellem_03,Fathi_04,Prieto_19}. Tidal forces exerted during close encounters or collisions distort and reshape galactic structure, and redistribute angular momentum across the galaxy, leading to gas inflows \citep{Mihos_1996,Barnes_1996,Hopkins_2006,Robertson_2006,Lagos_2018}. Moreover, density waves which create spiral arms can exert torques to the gas and stars \citep{Roberts_1975,Kendall_2008}, affecting the overall distribution of angular momentum in the galaxy, and causing gas inflows \citep{Elmegreen_1999,Grosbol_Patsis_Pompei_04}. Dynamical friction is another pivotal mechanism that can cause infall. As satellites, globular clusters or giant molecular clouds orbit within/around a galaxy, the dynamical friction against the galaxy's stellar population can slow these bodies down, consequently leading to angular momentum exchange between these bodies and the galaxy's medium \citep{Combes_2003,D'Onghia_2006,Jogee_2006,Dutton_2012}. 

Gas inflow can induce the secular evolutionary processes that slowly shape the galaxy’s structure over time, promoting the reshaping of disk profiles \citep{Shlosman_1993,Navarro_1995}, the formation and growth of central substructures such as gaseous and stellar disks and rings \citep{Shlosman_1989,Athanassoula_1992,Heller_1994,Buta_Combes_1996,Martini_03,Storchi-Bergmann_03,Fathi_05,Kim_12,Sormani_15}, and triggering starbursts in nuclear rings \citep{Heller_1994, Mazzuca_08}. Furthermore, gas reaching the nucleus can contribute to nuclear feeding \citep{Heller_1994,Park_2017}. Another mechanism that can introduce a dimension of complexity to gas dynamics is magnetic loops; large-scale structures of magnetic fields that arch above the galactic plane. These structures can influence the regulation of galactic magnetic fields \citep{Krause_1989,Beck_2013}, gas dynamics \citep{Liu_2023}, star formation \citep{McKee_07} and the nuclear feeding cycle \citep{Hu_22}; their influence can also interfere with other galactic structures, such as bars \citep{Abbate_20}. 

One of the most established mechanisms causing gas inflows involves shocks driven by bars. In the local universe, disc galaxies commonly host non-axisymmetric central stellar structures, with approximately two-thirds of this population hosting bars \citep{Eskridge_2000,Menendez_07,Gadotti_2009,Buta_2015}. Simulations have revealed that the bars play a crucial role in transporting gas from the outer regions to the inner kiloparsec of galaxies \citep{Athanassoula_1992,Garcia-Burillo-05,Fragkoudi_16}. Torques generated by the bar pull gas clouds which can then collide with large relative velocities, causing shocks in gas. Angular momentum transfer in shocks results in inflow of gas towards the inner regions of the galaxy. Once a steady-state flow is established, cloud collisions and therefore shocks in gas, occur at specific locations, which can extend over significant portions of a galaxy, \textcolor{black}{for several hundreds of parsecs}. Here we call such structures \textcolor{black}{\textit{extended shocks}}. Numerical models indicate that gas inflows occur in extended shocks, usually of either straight or spiral shape \citep{Athanassoula_1992,Maciejewski_2004b}. However, despite extensive research, it remains unclear whether extended shocks provide the dominant mechanism for gas inflow to galactic nuclei or whether the inflow there has a different origin.
 
\textcolor{black}{Shock fronts in gas are surfaces across which gas characteristics undergo abrupt changes: gas density and temperature increase, and its velocity is reduced, according to the jump conditions \citep{LandauLifshitzFluid_6}. As such velocity jumps are present in extended shocks throughout their extent, extended shocks appear in observations as coherent kinematic structures \citep{Zurita_04, Fathi_05, Storchi-Bergmann-07}.}
Our objective in this work is to study the gas kinematic maps \textcolor{black}{of the innermost galactic regions, within $\sim$1\,kpc radius, in search of the signatures of coherent kinematic structures that display straight or spiral morphology consistent with that associated with the shocks in models of gas dynamics.} \textcolor{black}{As coherent structures can be missed in incomplete kinematic maps derived from clumpy emission, we rely on gas tracers that provide closest to complete coverage of the inner kiloparsec -- \ha and \nii emission from ionised gas, observed with the \textit{Multi Unit Spectroscopic Explorer} (MUSE) is the optimal choice for coverage and resolution. Our findings allow us} to determine whether extended shocks are the prevalent mechanism of gas inflows to the galactic nuclei. We build on our pilot study presented in \citet{Kolcu_23}, where, on the example of NGC\,1097, we introduced our methods to highlight and identify the signatures of extended shocks in kinematic maps. In this work, we refine and apply these methods to all galaxies from the \textit{Composite Bulges Survey} (\textit{CBS}, PI: P. Erwin; \citealp{Erwin_21,Ashok_23,Kolcu_23}), and to two other galaxies that satisfy the selection criteria of the CBS sample, for which data with sufficient emission line signal are available.  When compared to previous work, suffering from instrumental limitations of former generation IFUs \citep{Fathi_05,Fathi_06,Storchi-Bergmann-07}, the level of details resolved with MUSE maximises the prospect of finding the coherent kinematic structures that can unveil extended shocks which are approaching the nucleus of galaxies.

The paper is organised as follows. In Sect.\,\ref{sec:Parent_sample and Data Sources} we describe the sample selection and the MUSE observations of the galaxies analysed in this study. In Sect.\,\ref{sec:data_extraction_MUSE}, we outline the procedure used to extract stellar and ionised-gas kinematics from the MUSE spectroscopy. In Sect.\,\ref{sec:method} we summarise the motivation behind the different techniques employed to identify signatures of extended shocks and highlights the complementarity of these approaches. In Sect.\,\ref{sec:common-features-seen-across-galaxies}, we present the collective analysis of the sample, drawing on the detailed kinematic results provided galaxy-by-galaxy in Appendix \ref{sec:notes-on-galaxies}. Specifically, we discuss the quantitative properties of the shocks (e.g. velocity jumps, radial extents) in Sect.\,\ref{sec:collective_quantitative_results_sample}, the common features evident across the sample in Sect.\,\ref{sec:common_features_seen_across_sample}, and the grouping of galaxies according to shock class and nuclear emission in Sects.\,\ref{sec:grouping-shock-class} and \ref{sec:grouping-nuclear-emission}, respectively. In Sect.\,\ref{sec:discussion} we explore correlations between the observed quantities and other galaxy properties, including: the relation between kinematic categories and the presence of large-scale bars (Sect.\,\ref{sec:correlation_between_kinematic_category_largescalebars}); correlations with inner bars (Sect.\,\ref{sec:shocks_nuclearbars_nuclearrings}); the connection between the strength of extended shocks and the gas velocity dispersion (Sect.\,\ref{sec:average_perturbation_against_shock_amplitudes}); and the dependence on host-galaxy mass (Sect.\,\ref{sec:shocks_mass}). Finally, Sect.\,\ref{sec:conclusion} provides a summary of the main results and conclusions. We provide supplementary plots used in different parts of our analysis in Appendix \ref{app:suplementary_sample_plots}.

\section{Sample Selection and MUSE Observations}
\label{sec:Parent_sample and Data Sources}

\subsection{Sample Selection}
\label{sec:CBS}
\textcolor{black}{The galaxy sample studied in this work comprises targets from the multi-wavelength imaging and spectroscopic campaign {{Composite Bulges Survey, {CBS}}} (PI: Peter Erwin; \citealp{Erwin_21,Ashok_23,Kolcu_23}) that show sufficiently strong ionised-gas emission in the MUSE data. We also include two additional galaxies that satisfy the CBS selection criteria but are not part of the CBS sample, and for which public MUSE observations are available.} CBS focuses on understanding of the formation and growth mechanisms of bulges, and the correlation with their host galaxy properties. Moreover, it aims to identify and characterise the different forms of nuclear stellar components within the inner 1--2\,kpc; from large-scale bars down to nuclear scale discs/rings/bars and nuclear star clusters \citep{Erwin_21,Ashok_2023}. 

The CBS sample is a mass- ($\rm M_*\geq10^{10}\,M_{\odot}$) and volume- ($D\leq20$ Mpc) limited set of 53 S0–Sbc galaxies with Galactic latitude $|b|>20^\circ$ and inclinations between $35^\circ$ and $60^\circ$ \citep{Erwin_21}. The sample has optical and near-infrared imaging with \textit{the Hubble Space Telescope (HST)}; \textcolor{black}{at the survey distance limit, the near-IR \textit{HST} F160W image quality (PSF FWHM $\approx0.15''$) corresponds to a physical resolution of $\sim$10--15\,pc, enabling us to resolve substructures within the inner bulge regions.} Among 53 galaxies in CBS, 46 of them have been observed with MUSE, from which observations of \textcolor{black}{35} galaxies are obtained specifically for the CBS and observations of 11 galaxies are sourced from different programmes, including the \textit{Time Inference with MUSE in Extragalactic Rings, {TIMER} Project} (PI:Dimitri Gadotti, \citealp{Gadotti_19}), the \textit{MUSE Atlas of Disks, MAD} (PI:Marcella Carollo, \citealp{Erroz-Ferrer_19}) and the \textit{Physics at High Angular resolution in Nearby GalaxieS Survey, {PHANGS}} (PI: Eva Schinnerer, \citealp{Emsellem_22}). 

Upon reviewing the available MUSE data for the 46 galaxies in CBS, to ensure the reliability of the gas kinematics, we restrict our analysis to those galaxies with detected emission lines in more than 50\% of the spaxels within central 40\arcsec$\hspace{-0.1cm}\times$40\arcsec ($\sim$1~kpc). This specific FOV covers the regions in which we search for kinematic signatures of extended shocks.  This threshold was chosen to ensure the presence of sufficient emission line signal. This approach led to the selection of 19 galaxies from CBS. \textcolor{black}{This criterion is a data-quality cut based on emission-line detectability across a spatial coverage, rather than on the measured velocities or dispersions. It should therefore not bias the kinematic measurements for galaxies that pass the cut, but it can introduce a selection bias by preferentially retaining systems with extended, high \textcolor{black}{signal-to-noise (SNR)} ionised-gas emission in the central $\sim$1\,kpc (e.g. higher gas content or lower effective obscuration) and excluding gas-poor, centrally dust-obscured, or very compact-emission systems. Our results should thus be interpreted as representative of the CBS subset with spatially extended, detectable ionised gas, not the full parent sample.}

To this sample we added 2 galaxies from the TIMER Project \citep{Gadotti_19}. Since the CBS dataset requires a minimum galaxy inclination of 35\degr, these galaxies, each with an inclination of 34\degr, were excluded from CBS, despite meeting all other criteria, such as distance and mass, and also surpassed the emission threshold we discussed in the previous paragraph. We decided to include these galaxies as the measurement error for inclination is at least on the order of 1\degr. We found no other galaxies at the limits of CBS selection criteria. As a result, our final sample comprises 21 galaxies. This is an almost complete sample of galaxies that satisfy the CBS selection criteria and have sufficient ionised emission, with the caveat that 7 out of 53 galaxies in the parent CBS sample have no MUSE observations.

From this sample, we selected NGC\,1097 as a target for the pilot study where we developed the methodology for identifying extended shock signatures through kinematic analysis, which is presented in \citet{Kolcu_23} (here after Paper IV). This methodology was then advanced and used for the remaining galaxies in our sample, whose analysis is presented in this work. More details on our methodology are provided in Sect.\,\ref{sec:method}. 

\subsection{MUSE observations of the sample}
\label{observations:sample}
Observations of all galaxies are done in MUSE Wide-Field Mode, WFM \citep{Bacon_10}. \textcolor{black}{MUSE, mounted on the UT4 of the Very Large Telescope \citep{Bacon_10}, with its large, 60\arcsec$\hspace{-0.1cm}\times$60\arcsec field of view (FOV), and fine, 0\farcs2$\times$0\farcs2 spatial sampling, allows capturing extensive portions of the central regions of nearby galaxies in great detail, across a rest frame wavelength range of 4650 to 9300\AA~which covers prominent emission lines. Over this range, the instrumental spectral resolution corresponds to an approximately wavelength-dependent FWHM of $\sim$2.5--2.9\,\AA\ (equivalently $R\simeq1770$ at 4800\,\AA\ increasing to $R\simeq3590$ at 9300\,\AA). Besides these capabilities, MUSE also offers high sensitivity, providing high SNR, effectively minimising the common need for binning. Furthermore, MUSE is equipped with cutting-edge adaptive optics (AO) technology, contributing significantly to the data acquisition by correcting for atmospheric distortions, improving the seeing and allowing us to achieve fine spatial resolutions}\footnote{\textcolor{black}{The delivered spatial resolution, quantified by the PSF FWHM is 0\farcs44-1\farcs21 across our sample (median 0\farcs7 arcsec; Table \ref{tab:targets-general}). This corresponds to $\sim$25-95\,pc at the target distances (median $\sim$60\,pc).}}.

In Table~\ref{tab:targets-general}, we summarise the general properties and observations of each galaxy studied in this work. \textcolor{black}{The spatial scale in pc per arcsec, for each galaxy is derived from the adopted distance listed in the table via the angular–physical conversion.} The spatial resolution (SR) is then estimated from the delivered image quality, quantified by the median point-spread function (PSF) full width at half maximum (FWHM; ``seeing") measured from the VLT guide star in the $V$ band during MUSE slow-guiding exposures, using $\mathrm{SR} = (\mathrm{spatial\ scale}) \times (\mathrm{PSF\ FWHM})$. \textcolor{black}{For a given PSF FWHM, the physical resolution therefore scales linearly with distance.}

However, during the observations of 10 galaxies obtained for CBS and for NGC\,1566 from MAD, MUSE's AO system is used. Therefore, for these galaxies, using the approach from \citet{Fusco_2020}, which accounts for AO monitoring, the PSF FHWMs have been reconstructed to obtain the corrected values. This method demonstrates significant improvements in the seeing for most galaxies, typically between 30--50\%. The PSF FHWMs recorded during observations, along with those improved by AO, are detailed in Table\,\ref{tab:targets-general}. This table also includes RC3 classification of each galaxy from \citet{deVaucouleurs_1991}, and the logarithmic stellar masses obtained from the Spitzer Survey of Stellar Structure in Galaxies (S4G), which have an approximate accuracy of 0.1\,dex \citep{Munoz-Mateos_13,Munoz-Mateos_2015}. For NGC\,7177, which is not included in S4G, the stellar mass is estimated from its $B$-band absolute magnitude in the HyperLEDA database, assuming the same 0.1\,dex uncertainty.

The large-scale images of each galaxy within our sample are provided in their respective sections in Appendix \ref{sec:notes-on-galaxies}, while their detailed analysis is explained. While images of IC\,2051, NGC\,289, NGC\,613, NGC\,1300, NGC\,1433, NGC\,1566, NGC\,4941 and NGC\,7513 are taken from The Carnegie-Irvine Galaxy Survey, CGS\footnote{The CGS sample images can be accessed via \url{https://cgs.obs.carnegiescience.edu/CGS/database_tables/sample0.html}} \citep{Ho_2011,Li_2011}, for the remaining galaxies in our sample, we use \textit{ugriz} images from SDSS\footnote{SDSS webpage can be accessed via the link \url{https://skyserver.sdss.org/dr16/en/home.aspx}} \citep{Strauss_2002,Blanton_2017}. 

\begin{table*}
\centering
\begin{threeparttable}
\caption{The galaxy sample studied in this work. NGC\,1097, whose analysis is presented in \citet{Kolcu_23} is included to this table for completeness.}
\label{tab:targets-general}
\setlength{\tabcolsep}{3pt}
\begin{tabular}{ccccccccccccc}
  \hline
  Galaxy ID & RC3 & $\rm \log_{10}(M_*/M_{\odot})$ & $\rm Vel_{sys}$ & Distance & Scale & Seeing & SR & Source & Proposal ID & Observing Period \\ 
   &  &  & $[{\rm km\,s^{-1}}]$ & $[{\rm Mpc}]$  & $[{\rm pc}/\arcsec]$ & $[\arcsec]$ & $[{\rm pc}]$ & & & \\
  (1) & (2) & (3) & (4) & (5) & (6) & (7) & (8) & (9) & (10) & (11) \\
  \hline
  IC\,2051   & SBbc      & 10.58 & 1741 & 19.9\textsuperscript{1} & 96.5 & 0.79           & 76.0 & CBS   & 0104.B-0404(A) & 104\\
  NGC\,289   & SBbc      & 10.63 & 1629 & 20.2\textsuperscript{1} & 97.9 & 0.70           & 69.0 & MAD   & 096.B-0309(A)  & 96 \\
  NGC\,613   & SBbc      & 11.09 & 1481 & 18.1\textsuperscript{1} & 87.8 & 1.00           & 88.0 & TIMER & 097.B-0640      & 97\\
    NGC\,1097  & SBb      &  11.24 & 1271 & 14.8\textsuperscript{1} & 71.7 & 0.80 & 58.0 & TIMER & 097.B-0640      & 97\\
  NGC\,1300  & SBbc      & 10.58 & 1573 & 19.6\textsuperscript{1} & 95.0 & 1.00           & 95.0 & TIMER & 097.B-0640      & 97\\
  NGC\,1433  & SB(r)ab   & 10.30 & 1074 & 18.6\textsuperscript{5} & 90.2 & 0.90           & 81.0 & TIMER & 097.B-0640      & 97\\
  NGC\,1566  & SABbc     & 10.58 & 1501 & 17.0\textsuperscript{1} & 82.4 & 0.56$^*$  & 46.0 & MAD   & 0100.B-0116(A)  & 100\\
  NGC\,3351  & SBb       & 10.49 &  779 & 10.0\textsuperscript{2} & 48.5 & 0.70           & 34.0 & TIMER & 097.B-0640      & 97 \\
  NGC\,3368  & SABab     & 10.72 &  882 & 10.5\textsuperscript{2} & 50.9 & 0.49$^*$  & 25.0 & CBS   & 0104.B-0404(A)  & 104 \\
  NGC\,3489  & SAB0      & 10.14 &  677 & 11.7\textsuperscript{4} & 56.7 & 0.52$^*$ & 30.0 & CBS   & 109.22VU.001    & 109\\
  NGC\,3626  & SAB0      & 10.59 & 1489 & 19.5\textsuperscript{3} & 94.5 & 0.48$^*$  & 46.0 & CBS   & 109.22VU.001    & 109 \\
  NGC\,4237  & SAbc      & 10.36 &  864 & 18.9\textsuperscript{3} & 91.6 & 0.44$^*$  & 40.0 & CBS   & 0104.B-0404(A)  & 104\\
  NGC\,4303  & SAB(rs)bc & 10.86 & 1562 & 17.0\textsuperscript{5} & 82.4 & 0.80           & 66.0 & TIMER & 097.B-0640      & 97\\
  NGC\,4321  & SABbc     & 10.93 & 1566 & 15.2\textsuperscript{2} & 73.7 & 0.86$^*$  & 64.0 & CBS   & 0104.B-0404(A)  & 104\\
  NGC\,4380  & SAb       & 10.32 &  952 & 15.9\textsuperscript{3} & 77.1 & 0.49$^*$  & 38.0 & CBS   & 0104.B-0404(A)  & 104\\
  NGC\,4457  & SAB0/a    & 10.58 &  880 & 16.5\textsuperscript{6} & 80.0 & 0.59$^*$  & 47.0 & CBS   & 109.22VU.001    & 109\\
  NGC\,4643  & SB0/a     & 11.03 & 1331 & 19.1\textsuperscript{1} & 92.6 & 0.90           & 84.0 & TIMER & 097.B-0640      & 97 \\
  NGC\,4941  & SABab     & 10.36 & 1113 & 15.6\textsuperscript{1} & 75.6 & 1.21           & 92.0 & MAD   & 096.B-0309(A)   & 96 \\
  NGC\,5248  & SABbc     & 10.67 & 1150 & 17.3\textsuperscript{1} & 83.9 & 0.70           & 59.0 & TIMER & 097.B-0640      & 97\\
  NGC\,7177  & SABb      & 10.44 & 1146 & 17.5\textsuperscript{1} & 84.8 & 0.89$^*$  & 76.0 & CBS   & 109.22VU.001    & 109\\
  NGC\,7513  & SBb       & 10.21 & 1566 & 19.8\textsuperscript{1} & 96.0 & 0.58$^*$  & 56.0 & CBS   & 109.22VU.001    & 109\\
  \hline
\end{tabular}
\begin{tablenotes}[para, flushleft]
\item Notes - (1) Galaxy; (2) RC3 classification \citep{deVaucouleurs_1991}; (3) Stellar mass from $\rm S^4G$; (4) Systemic velocity from NED; (5) Distance sources: 1=\,Virgocentric-corrected redshift from HyperLEDA with $H_0=72$; 2=\,Cepheids (metallicity-corrected) \citep{Freedman_01}; 3=\,SBF \citep{Cantiello_18}; 4=\,SBF \citep{Tonry_01} with correction from \citet{Mei_05}; 5=\,TRGB \citep{Anand_21}; 6=\,Virgo Cluster distances; (6) Spatial scale; (7) PSF FWHM (seeing), star denotes AO-corrected values following \citet{Fusco_2020}; (8) Spatial resolution; (9) IFU data source; (10) Proposal ID; (11) ESO observing period.
\end{tablenotes}
\end{threeparttable}
\end{table*}

\section{Extraction of the kinematic data}
\label{sec:data_extraction_MUSE}
To derive emission line fluxes and to extract stellar and ionised gas kinematics from MUSE data, we employed the Data Analysis Pipeline\footnote{\texttt{DAP} can be accessed at: \url{https://gitlab.com/francbelf/ifu-pipeline}} (\texttt{DAP}) of the PHANGS collaboration \citep{Emsellem_22}, which is an implementation based on the GIST pipeline \citep{Bittner_19}, an exceptionally functional software of modular structure. \texttt{DAP} employs the commonly used penalised pixel-fitting (\texttt{PPXF}) code \citep{Cappellari-emsellem-2004} to perform both stellar and emission line analysis. It can also perform stellar population analysis but to optimise the processing time we discarded this module, since it was outside of the scope of this work. 

\subsection{Extracting stellar line of sight velocity distribution moments}

We obtain the stellar kinematics in the form of line-of-sight velocity distribution moments (LOSVD) -- mean velocity (V), velocity dispersion ($\sigma$), and higher order Gauss-Hermite moments ($h3$ and $h4$).

The data are re-sampled on a logarithmic wavelength axis and then spatially binned with Voronoi tessellation \citep{Cappellari_Copin_03}. For the spectral fitting, we use a nominal rest-frame wavelength range of 4800--7000\AA. By using \texttt{PPXF}, we convolve EMILES single-stellar population (SSP) models \citep{Vazdekis_15} with the LOSVD, which are then fitted to each spectrum. The best-fitting parameters are determined by $\chi^2$ minimisation in pixel space. To exclude low-quality spaxels dominated by noise, we impose a \textcolor{black}{minimum SNR threshold of 3, consistent with commonly adopted IFU analysis setups (e.g. \citealp{Bittner_19}), where spaxels below the isophotal detection limit (typically $\langle{\rm SNR}\rangle \approx 3$) are excluded to avoid low-surface-brightness systematics.} Additionally, each Voronoi bin is targeted to achieve a SNR of 40, a commonly used value in the literature that balances the accuracy of the extracted measurements with spatial resolution \citep{van_der_Marel_1993}.

We employed 8$^{th}$ order multiplicative Legendre polynomials in \texttt{PPXF} to fit the stellar continuum, which corrects inaccuracies in the spectral calibration and eliminates the need to provide dust reddening curves \citep{Cappellari_2017} in fitting. While additive polynomials can reduce discrepancies between stellar templates and absorption lines, and remove artefacts caused by sky subtraction \citep{Cappellari_2017}, they can also alter the equivalent width of the Balmer absorption lines and introduce non-physical corrections to line fluxes \citep{Emsellem_22}. Therefore, we chose not to include additive polynomials in our fitting process.

\begin{figure}
        \hspace{-0.25cm}
        \includegraphics[width=0.505\textwidth]{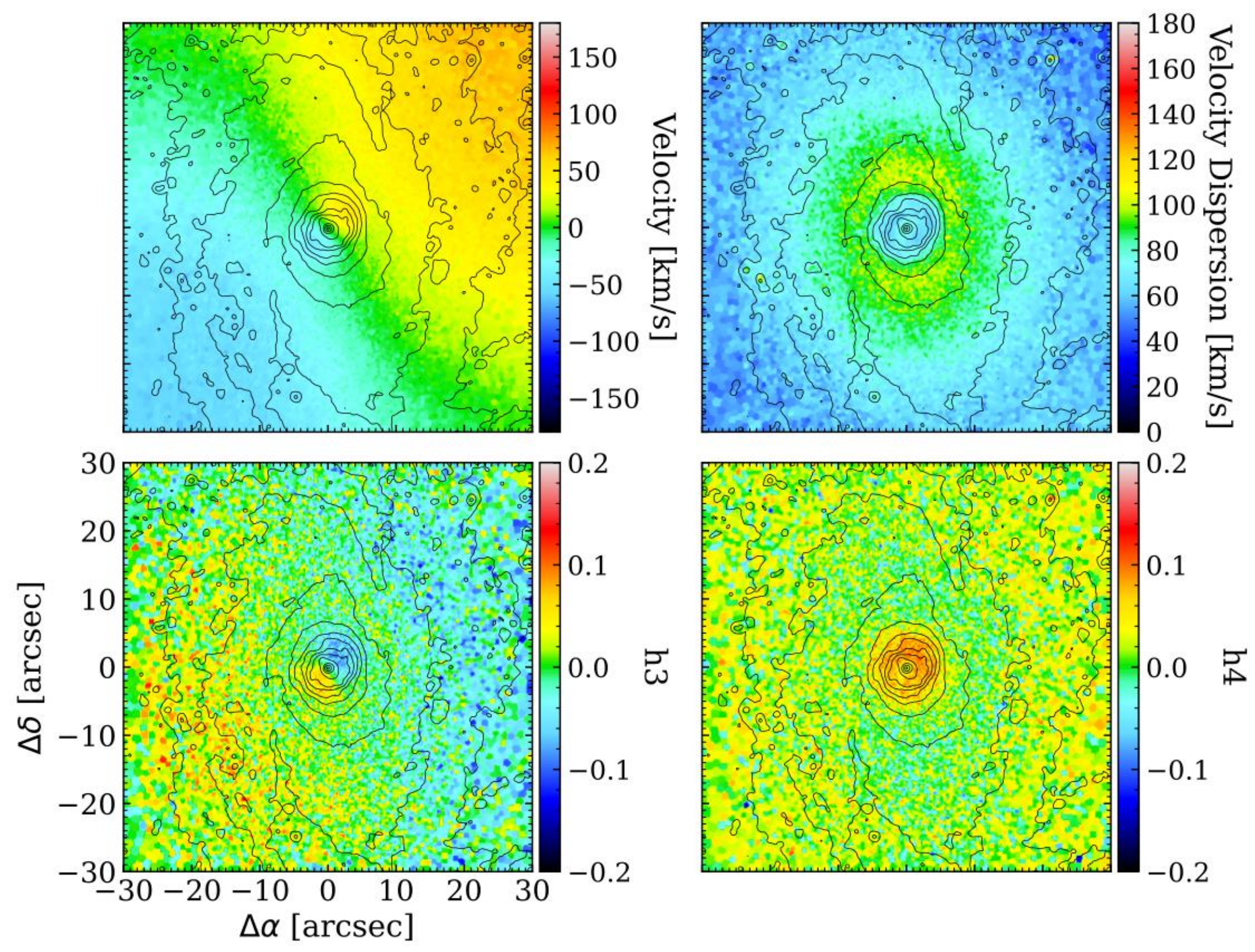}
        \caption{ Stellar kinematic maps of NGC\,4303. Panels show stellar velocity (top left), stellar velocity dispersion (top right), $h3$ (bottom left) and $h4$ (bottom right) moments of LOSVD. The black contours represent the continuum flux. Galaxy centre is at $\Delta\alpha$,$\Delta\delta$=[0,0].}
        \label{fig:NGC4303_stellar_conscsie}
\end{figure}

The maps of mean velocity (V), velocity dispersion ($\rm \sigma$) and higher order Gauss-Hermite moments $h3$ and $h4$ for a representative galaxy from our sample, NGC\,4303, are shown in Fig.\,\ref{fig:NGC4303_stellar_conscsie}, with the continuum flux overplotted. The mean velocity in each spaxel is corrected for the galaxy's systemic velocity, taken from the NASA/IPAC Extragalactic Database (NED\footnote{NED can be accessed via \url{https://ned.ipac.caltech.edu/}}), and provided in Table\,\ref{tab:targets-general}. The centre of galaxy is identified with the brightest spaxel in its nucleus and is located at $\Delta\alpha, \Delta\delta$ = [0, 0] on each map. As this paper does not focus on stellar kinematics, we do not present the stellar kinematic maps of the remaining galaxies in our sample unless they are essential for understanding gas dynamics.

\subsection{Extracting ionised gas line of sight velocity distribution moments for multicomponent gas}
\label{sec:method:kinematic-extraction-secondary-components}
To obtain the ionised gas kinematics, we initially fit single Gaussian functions to the emission lines while simultaneously modelling the stellar continuum. The emission-line fitting is carried out on individual spaxels to preserve spatial resolution. During these fits, the stellar continuum is included but its kinematic parameters (velocity and dispersion) are fixed to the values previously determined from Voronoi-binned stellar kinematic fits. \textcolor{black}{For each spectrum, pPXF performs a weighted $\chi^2$ minimisation using the per-pixel noise vector. Formal statistical uncertainties on gas kinematics and line fluxes are obtained from the covariance of the best-fitting solution (pPXF/GIST pipeline/Data Analysis pipeline), and thus depend on the line SNR and the spectral resolution through the adopted line profiles and instrumental broadening. We mask wavelength intervals affected by strong skylines and fit the stellar continuum and emission lines simultaneously to reduce biases from sky residuals and template mismatch; such systematics are not fully reflected in the formal error estimates}

We gathered the emission lines into three main groups, and among each group, the kinematics are shared. We left the emission line fluxes free from each other, except for the constraints from atomic physics \citep{Osterbrock_book}. The emission lines are grouped as follows:

\begin{enumerate} [align=left, leftmargin=*]
\item Balmer lines: $\rm H\beta\lambda4861$, $\rm H\alpha\lambda6562$ (hereafter $\rm H\beta$ and \ha)
\item Low-ionisation lines: [N I]$\lambda\lambda$5197,5200; [N\,II]$\lambda$5754; [O I]$\lambda\lambda$6300,64; [N\,II]$\lambda\lambda$6548,83;  [S II]$\lambda\lambda$6717,31
\item High-ionisation lines: [O\,III]$\lambda\lambda$4959,5007; [S III]$\lambda$6312
\end{enumerate}

The wavelengths singled out in each emission line group are taken from the National Institute of Standards and Technology (NIST)\footnote{ NIST can be accessed from https://www.nist.gov/pml/atomic-spectra-database.}.

\subsubsection{Fitting emission lines with multiple Gaussians}
\label{sec:excess_masking}
As we presented in Paper IV, our pilot study of NGC\,1097 revealed a significant limitation of the standard approach of fitting emission lines with single Gaussians, when the line contains emission coming from different sources along the line of sight (LOS), such as from the disk and from the AGN outflows. When one attempts to fit such a line with a single Gaussian, the resultant fit is poor, which can artificially increase the velocity dispersion and alter the velocity. Such inaccuracies could significantly impact our work, particularly when constructing diagnostic maps of residual velocity and velocity differences, in which we search for the kinematic signatures of extended shocks (see Sect.\,\ref{sec:method}).

Therefore, for spaxels that show unusually high velocity dispersion, compared to their surroundings, for any group of emission lines, we carefully examined the spectra for signs of multiple sources of emission which are expected to show up as an \textit{excess emission} from the weaker source superimposed on the Gaussian profile of the emission from the stronger source. Discovering such emission signature prompts us to apply the \textit{secondary component fitting approach}, introduced in Paper IV. When emission from two distinct sources is present along the LOS, fitting a secondary Gaussian component allows to separate out the emission attributed to excess sources and thus get a better estimate of the kinematics of the main source.

\textcolor{black}{Galaxies studied in this work} exhibited excess emission in their spectra that displayed more complex line features than those observed in NGC\,1097's spectra in our pilot study. Therefore, we have found it necessary to tailor certain aspects of the secondary component fitting to accommodate the unique characteristics of each individual galaxy, which we explain below.

During this tailored emission line fitting process, we start with the same assumption that is made in the analysis of NGC\,1097. We assume that the excess emission to which the secondary component is fitted comes from the same emission line to which the main component is fitted but either blue or redshifted with respect to the main component. If the excess emission is present in more than one emission line, we use the following configuration within DAP to incorporate secondary components to the emission line fitting: if excess emission is present in emission lines from different groups defined above, the kinematics of secondary components are determined independently for each group. For example, in NGC\,4303 (see Appendix\,\ref{sec:notes-on-galaxies}, Fig.\,\ref{fig:NGC4303_mask}), a secondary component fitted to \nii excess emission and a secondary component fitted to \ha excess emission are independent from each other. However, when excess emission is significant in more than one emission line within the same emission line group, such as [N\,II] doublets, the kinematics and the line ratios of the fitted secondary components are tied together, analogous to the approach used for tying the main component emission.

\textcolor{black}{Although we have tried fitting the secondary components for various emission lines that show excess, only the one fitted to \nii changed the derived kinematics. In Paper IV, fitting of the secondary component to \nii not only influenced the kinematics derived for \nii but also reduced the \ha velocity dispersion by $\sim 20\%$ and shifted its LOS velocity by roughly 20\kms. In this work, we find that in the three galaxies in our sample, in which a significant excess emission is present, including secondary component to \nii reduced the velocity dispersion of \nii by $\sim 15-25\%$ and shifted the LOS velocity by 20\kms up to 60\kms in some spaxels at certain cases (see NGC\,4303, Appendix\,\ref{sec:NGC4303_notes}). In these three galaxies, unlike in NGC\,1097 shown in Paper IV, \ha kinematics were not significantly affected in by fitting of the secondary component. Moreover,}  none of the galaxies in our sample showed significant excess emission in \oiii, even in the regions with high \oiii velocity dispersions. For instance, slightly non-Gaussian \oiii profiles seen in NGC\,4303's spectra are caused by a weak excess emission but including a secondary component to the fitting did not significantly influence the derived measurements. \textcolor{black}{This suggests that \oiii measurements from single Gaussian fits are in general reliable.} Thus, to simplify the analysis and to reduce the number of free parameters in the fit, we omitted including a secondary component for \oiii in these cases, this decision is elaborated further in Appendix\,\ref{sec:notes-on-galaxies}.

\subsubsection{Creating LOSVD moments' maps of multicomponent gas}
Although excess emission can significantly affect the measured velocity and velocity dispersion values of the main component, it does so for a small number of spaxels, generally grouped in a specific region of a galaxy. To identify which spaxels are most significantly affected by excess emission, we first analysed the differences between measurements from single Gaussian fits and those including secondary components. We focused on detecting large changes in the velocity and velocity dispersion measurements of the main emission line component caused by the introduction of a secondary Gaussian component. As in Paper IV, in order to construct the most accurate kinematic maps of the main emission component, we created masks, where these changes exceed certain magnitude. Within the mask, kinematic measurements for the main component were taken from the fits that included a secondary component. Outside the mask, kinematic measurements are made by fitting a single Gaussian to each emission line.

\textcolor{black}{As the excess emission significantly affected only the measurements of \nii, when establishing the masking criteria for the affected galaxies, we only used the \nii velocity shift and velocity dispersion change conditions.} We refined the masking criteria employed in Paper IV by placing the following two conditions, which have to be satisfied for a spaxel to be included in the mask:

\begin{enumerate}[label=\roman*., align=left, leftmargin=*]
    \item there is a significant ($\geq$10\kms) \nii velocity shift or a substantial ($\geq$15\%) decrease in \nii velocity dispersion between the single Gaussian fit and the fit including the secondary component
    \item the velocity dispersion of the secondary \nii component is not excessively high, ensuring that the secondary component is a representation of real emission rather than noise.
\end{enumerate}

For each of the three affected galaxies, the specific values chosen for the conditions listed above are detailed in their respective subsections in Appendix\,\ref{sec:notes-on-galaxies}. Detailed information on the fitting approach of the secondary component, mask conditions, and further kinematic analysis for each of the three galaxies exhibiting excess emission in their spectra is provided in their respective subsections in Appendix\,\ref{sec:notes-on-galaxies} (Sects.\,\ref{sec:NGC1433_notes}, \ref{sec:NGC4303_notes} and \ref{sec:NGC4321_notes}). 

\begin{figure*}
    \centering
    \includegraphics[width=1\textwidth]{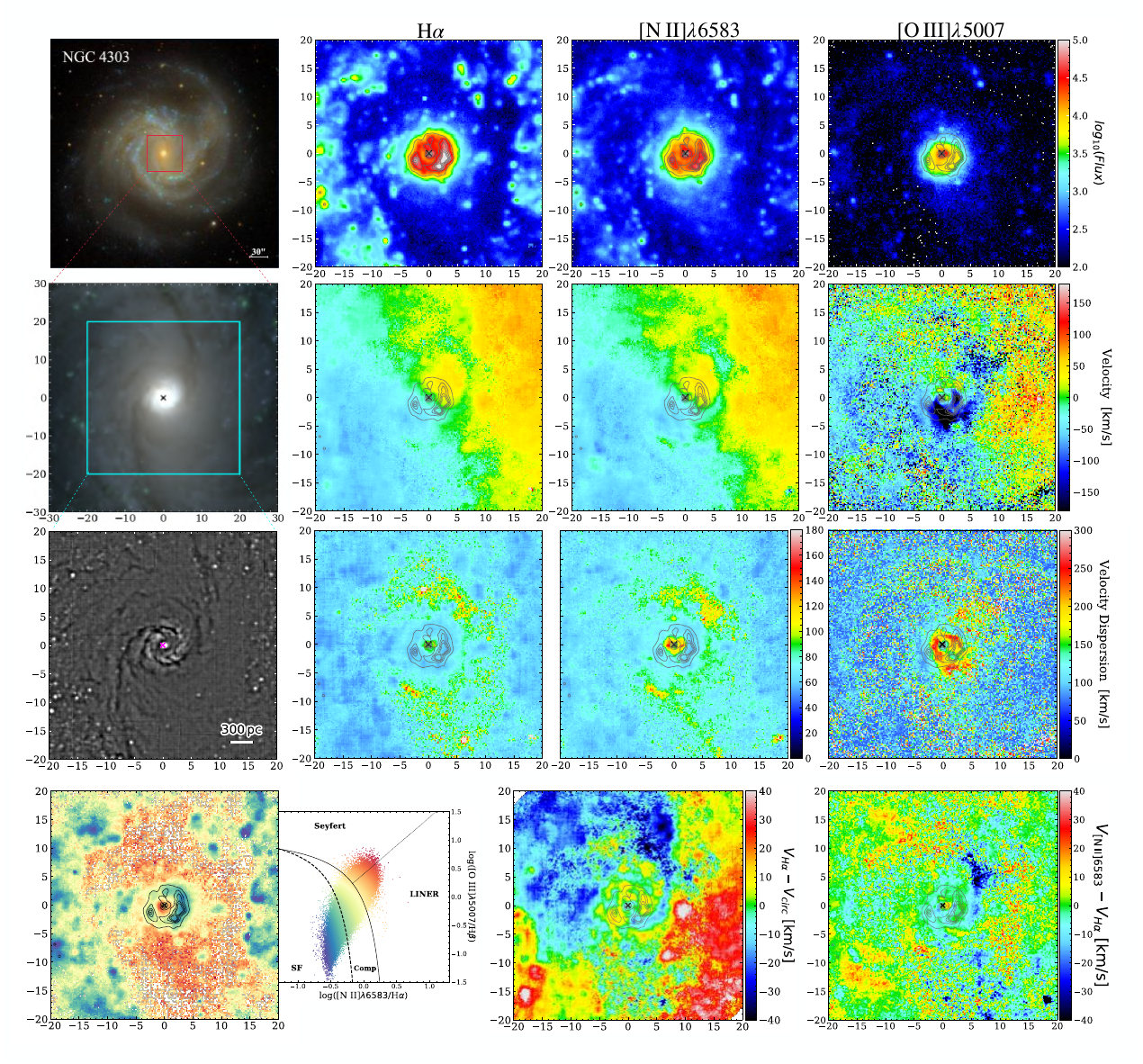}
    \caption{
    Images and diagnostic maps of NGC\,4303.  Unless stated otherwise, the x and y axes give positional offsets from the galaxy centre ($\Delta\alpha$, $\Delta\delta$) in arcseconds. The first three panels in the left-hand column show a large-scale image (taken either from CGS or SDSS) with the MUSE field of view marked as red box (top), a MUSE-based colour image with $40\arcsec \times 40\arcsec$ region marked with a cyan box (middle), and an unsharp mask image enclosed in cyan box region, highlighting dust features (bottom). All other panels show the region marked with a cyan box: top row -- H$\alpha$, \nii, \oiii flux maps; second row -- velocity maps; third row--velocity dispersion maps. The bottom row shows the BPT map with its diagnostic diagram (see Sect.\,\ref{met:bpt}), the residual velocity map (see Sect.\,\ref{met:residual_velocities}), and the velocity difference map (see Sect.\,\ref{met:velocity_differences}). The gray contours trace the \ha flux.}
    \label{fig:NGC4303}
\end{figure*}

For each galaxy in our sample, the flux, velocity and velocity dispersion maps across the three emission line groups are presented in the nine top-right panels of Figures A1-A20 in Appendix\,\ref{sec:notes-on-galaxies}. In the same panels of Fig.\,\ref{fig:NGC4303}, these results are presented for the example of NGC\,4303. For comparison, the kinematic maps of the three galaxies with excess emission, derived from single Gaussian fitting are presented in Fig.\,\ref{fig:NGC1433_NGC4303_NGC4321_R1} of Appendix\,\ref{app:suplementary_sample_plots}. The mean velocity in each spaxel is corrected for the galaxy’s systemic velocity, taken from the NASA/IPAC Extragalactic Database (NED) and provided in Table\,\ref{tab:targets-general}. The centre of each galaxy is identified with the brightest spaxel in continuum emission in its nucleus and is located at $\Delta\alpha, \Delta\delta$ = [0, 0] on each map.

The refined fitting approach presented above significantly enhances the accuracy of the kinematic measurements extracted from MUSE data cubes, thereby minimising errors associated with excess emission in the diagnostic maps we construct to search for the kinematic signatures of extended shocks. Consequently, this approach substantially improves the quality of the analysis and enhances the robustness of our findings.

\section{Methodology}
\label{sec:method}

\textcolor{black}{Orbital theory and hydrodynamical simulations indicate that shocks can play an important role in driving gas inwards towards galactic nuclei \citep{Athanassoula_1992, Maciejewski_2004b}. Observationally, however, identifying such shocks is not straightforward. Although dust morphology can reveal structures that are suggestive of shock compression  \citep{Martini_1999,Pogge_02}, morphology alone cannot determine whether these features truly correspond to shocks in the gas or instead arise from other processes unrelated to inflow. Kinematic information is therefore essential, since shocks are expected to produce rapid changes in velocity and hence extended signatures superimposed on the overall rotation pattern of the disc. 
\\ \\
In Paper IV, we showed that residual velocity maps are a powerful diagnostic of such structures, but also that relying on a single method may be insufficient, because residual maps can be affected by uncertainties in the disc model and by vertical gas motions. For this reason, we combine complementary diagnostics based on kinematics, dust morphology, and ionisation maps in order to identify and characterise signatures of extended shocks in the central regions of our galaxies. The subsections that follow describe the techniques used in each approach, explain their complementarity, and summarise the logic of our identification strategy through a decision tree.}

\subsection{Techniques for identifying and characterising signatures of extended shocks}
\subsubsection{A search for coherent kinematic structures in deviations from circular motion}
Signatures of extended shocks are best highlighted in \textbf{residual velocity maps} \label{met:residual_velocities}\textcolor{black}{created by fitting to the observed velocity field an axisymmetric rotating disk model, and then subtracting the velocity field of such model, which is referred as $V_{circ}$ in Figs.\,\ref{fig:NGC4303} and \ref{fig:ic2051_4by4}--\ref{fig:NGC7513}, from the observed velocity field.} Coherent structures in the residual velocity map that exhibit large magnitudes disclose the presence of extended shocks because shocks in gas cause velocity discontinuities.

For each galaxy, we model a thin flat rotating disk in circular motion by fitting the observed velocity field of \ha, using the \texttt{Kinemetry} software \citep{Krajnovic_06}. \texttt{Kinemetry} is specifically designed for IFU data and performs a harmonic expansion along the best-fitting ellipses. The residual velocities highly depend on the disk orientation parameters defined in the disk model: the Position Angle (PA) of the line of nodes (LON) and the inclination (\textit{i}). If the LOS velocity of a wrongly fitted disk model is subtracted from the observed LOS velocity, artefacts will arise in the residual velocity map \citep{van_der_Kruit-Allen-1978}  and will obscure the real features coming from noncircular motion and velocity jumps. Therefore, to ensure the artefacts in the residual velocity field are not misinterpreted for real features, one must be careful while fitting the disk and examining the resultant maps.

As we explained thoroughly in Paper IV, to accurately model the thin flat rotating disk, one can perform three \texttt{Kinemetry} fits, by gradually fixing the global disk orientation parameters. The first fit allows all ellipse parameters to vary, but we showed that the derived radial distributions of parameters do not significantly vary between the runs with the centre fixed or free. Thus, to optimise the analysis we omitted this first Kinemetry fit and we ran \texttt{Kinemetry} for a fixed centre and the additive constant in the expansion being equal to the velocity measured at the centre.
We fix the centre at the photometric centre of each galaxy based on the brightest spaxel in its nucleus. 

From the ellipse parameter distributions obtained in this initial fit, we deduce optimal values for the PA of the LON and \textit{i}, which can accurately describe the galaxy disc, particularly within the inner kpc regions. Lastly, we construct our final disk model with centre, PA of LON and \textit{i} fixed to those deduced based on the previous fits. Due to insufficient data within ellipses in the inner 2\arcsec, and the influence of the AGN for AGN host galaxies, we consider the estimated parameters within this radius as unreliable and therefore discard them.
\begin{figure*}
\centering
    \includegraphics[width=1.02\textwidth]{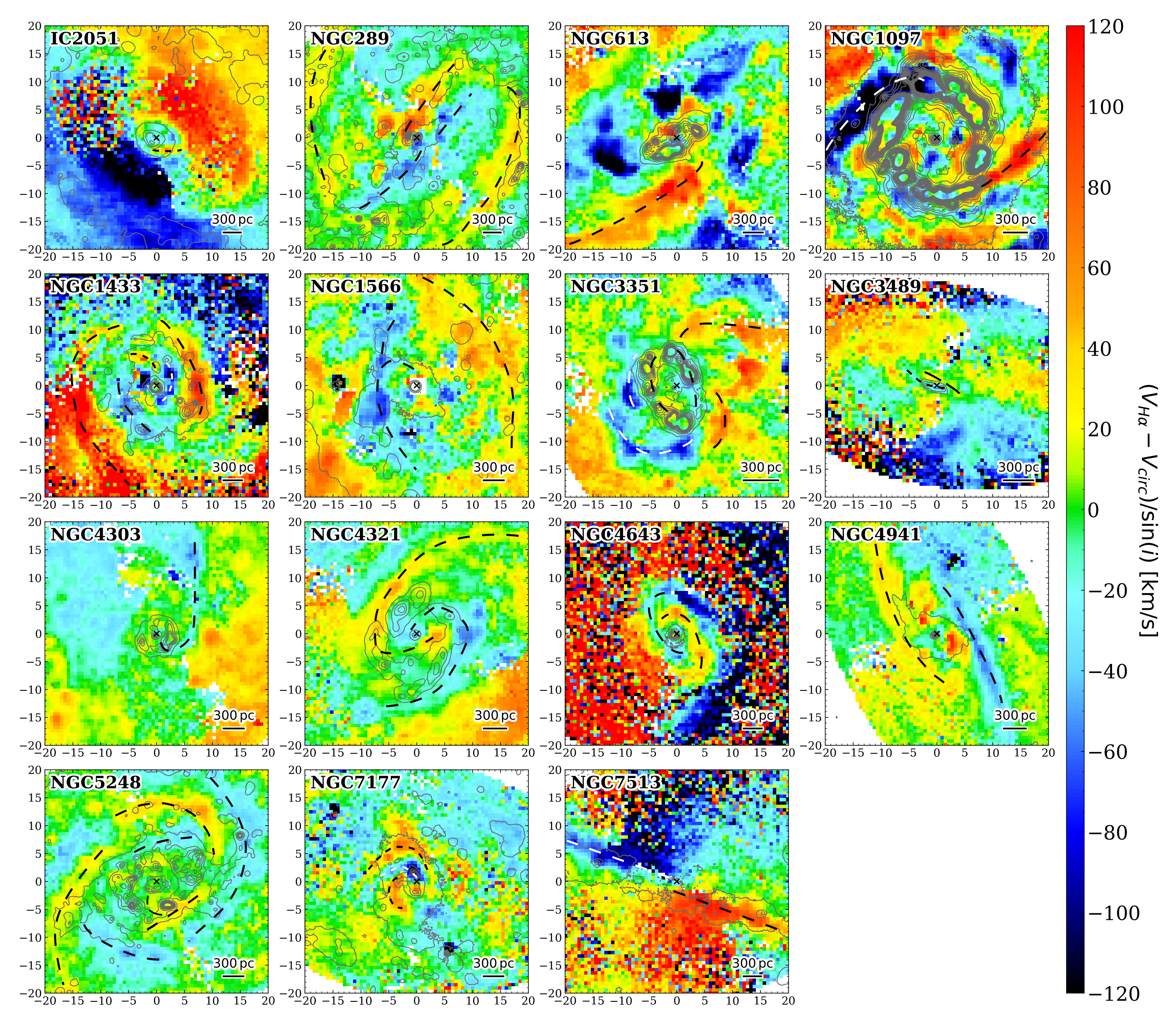}
    \caption{\textcolor{black}{Residual velocity maps of galaxies exhibiting coherent extended structures (see Sect.~\ref{met:residual_velocities}). Spaxels with residual velocity measurements below $2\sigma$ significance are masked. The galaxy centres are marked with an ``x". Black/white dashed lines indicate the approximate locations of the extended kinematic features. The value of each spaxel is the average over $3\times3$ spaxels' box centred on it. The grey contours represent the \ha emission.}}
    \label{fig:Sample_extended_structures_perturbed_galaxies_residualvel}
\end{figure*}

\begin{figure*}
\centering
    \includegraphics[width=1\textwidth]{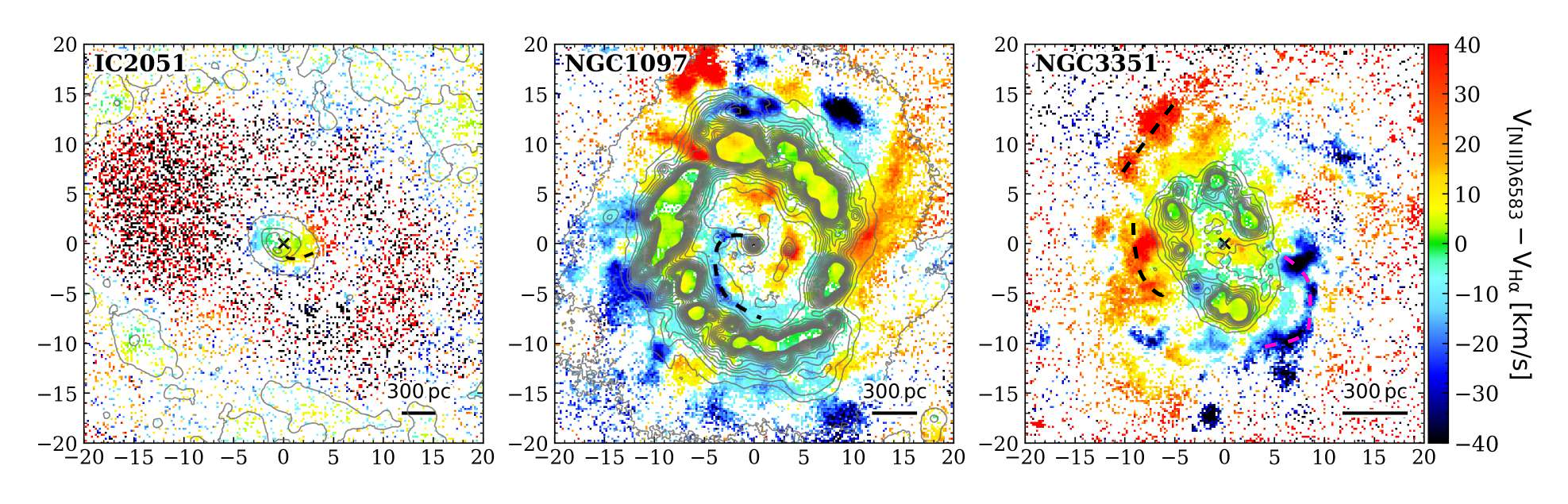}
    \caption{\textcolor{black}{Velocity difference maps of galaxies exhibiting coherent extended structures. Spaxels with residual velocity measurements below $2\sigma$ significance are masked. The galaxy centres are marked with an ``x". Black/magenta dashed lines indicate the approximate locations of the extended kinematic features. Grey contours show the \ha flux distribution. These maps were used to support and refine the perturbation classifications assigned to the sample (see Sects.\,\ref{met:velocity_differences} and \ref{sec:met:criteria}).}}
    \label{fig:Sample_extended_structures_perturbed_galaxies_velocitydifference}
\end{figure*}

We note that our application of \texttt{Kinemetry} assumes that the gas flow is intrinsically circular. For galaxies whose gas flow is dominated by local (e.g., stellar outflows) motions, the distributions of PA of the LON and \textit{i} derived from \texttt{Kinemetry} will show large fluctuations and settling on optimal values based on such fluctuated distributions will be challenging. Consequently, wrongly fitted parameters in the disk model will raise certain harmonic terms in residual velocity maps. For instance, an incorrect inclination can lead to $m$=3 harmonic terms, while an incorrect PA of the LON can result in $m$=1 harmonic terms \citep{van_der_Kruit-Allen-1978}. We  show and explain this effect in Sects.\,3.1 and 4.1 of Paper IV. If not taken into account with care, such artefacts can be mistakenly interpreted as real features.

Conversely, identifying the artefacts in residual velocity maps can guide us to determine the disk parameters that need to be adjusted. Thus, for galaxies whose disk orientation is not well constrained by \texttt{Kinemetry}, we first define an initial PA of the LON and \textit{i} based on the parameters computed from large-scale studies. Next, we alter these initial parameters within a range deduced from the fluctuated distributions given by \texttt{Kinemetry}. With this approach, we construct different disk models and the corresponding residual velocity maps, in which we search for the signs of artefacts. As for the optimal disk orientation parameters, we settle on the values that minimise the strength of the artefacts in the residual velocity maps. After the rotating disk in circular motion is modelled, we subtract its LOS velocity from the observed LOS velocity to construct the residual velocity map. 

\textcolor{black}{Residual velocity maps showing extended structures are presented in Fig.\,\ref{fig:Sample_extended_structures_perturbed_galaxies_residualvel}, where the deprojected residual velocity values (residual velocity corrected for $\sin i$) are displayed. In these maps, we search for coherent structures that extend over large spatial scales, similar to those observed in bars and nuclear spirals \citep{Zurita_04, Fathi_05, Blasco-Herrera_2013, Schnorr-Muller_2016, Schnorr-Muller_2017}. The figure includes only galaxies in which such extended coherent structures are identified (which are outlined with dashed lines), and therefore serves as a representative example of the features targeted in our analysis. However,} it is important to note that not all extended structures in the residual velocity map are associated with extended shocks in gas. \textcolor{black}{To emphasise, the galaxies whose deprojected residual maps are presented in Fig.\,\ref{fig:Sample_extended_structures_perturbed_galaxies_residualvel}, are later grouped under different kinematic perturbation categories and not all structures are associated with extended shocks.} Shock conditions imply a large change in velocity \citep{Athanassoula_1992,Maciejewski_2004b}. Thus, besides their coherent appearance, these structures should also have large residual velocity magnitudes, indicating large velocity jumps characteristic for shocks. For instance, coherent structures in residual velocity field presented in \citet{Davies_09} which are associated with extended shocks, exhibit the LOS residual velocities of magnitudes 30\kms or higher, corresponding to deprojected velocity magnitudes of $\sim$40\kms or higher. These magnitudes are indicative of a nuclear spiral shock that causes a velocity jump of approximately 70\kms between the minimum and maximum velocity as shown in \citet{Maciejewski_2004b}. As residual velocities describe the oscillations around the mean, their amplitude is half of the velocity jump. Furthermore, \citet{Athanassoula_1992} also resolves a velocity jump of $\sim$100\kms when extended shocks are present, indicating deprojected velocity magnitudes around $\sim$50\kms in extended shocks. In order to determine whether the coherent structures found in the velocity residual maps can be caused by extended shocks in gas, we examine the magnitude of velocity residuals in them. These results are presented in {\textcolor{black}{Sect.\,\ref{sec:determining_velocity_jumps}}.

\subsubsection{Coherent structures in maps of velocity difference between distinct emission lines}
\label{met:velocity_differences}
In Paper IV, we introduced a complementary approach to finding extended shocks in gas. It is based on the simple idea of subtracting the velocity fields of two different tracers. Thus in this approach, we construct \textbf{the velocity difference maps} by subtracting the LOS velocity of \ha from the LOS velocity of \nii. For each galaxy, these are presented in Appendix\,\ref{sec:notes-on-galaxies}, and examples are given in the bottom-right panel of Fig.\,\ref{fig:NGC4303} \textcolor{black}{and in Fig\,\ref{fig:Sample_extended_structures_perturbed_galaxies_velocitydifference} for different galaxies}. This method purely depends on the observed data and involves no contribution from kinematic or mass models. Therefore, the constructed velocity difference maps do not reflect artefacts from any underlying model and only highlight features present in the data.

\textcolor{black}{This approach stems from the notion that the line profile recorded at any spaxel of a galaxy dataset is a convolution of emission from all regions of gas along the LOS towards this spaxel, and each region may have different physical conditions and may move with respect to the others. If physical conditions in these regions differ, then the relative contribution of each region to the emission along this LOS may differ between different emission lines.} Thus, the integrated line profile, and therefore the LOSVD recorded in each spaxel may vary between the emission lines. As a result, the evaluated value of its first moment (velocity) may also differ depending on the tracer. Areas upstream and downstream of a shock in gas are a good example of two regions of gas that differ in physical conditions and move with respect to one another. As jump conditions in the shock impose a velocity difference between these two regions, for a spaxel which records emission from pre-shock and post-shock regions, the derived velocities may vary among the tracers. When extended shocks are present, then in many adjacent spaxels the velocity difference takes consistently positive (or negative) values, and therefore extended shocks may give rise to continuous structures in velocity difference maps. 

\textcolor{black}{Note that when this scenario is applied to our data, the extended shocks will not always result in a velocity difference between the tracers. No velocity difference will be observed if different physical conditions in pre- and post-shock gas do not affect emission from the tracers, or if there is no significant difference in the LOS velocity between pre- and post-shock gas. Therefore, this method of finding extended shocks, although not dependent on galaxy models, can only be used in addition to others methods. In particular, velocity residuals, which are departures from circular motion in the disc plane, can be distorted in the presence of vertical motions, while the velocity differences do not assume a rotating disc. Note also that while the amplitude of the velocity residuals has a clear interpretation, interpreting the amplitude of velocity differences would require detailed modelling of emission from shock-excited plasma. \textcolor{black}{Examples of extended structures found in velocity difference maps are outlined in Fig.\,\ref{fig:Sample_extended_structures_perturbed_galaxies_velocitydifference} with dashed lines.}}

\subsubsection{Determining the role of SF and AGN using BPT maps}
\label{met:bpt}
To complement the kinematic analysis explained in the previous subsections, we study types of ionisation mechanisms prevailing in the structures that we find in the central regions of galaxies. \textcolor{black}{The diffuse ionized emission that we detect originates largely from the diffuse ionized gas, sustained by AGN, SF and shocks in the turbulent ISM.} To distinguish the dominating source of ionisation, such as AGN or star formation regions, from optical emission lines we constructed the Baldwin, Phillips, Terlevich (BPT) diagrams \citep{Baldwin_1981} for each galaxy within our sample. Each point on the BPT diagram corresponds to one spaxel in the data, and the position of each point is determined by fluxes measured for that spaxel. In order to exclude spaxels that are predominantly influenced by the noise, we implemented a threshold amplitude over noise (AON) \citep{Sarzi_06} of \hb, \ha, \nii and \oiii emission lines. Spaxels with an AON for any of the four emission lines below the defined threshold were discarded. We avoid using the same AON value for all targets when constructing the BPT diagrams, as it could lead to a loss of spaxels with useful information if not applied carefully. Therefore, an arbitrary AON threshold is applied in order to prioritise a balance between maximizing the number of the displayed spaxels with information and minimising noise levels in each map.

To differentiate between ionisation driven by star-forming regions and AGN, we used the \citet{Kewley_01a} (hereafter Kewley+01) relation which indicates what line ratios can be produced by stellar photoionization, as for points above the Kewley+01 line emission likely originates from shocks or AGN. We note that the Kewley+01 relation does not account for the composite emission with ionisation from both SF and AGN. Therefore, we also include the modified Kewley+01 line from \citet{Kauffmann_03} (hereafter Kauffmann+03) relation, so that the region between Kaufmann+03 and Kewley+01 lines is the location of the composite emission, and the points below the Kaufmann+03 line show ionisation from pure SF. We applied a further separation in the AGN region by using the \citet{Cid_Fernandes_10} relation, which divides the region into two: low-ionisation nuclear emission-line region (LINER) and Seyfert region.

As explained above, each point on the BPT diagram represents a spaxel in the data. These points are colour coded according to their distance from the Kewley+01 line. We imposed the same colour code on the spatial maps of galaxies, producing the \textbf{BPT maps} to highlight different ionisation mechanisms in every location of each galaxy. Note that our BPT maps do not differentiate between LINER and Seyfert excitation, but the accompanying BPT diagram indicates which AGN excitation mechanism dominates overall. The BPT diagram and BPT map of each galaxy we studied in this work are presented in Appendix\,\ref{sec:notes-on-galaxies}, and examples of them are given in the bottom-left panel of Fig.\,\ref{fig:NGC4303}. Examining the velocity difference and residual velocity maps together with the BPT maps helps us in interpreting the origin of the observed coherent kinematic structures. 

\subsubsection{Studying the dust morphology in unsharp mask images}
\label{method:dustmorphology}

\textcolor{black}{Gas and dust are generally well mixed in the interstellar medium, so large-scale shocks that compress the gas are also expected to compress and redistribute the dust along the same structure. In this context, the alignment of coherent kinematic structures with large residual velocity magnitudes and dust lanes strengthens the interpretation that extended shocks are present. In particular, when the kinematic signature appears fragmented or partially discontinuous, coherent dust lanes seen in unsharp-masked images can provide useful supporting evidence that separated kinematic segments may trace different parts of the same larger-scale shock structure, rather than unrelated local disturbances.} However, it is important to note that dust morphology alone is not sufficient to build a definitive argument for the presence of extended shocks. This is because the coherent morphology of dust structures can also arise from the differential rotation, unrelated to shocks in gas.

To study the morphology of dust structures, we constructed \textbf{{unsharp mask images}} using a series of image processing techniques applied to MUSE images. Initially, we created red MUSE images by isolating emissions from redder wavelengths ($\sim$7800--9350\AA). We then applied a Gaussian blur technique to the red images. This step is crucial as it reduces noise and smoothens the details in the images, enhancing the signal-to-noise ratio for subsequent analysis. For each galaxy, we adjusted the width of the Gaussian to emphasise dust structures within the innermost regions. While this application made inner dust structures clearer, it also weakened the visibility of dust lanes along the bars of galaxies, which ideally would require a wider Gaussian. Subsequently, we applied the unsharp masking technique, which involves subtracting the blurred version of the image from the original, to produce images that accentuate the edges and fine details of the dust structures.

Although this unsharp masking approach was not included in our pilot study of NGC\,1097 in Paper IV, we examined there the dust structures inwards from the nuclear ring using a NIR J-band image from \citet{Prieto_05}. While for NGC\,1097 we relied on NACO observations, such observations were not available for other galaxies of this sample. Therefore, to study dust structures in the remaining galaxies we used MUSE observations. Alongside other diagnostic maps, the unsharp mask image of each galaxy we studied in this work is presented in Appendix\,\ref{sec:notes-on-galaxies}, with an example given in the first panel of the third row of Fig.\,\ref{fig:NGC4303}. The arrangement of Fig.\,\ref{fig:NGC4303}, and of analagous figures in Appendix\,\ref{sec:notes-on-galaxies} for the remaining galaxies of the sample, allows the reader to identify structures common in different approaches and facilitates following of the analysis of individual galaxies.

\subsection{Complementarity of approaches used to identify signatures of extended shocks}
\label{sec:complementing_approaches_for_shocks}
As explained in Sect.\,\ref{sec:method}, we used various methodologies to reveal and study the coherent kinematic structures that might indicate the presence of  shocks in gas. It is essential to emphasise that relying solely on one method might not be sufficient to definitively confirm or dismiss the presence of shocks.

For instance, in the residual velocity maps, which is the classical approach used in various other studies (e.g., \citealp{Fathi_05,Fathi_06,Izumi_13}), some coherent kinematic structures may be artefacts of subtracting the model of circular disc with incorrect orientation \citep{van_der_Kruit-Allen-1978}, and estimating the accurate orientation of a disk model might be challenging when global non-circular motions dominate the gas flow (see Paper IV). Such artefacts can obscure real coherent features present in the data.  In addition, even in the absence of coherent kinematic structures in the residual velocity maps, shocks can still be deduced from extended regions of significant velocity difference between distinct emission lines, such as \ha and \nii. 

Even though we expect to see alignments between the structures observed in residual velocity maps and those in velocity difference maps, it is important to note that these two methods have different sensitivities to the vertical motion of the gas. The residual velocity maps depend on a model, which assumes gas in circular motion confined to a thin disk. This means that if the gas has some vertical motion, like in outflows, coherent structures may appear in the residual velocity map. However, if such outflow is seen in both \ha and \nii emission, no structures may be seen in the velocity difference maps. This effect is seen in the diagnostic maps of NGC\,3351 (see Fig.\,\ref{fig:NGC3351} in Appendix\,\ref{sec:NGC3351_notes}). While the residual velocity map reveals the coherent ``L" shape structure of the transverse dust lane of the galaxy, which is associated with outflowing gas from the galaxy's nuclear ring \citep{Leaman_19}, this structure is absent in the velocity difference maps, which show a pair of red and blue lanes consistent with the classical picture of gas inflow in a bar.

As the velocity residuals and differences might be disguised in the presence of strong SF or AGN activity that dominates the velocity field, to complement our kinematic analyses, we studied the BPT map of each galaxy \textcolor{black}{to pinpoint locations where gas kinematics may be affected by SF or AGN.} However, we note that this approach can be limited if irradiation from the AGN dominates the central regions of the galaxies \textcolor{black}{(see e.g. NGC\,3489 in Sect.\,\ref{sec:NGC3489_notes}).}

Finding coherent dust structures in the unsharp mask images can provide additional support for the existence of  shocks, as shocks are expected to enhance the dust density. However, it is not expected for coherent structures seen in either residual velocity or velocity difference maps to align with dust structures over large azimuthal extents. This is because velocity projection convolves the in-plane velocity structure with a sinusoidal modulation along the azimuth, resulting in ``one-less" kinematic multiplicity of extended structures \citep{Canzian_1993,Schoenmakers_1997}. Thus for example a velocity jump in a two-arm morphological
spiral will appear as a one-arm kinematic spiral of different pitch angle, hence the morphological and kinematic spirals cannot overlap over
large azimuthal extents, despite well overlapping in e.g. bar lanes, which are mostly radial and therefore do not extend over large azimuth. We expect good alignment for structures extending up to 90\degr in azimuth. On the other hand, if the region has low optical depth (small amount of dust), the presence of continuous dust structure may not be apparent at all even when kinematic structures are evident. 

Continuous dust structures can also originate from molecular clouds stretched by differential rotation, indicating that not all coherent dust filaments necessarily have a shock origin. One might also argue that dust obscuration in coherent filaments could produce coherent structures in residual velocity maps even in the absence of such structures in the actual gas velocity. This is possible in an optically thick disc, whose rotation depends on the height above the disc, because along the line of sight containing the dust structure emission may be dominated by regions located higher above the disc, and therefore rotating with different velocity. The velocity difference maps would not be affected by such obscuration as strongly as velocity residual maps since \ha and \nii are at similar wavelengths and would be affected similarly by the dust obscuration. Coherent structures in residual velocity maps caused by obscuration can closely resemble those resulting from shocks, including the misalignment caused by one-less kinematic multiplicity compared to the morphological one. However, the magnitude of coherent structures in residual velocity maps caused by dust obscuration should be rather small, proportional to the projected velocity difference at smaller and larger heights above the disc. 

On the other hand, the magnitude of coherent structures in the residual velocity maps resulting from extended shocks can be large. For example, studying the observations of ionised gas, \citet{Zurita_04} revealed structures of extended shocks along the bar of galaxy NGC\,1530, characterised by positive and negative residual velocities along the LOS with magnitudes up to 120\kms. These magnitudes would correspond to even stronger shocks and larger velocity jumps along the galaxy's bar than those determined from models. Moreover, within the innermost regions of the NGC\,1530, their residual velocity map reveals distinct structures, with positive and one with negative velocities, both with magnitudes of $\sim$40\kms. These structures coincide with nuclear dust spirals visible in their $J-K$ colour images of the galaxy, and are interpreted as evidence of spiralling inflow to galaxy's nucleus. Their observational findings show an excellent alignment with their kinematic model of the gas flow and velocities along the dust spirals. Similar structures associated with spiral shocks and gas inflows to the nucleus, are also seen in other galaxies, such as NGC\,5236 \citep{Fathi_2008} and UG\,10200 \citep{Blasco-Herrera_2013}, with residual velocity magnitudes of $\sim$30--50\kms, and in NGC\,3081 and NGC\,1358 \citep{Schnorr-Muller_2016,Schnorr-Muller_2017} with magnitudes of $\sim$50\kms.

Our analysis highlights that not all diagnostic methods necessarily reveal extended shocks in gas simultaneously, emphasising the importance of employing multiple complementary approaches when searching for such features. Although the residual velocity and velocity difference methods are each influenced by distinct factors, and a perfect spatial correspondence between their coherent structures is not always expected, in certain cases they exhibit good alignment. For instance, in NGC\,4303 (see Fig.\,\ref{fig:NGC4303} and Appendix \ref{sec:NGC4303_notes}) both methods show aligning structures. The residual velocity map of NGC\,4303 reveals a coherent structure extending continuously from the north to the south, exhibiting large deprojected velocities of approximately 50\kms, downstream from the northern dust lane along the bar. This structure extends past the nuclear ring and directly reaches the nucleus of the galaxy. Similarly, the velocity difference map of the galaxy reveals a coherent structure that is almost spatially aligned with the one observed in the residual velocity map along the bar. These structures are associated with extended shocks in the gas.

\subsection{\textcolor{black}{Criteria for identifying extended shocks}}
\label{sec:met:criteria}
\textcolor{black}{As described in the previous subsections, we use a combination of complementary diagnostics to identify and interpret signatures of extended shocks in the centres of galaxies. The residual velocity (Sect.\,\ref{met:residual_velocities}) and velocity difference maps (Sect.\,\ref{met:velocity_differences}) serve as the primary identification maps, since extended shocks are expected to produce coherent kinematic structures associated with rapid changes in velocity. However, because these two approaches are sensitive to different effects and need not reveal exactly the same structures, a candidate feature is not required to appear identically in both maps. Once we identify candidate extended kinematic features in one or both maps, we use the dust morphology (Sect.\,\ref{method:dustmorphology}) and BPT maps (Sect.\,\ref{met:bpt}) to assess whether the feature is consistent with an extended shock and to interpret possible discontinuities, contamination, or alternative explanations.
\\ \\
Our criteria for identifying extended shocks are therefore as follows.}

\begin{enumerate}[label=\roman*., align=left, leftmargin=*]    \item \textcolor{black}{\textbf{Initial identification in the kinematic maps.}
    We identify candidate signatures of extended shocks by examining the residual velocity and velocity difference maps for spatially extended, coherent features--as outlined in Figs.\,\ref{fig:Sample_extended_structures_perturbed_galaxies_residualvel} and \ref{fig:Sample_extended_structures_perturbed_galaxies_velocitydifference}. A candidate must be visible in at least one of these maps, while detection in both strengthens the identification. This approach accounts for the fact that the residual velocity map can be influenced by uncertainties in the disc model and by vertical motions, whereas the velocity difference map is model-independent and responds differently to gas traced by distinct emission lines but rarely produces strong signal. The identified features must be extended and elongated rather than compact or patchy. In practice, we look for quasi-linear or spiralling structures that can be traced across multiple adjacent spaxels -- corresponding to at least is hundreds of parsecs - and whose length is clearly larger than their width. These structures may appear either as continuous features or as fragmented segments; in the latter case, they are retained if the segments plausibly trace a larger-scale structure based on their orientation, geometry, and overall continuity across the map.}

    \item \textcolor{black}{\textbf{Velocity amplitude criterion.}
    Once a coherent extended kinematic feature has been identified, we examine its deprojected residual velocity to determine whether it is consistent with the magnitudes expected for shocks. Spatial coherence alone is not sufficient (see e.g. NGC\,4237, Sect.\,\ref{sec:NGC4237_notes}): the feature should also show a velocity amplitude that is compatible with extended shocks. This criterion is essential because not all extended features in the kinematic maps correspond to shocks, whereas large residual magnitudes are expected for genuine shock signatures.}

    \item \textcolor{black}{\textbf{Presence of a counterpart feature.}
    A particularly strong indication of an extended shock signature is the presence of a corresponding feature on the opposite side of the centre with an opposite velocity sign, as expected in shocks driven by bars or in nuclear spiral shocks \citep{Davies_09}. This is especially valuable when the candidate structure is not fully continuous, as paired features can support the interpretation that they trace a single large-scale shock-related pattern rather than unrelated local disturbances. However, this criterion is not always satisfied. The detectability of counterpart features depends on projection effects, the three-dimensional geometry of the system, and the complexity of the gas flows. Asymmetries in the gas distribution, local perturbations, or obscuration can result in only one side of the structure being clearly visible. Therefore, while the presence of a counterpart feature strengthens the classification, its absence does not preclude the identification of an extended shock signature.}
        
    \item \textcolor{black}{\textbf{Support from dust morphology.}
    For candidate features, we also examine the unsharp-masked image for aligned dust lanes. Such alignment strengthens the interpretation in terms of extended shocks, since shocks are expected to compress gas mixed with dust. This is particularly useful in more fragmented cases, where the dust morphology can help assess whether disconnected kinematic segments are still part of a single larger-scale structure. However, dust morphology is used only as supporting evidence: dust lanes alone cannot demonstrate the presence of a shock, and agreement between dust and kinematic structures is not always expected because of projection effects and because coherent dust structures may also arise for reasons unrelated to shocks.}
    
\end{enumerate}

\textcolor{black}{We apply the criteria above in the form of the following decision tree: a feature is considered a candidate for an extended shock when it appears as an extended coherent structure in at least one kinematic diagnostic map and exhibits deprojected residual velocities consistent with those expected for shocks. The presence of a counterpart feature strengthens the case. Features that satisfy only part of these conditions, or whose interpretation may be complicated by star formation or AGN, are treated more cautiously with the help of dust and BPT maps.  Fig.\,\ref{fig:Sample_extended_structures_perturbed_galaxies_residualvel} and Fig.\,\ref{fig:deprojected_velocity_residuals}  present the deprojected residual velocity magnitudes for the galaxies in our sample. In Fig.\,\ref{fig:Sample_extended_structures_perturbed_galaxies_residualvel}, we highlight the features that we identify as extended, in Fig.\,\ref{fig:deprojected_velocity_residuals} we present the kinematic maps in which we did not indentify extended features.}

\textcolor{black}{Once a kinematic structure has been identified, following the decision tree presented above, and associated with extended shocks, the BPT information provides additional physical context, helping to assess whether discontinuities in the structure coincide with star-forming regions or whether the observed kinematics may be influenced by AGN-driven outflows. In this sense, the BPT maps aid the interpretation of the nature and morphology of the candidate feature, but do not by themselves define the presence of an extended shock signature. Maintaining this distinction between identification (based on kinematics) and characterisation (informed by ionisation diagnostics) is essential, as BPT information is most powerful once the underlying kinematic structure has already been established. In this way, our decision tree combines model-dependent and model-independent kinematic diagnostics with supporting morphological and ionisation information, while keeping the identification stage distinct from the characterisation stage. }

\section{Collective analysis of galaxies in the sample}
\label{sec:common-features-seen-across-galaxies}

Each galaxy of our sample is analysed individually in its subsection of Appendix\,\ref{sec:notes-on-galaxies}, which gives the full set of kinematic maps, diagnostic maps, and images in figures analogous to Fig.\,\ref{fig:NGC4303}, which we present here for the example of NGC\,4303. A detailed explanation of each panel in these figures is provided in the beginning of Appendix\,\ref{sec:notes-on-galaxies}. These figures visualise the approaches that we take when identifying extended shocks, and the maps that we use for that. In this section, we gather the quantitative results of our analysis and provide an overview of our findings on the common kinematic characteristics observed in galaxies across our sample, based on their analyses provided in Appendix\,\ref{sec:notes-on-galaxies}.

\subsection{Collected quantitative results for galaxies in the sample}
\label{sec:collective_quantitative_results_sample}

\subsubsection{Orientation of disk model for residual velocity fits}
The parameters and approach taken to construct the disk model of each galaxy are individually provided in their respective subsections in Appendix\,\ref{sec:notes-on-galaxies}. 
The PA of LON and \textit{i}, derived for each galaxy using the appraoch given in \textcolor{black}{Sect.\,\ref{met:residual_velocities}}, are summarised in Table\,\ref{tab:Disk_orientation_parameters}, with the large-scale parameters provided for comparison.

\begin{table}
\caption{Large-scale and Kinemetry-inferred disk orientation parameters. Superscripts denote the literature sources of the large-scale parameters.}
\label{tab:Disk_orientation_parameters}
\centering
\setlength{\tabcolsep}{4pt}

\begin{tabular}{lcccc}
\hline
Galaxy ID & \multicolumn{2}{c}{PA (deg)} & \multicolumn{2}{c}{Inc. (deg)} \\
       & Kinemetry & Large scale & Kinemetry & Large scale \\ 
\hline
IC\,2051   & 60   & 72$^{1}$        & 41 & 55$^{1}$ \\
NGC\,289   & 125  & 130$^{2}$       & 46 & 46$^{2}$ \\
NGC\,613   & 110  & 115$^{3}$       & 37 & 37$^{3}$ \\
NGC\,1300  & 107  & 87$^{4}$        & 35 & 35$^{4}$ \\
NGC\,1433  & 19.7 & 19.7$^{5}$      & 29 & 29$^{5}$ \\
NGC\,1566  & 40   & 35$^{5}$        & 37 & 31$^{5}$ \\
NGC\,3351  & 23   & 10$^{6}$        & 46 & 46$^{6}$ \\
NGC\,3368  & 172  & 172$^{6}$       & 46 & 50$^{6}$ \\
NGC\,3489  & 80   & 71$^{6}$        & 60 & 58$^{6}$ \\
NGC\,3626  & 163  & 163$^{1}$       & 53 & 47$^{1}$ \\
NGC\,4237  & 107  & 106$^{7}$       & 53 & 53$^{7}$ \\
NGC\,4303  & 135  & 132$^{5}$       & 42 & 23.5$^{5}$ \\
NGC\,4321  & 155  & 152$^{8}$       & 36 & 32$^{8}$ \\
NGC\,4380  & 161  & 126$^{9}$       & 59 & 56$^{9}$ \\
NGC\,4457  & 80   & 66$^{9}$        & 20 & 22$^{9}$ \\
NGC\,4643  & 58   & 53$^{6}$        & 43 & 38$^{6}$ \\
NGC\,4941  & 25   & 21$^{10}$       & 60 & 48$^{10}$ \\
NGC\,5248  & 114  & 114$^{8,11}$   & 44 & 45$^{8,11}$ \\
NGC\,7177  & 90   & 83$^{6}$        & 46 & 48$^{6}$ \\
NGC\,7513  & 105  & 105$^{9}$       & 46 & 46$^{9}$ \\
\hline
\end{tabular}

\vspace{3pt}
\begin{flushleft}
Notes -- sources of large-scale parameters:  
1--\citet{Salo_15},  
2--\citet{Walsh_1997},  
3--\citet{Bajaja_1995},  
4--\citet{Lindblad_1997},  
5--\citet{Lang_20},  
6--\citet{Erwin_08},  
7--\citet{Chemin_06},  
8--\citet{Haan_08},  
9--\citet{Munoz-Mateos_2015},  
10--\citet{Gutierrez_11},  
11--\citet{Haan_09}.
\end{flushleft}
\end{table}

\subsubsection{Determination of velocity jumps in shocks}
\label{sec:determining_velocity_jumps}
In order to ensure that the coherent structures seen in the residual velocity maps are indeed due to extended shocks, we examined the deprojected residual velocity magnitudes in these structures. Working under the assumption that the gas moves in the plane of the galaxy disc, we obtain the deprojected velocities using the formula $V_{true} = V_{LOS}/sin(i)$, where $V_{true}$ is the deprojected velocity, $V_{LOS}$ is the observed LOS velocity, and $i$ is the inclination angle of the galaxy. We record the highest and lowest magnitudes, for both positive and negative velocities, along the coherent structures, which could indicate extended shocks. To reduce the level of noise in the residual velocity maps we applied box filter convolution, so that the value at each spaxel is the average over $3\times3$ spaxels' box centred on it. Maps of the deprojected residual velocity, created following this approach, are presented in \textcolor{black}{Fig. \ref{fig:Sample_extended_structures_perturbed_galaxies_residualvel} for a subset of galaxies for which extended structures are detected and in Appendix\,\ref{app:suplementary_sample_plots}, Fig.\,\ref{fig:deprojected_velocity_residuals} for the remaining galaxies.} The recorded velocity magnitudes in extended structures are explicitly stated for each galaxy in their respective subsections in Appendix\,\ref{sec:notes-on-galaxies}, and also listed in \textcolor{black}{Table\,\ref{tab:galaxies_shock_classes}}. For completeness, in this table we also included the target of our pilot study, NGC\,1097.

In the majority of galaxies that we studied, the highest deprojected residual velocity magnitudes, recorded along the coherent structures associated with extended shocks, are $\sim$50\kms or higher. These magnitudes are similar to the magnitudes observed in other galaxies, which are associated with spiral shocks extending to the nucleus \citep{Zurita_04,Davies_09,Fathi_2008,Blasco-Herrera_2013,Schnorr-Muller_2016,Schnorr-Muller_2017}. Particularly, for galaxies like NGC\,1433 and NGC\,4643, the deprojected velocity magnitudes exceed 100\kms, indicating much stronger shocks and larger velocity jumps, even larger than the jumps of $\sim$100\kms seen in models by \citet{Athanassoula_1992} and of $\sim$70\kms by \citet{Maciejewski_2004b}.

On the other hand, the coherent structures observed in the residual velocity maps of the two unbarred galaxies in our sample, NGC\,4237 and NGC\,4380, have deprojected velocity magnitudes of only 20\kms. This indicates that these structures are either caused by obscuration (see \textcolor{black}{Sect.\,\ref{sec:complementing_approaches_for_shocks})} or by a low-amplitude wave propagating in gas, below the threshold of steepening into a shock. \textcolor{black}{This amplitude of the velocity residual is several times smaller than the magnitudes seen in the other galaxies of our sample (see also Fig.\,\ref{fig:shocks_nuclearbar}).} 

Note that in the majority of galaxies that we studied, the lowest residual velocity magnitudes recorded along the coherent kinematic structures associated with extended shocks, have values similar those in the unbarred galaxies of our sample. This may indicate that in the regions where we recorded the lowest magnitudes, the shocks 
wane into mere waves propagating in gas, but this can also be the effect of noise in the data, still noticeable after binning.

\subsubsection{Radial extent of the shocks}
\label{sec:radialextent_of_shocks}
For each galaxy in which signatures of extended shocks are identified, Appendix\,\ref{sec:notes-on-galaxies} describes in detail how deeply inward and how far outward the shock signatures can be traced within the disc. \textcolor{black}{Table\,\ref{tab:galaxies_shock_classes}} (5th column) summarises the smallest and the largest deprojected galactocentric radii of these features, converted to physical units (pc) using the distance to each galaxy. 

The deprojected radii listed in the 5th column of Table\,\ref{tab:galaxies_shock_classes} show that extended shock signatures are detected across large radial ranges, reaching inward to radii below 100 parsecs, and extending out to more than 3\,kpc, in some cases to the edge of the MUSE FOV. In most systems, the furthest radii lie between $\sim 1$ and 3\,kpc, demonstrating that the shocks often reach well into the galactic disc, with several galaxies -- most notably NGC\,613, NGC\,4303, NGC\,4321, NGC\,4941 and NGC\,5248 -- showing shock-related features beyond 2\,kpc.

The smallest galactocentric radii of extended shocks signatures indicate that these structures can penetrate deep into the inner regions: in 7 galaxies they are around 100\,pc, the value similar to the typical resolution of the observations. In three galaxies in our sample -- NGC\,613, NGC\,4457 and NGC\,4941, strong outflows from the AGN obscure the features within the innermost regions of the diagnostic residual velocity and velocity difference maps. Consequently, the signatures of extended shocks in these galaxies cannot be traced further inward to the same radii as in the systems without strong AGN outflows, and the measured closest radii -- typically around $\sim 600$\,pc -- likely underestimate how close to the galactic nucleus the shocks in these galaxies actually extend. Moreover, we note that the innermost few arcseconds of NGC\,1566 are discarded due to saturated spaxels and the signatures of extended shocks in this galaxy extend down to 480\,pc, which is also likely underestimated. In the remaining 3 galaxies -- IC\,2051, NGC\,289 and NGC\,3351 -- whose central regions do not appear affected by the AGN outflows, extended shocks reach inward only to $\sim$300\,pc. In these three galaxies the innermost radius at which shocks can be reliably traced coincides with the location of the nuclear rings, suggesting that the rings may act as a structural barrier that prevents shock signatures from propagating further into the nucleus.

\begin{table*}
    \caption{The individual findings for the sample galaxies studied in this work. Columns: (1) index used in figures to represent each galaxy; (2) galaxy ID; (3) nuclear emission type (see Sect.\,\ref{sec:grouping-nuclear-emission}); (4) kinematic category (see Sect.\,\ref{sec:grouping-shock-class}); (5)  \textcolor{black}{ranges of deprojected galactocentric radii of the signatures associated with extended shocks (see Sect.\,\ref{sec:radialextent_of_shocks})}; (6) minimum and (7) maximum positive/negative deprojected velocity magnitudes in extended structures (see Sect.\,\ref{sec:determining_velocity_jumps}); (8) average \ha\ velocity dispersion (Sect.\,\ref{sec:determining_averageHalpha_dispersion}); (9) inner bar presence; (10) nuclear ring presence -- ``N'': absence, ``Y'': presence, ``Y?'': potential presence, ``?'': uncertain classification.}
    \label{tab:galaxies_shock_classes}
    \setlength{\tabcolsep}{5pt}
    \centering
    \resizebox{\linewidth}{!}{%
\begin{tabular}{ccccccccccc}
\hline
\textbf{Index} & \textbf{Galaxy} & \textbf{Nuclear} & \textbf{Kinematic} & \textbf{Radial Range} & \textbf{Min Vel.} & \textbf{Max Vel.} & \textbf{$<{\sigma}>$} & \textbf{Inner} & \textbf{Nuclear} \\
& \textbf{ID} & \textbf{Emission} & \textbf{Category} & \textbf{[pc]} & \textbf{[km/s]} & \textbf{[km/s]} & \textbf{[km/s]} & \textbf{Bar} & \textbf{Ring} \\
(1) & (2) & (3) & (4) & (5) & (6) & (7) & (8) & (9) & (10) \\
\hline
1  & IC\,2051    & Composite & Ext. Shocks           & 320--820   & $-15, 20$   & $-40, 65$   & $75.7 \pm 3.8$   & N  & Y \\
2  & NGC\,289    & Composite & Ext. Shocks           & 300--2150  & $-20, 15$   & $-65, 90$   & $72.4 \pm 0.9$   & N  & N \\
3  & NGC\,613    & Composite & Ext. Shocks+Outflow   & 650--3010  & $-30, 40$   & $-90, 110$  & $101.1 \pm 5.6$  & N  & Y \\
4  & NGC\,1097   & LINER     & Ext. Shocks           & 100--2660  & $-50, 40$   & $-130, 150$ & $80.7 \pm 4.6$   & Y  & Y \\
5  & NGC\,1300   & LINER     & Other Perturb.        & --          & --           & --           & $70.5 \pm 3.8$   & N  & Y \\
6  & NGC\,1433   & LINER     & Ext. Shocks           & 140--2160  & $-20, 25$   & $-116, 115$ & $74.6 \pm 3.0$   & Y  & Y \\
7  & NGC\,1566   & --        & Ext. Shocks           & 480--1860  & $-20, 20$   & $-80, 70$   & $100.4 \pm 11.4$ & Y? & Y? \\
8  & NGC\,3351   & SF        & Ext. Shocks           & 270--560   & $-25, 12$   & $-95, 50$   & $81.0 \pm 9.7$   & N  & Y \\
9  & NGC\,3368   & LINER     & Other Perturb.        & --          & --           & --           & $103.8 \pm 1.9$  & Y  & N \\
10 & NGC\,3489   & Seyfert   & Ext. Shocks           & 140--880   & $-12, 8$    & $-40, 30$   & $69.8 \pm 1.0$   & N  & ? \\
11 & NGC\,3626   & LINER     & Other Perturb.        & --          & --           & --           & $74.0 \pm 4.2$   & Y  & N \\
12 & NGC\,4237   & Composite & Unperturbed           & --          & $-5, 5$     & $-20, 18$   & $54.1 \pm 0.2$   & N  & N \\
13 & NGC\,4303   & LINER     & Ext. Shocks           & 120--2200  & $-15, 10$   & $-50, 20$   & $69.3 \pm 4.2$   & Y  & Y \\
14 & NGC\,4321   & LINER     & Ext. Shocks           & 130--2390  & $-15, 15$   & $-55, 55$   & $66.9 \pm 1.9$   & Y  & Y \\
15 & NGC\,4380   & LINER     & Unperturbed           & --          & $-4, 9$     & $-13, 20$   & $53.7 \pm 0.3$   & N  & Y? \\
16 & NGC\,4457   & LINER     & Other Perturb.        & --          & $-35, 45$   & $-125, 120$ & $86.4 \pm 1.3$   & Y  & N \\
17 & NGC\,4643   & LINER     & Ext. Shocks           & 130--2070  & $-20, 20$   & $-110, 70$  & $71.2 \pm 0.6$   & N  & N \\
18 & NGC\,4941   & Seyfert   & Ext. Shocks+Outflow   & 620--3320  & $-20, 15$   & $-60, 45$   & $95.3 \pm 11.8$  & Y  & N \\
19 & NGC\,5248   & Composite & Ext. Shocks           & 90--2830   & $-15, 20$   & $-50, 60$   & $62.6 \pm 1.5$   & N  & Y \\
20 & NGC\,7177   & Composite & Ext. Shocks+Outflow   & 150--1100  & $-15, 20$   & $-70, 60$   & $82.5 \pm 1.2$   & N  & N \\
21 & NGC\,7513   & LINER     & Ext. Shocks        & 100--2080          & $-30, 30$           & $-80, 100$           & $59.3 \pm 0.5$   & N  & N \\
\hline
\end{tabular}}
\end{table*}

\subsection{Common features seen across the sample}
\label{sec:common_features_seen_across_sample}
One of the striking examples of  shocks in gas that yield inflows to the galactic nucleus is revealed within the kinematic fields of {NGC\,5248} \textcolor{black}{(see Fig.\,\ref{fig:Sample_extended_structures_perturbed_galaxies_residualvel}, last row/first panel and Fig.\,\ref{fig:NGC5248})}. The residual velocity map of that galaxy exhibits a coherent continuous structure, manifested as a spiral pattern with a red (with positive velocities) and a blue (with negative velocities) arm. These structures appear as counterparts of one another, with a pattern that is almost a textbook example of extended shocks in gas, similar to those expected from models \citep{Davies_09}. Inner to the star-forming ring of the galaxy, the observed structures become more patchy yet there are indications of some parts of these structures extending to the innermost regions. Similar counterpart structures with positive and negative velocities, extending over large range of radii are also seen in the residual velocity maps of {NGC\,1566}, {NGC\,4321}, {NGC\,4643} and \textcolor{black}{NGC\,7513} (see Figs.\,\ref{fig:NGC1566}, \ref{fig:NGC4321}, \ref{fig:NGC4643} and \ref{fig:NGC7513}). In these galaxies, the coherent continuous structures exhibit large deprojected residual velocity magnitudes ranging approximately from 60 to 100\kms. The residual velocity map of {NGC\,4303}, shown in Fig.\,\ref{fig:NGC4303}, also reveals a coherent structure, extending continuously from the north of the frame southwards, exhibiting large deprojected velocities of approximately 50\kms, downstream from the northern dust lane along the bar. However, there is no clear counterpart with positive velocities.

In some galaxies, such as {IC\,2051} and {NGC\,3489}, see Figs.\,\ref{fig:ic2051_4by4} and \ref{fig:NGC3489}, the resolution of structures in central regions is limited due to galaxy distance and the inclined geometry of the system. However, we can discern aligning structures across our multiple approaches, exhibiting large deprojected velocity magnitudes around $\sim$50\kms, that are indicative of extended shocks.

Within the inner kpc region of some galaxies, such as in {NGC\,289}, {NGC\,1433} and {NGC\,3351}, several extended structures are observed in the residual velocity map, exhibiting deprojected velocity magnitudes of 90\kms or higher (see Figs.\,\ref{fig:NGC289}, \ref{fig:NGC1433} and \ref{fig:NGC3351} respectively). However, unlike those seen in NGC\,5248, these structures do not extend continuously from the outer regions of the galaxy towards the nucleus. Instead, they appear as discontinuous pieces. For instance, the extended structures in NGC\,1433 are located within the nuclear disk, where strong SF occurs. These structures mostly are located downstream from the curving dust lanes in the nuclear disk. In NGC\,3351, these structures are found in and inner to the nuclear ring and just outside it, where strong SF and stellar outflows occur. Similarly, the structures in the central regions of NGC\,289 are also located in areas dominated by SF. In all there galaxies, where extended structures occur in the residual velocity map, the velocity differences also show enhancements. In these three galaxies, continuous dust lanes extend from the outer regions towards the nucleus, while kinematic structures are rather discontinuous. Based on the alignments across multiple methodologies, we associate the observed discontinuous yet extended structures with extended shocks in the gas. We hypothesise that due to strong star-forming activities, the self-gravity of the gas is strong enough to disrupt the continuity of extended shocks in the gas, leaving behind disconnected pieces. In contrast, the dust is less affected by these disturbances and maintains its spiral form as it propagates towards the nucleus.

The kinematics within the innermost regions of three galaxies in our sample, namely {NGC\,613}, {NGC\,4941} and \textcolor{black}{NGC\,7177} are strongly dominated by AGNs. This dominance is primarily evident in the velocity dispersion and BPT maps of these galaxies (see Appendix\,\ref{sec:NGC613_notes}, \ref{sec:ngc4941_notes} and \ref{sec:ngc7177_notes}). The residual velocity maps of these galaxies exhibit extended shock signatures in their bar regions, with large deprojected velocity magnitudes ranging from roughly 60\kms to 100\kms. However, strong AGN outflows which are dominating the kinematics in the innermost regions, prevent the tracing of shock signatures to the nucleus.

Furthermore, the kinematic fields of several other galaxies in our sample, namely {NGC\,1300, NGC\,3368, NGC\,3626 and NGC\,4457}, shown in Figs.\,\ref{fig:NGC1300}, \ref{fig:NGC3368}, \ref{fig:NGC3626} and \ref{fig:NGC4457} respectively, are strongly disturbed, but we cannot associate them  with any mechanism of inflow. They seem rather related to AGN or stellar outflows, or we cannot identify the source of the disturbance. Standing as one of the most intriguing galaxies within our sample is NGC\,3626. Throughout the observed region, the gas counterrotates with respect to the stars, yet the gas reaching the central regions collapsed to form a boxy ring structure within the inner $\sim$15\arcsec (see Appendix\,\ref{sec:NGC3626_notes}). However, the origin and motion of gas in this galaxy are complex and there is no evidence of extended shocks to the nucleus. 

Lastly, the only two unbarred galaxies in our sample, NGC\,4237 and NGC\,4380, have relatively unperturbed kinematic fields and show no evidence of extended shocks. Their residual velocity maps typically exhibit low deprojected velocity magnitudes, not exceeding $\sim$20\kms. No prominent structures are seen in the velocity difference maps either. These results suggest the absence of shocks in these systems. Any mild perturbations may instead arise from obscuration effects, variations in velocity with height, or wave patterns generated by lopsidedness. 

\subsection{Grouping of galaxies based on the observed signatures of extended shocks: prevalence of shocks in the sample}
\label{sec:grouping-shock-class}

As we noticed in Sect.\,\ref{sec:common_features_seen_across_sample}, the individual examinations of each galaxy (presented in Appendix\,\ref{sec:notes-on-galaxies}) reveal similar patterns in several cases. This allowed us to classify the galaxies into four distinct categories, based on the extended shock signatures identified from their diagnostic maps. We further refer to this classification as the \textit{kinematic category}.

The first kinematic category is defined as \textcolor{black}{\textit{\textbf{extended shocks}}. Galaxies in this category exhibit compelling shock signatures that extend over large distances into the inner parsec regions and are characterised by high residual velocities or significant velocity differences. These extended shock signatures can appear as coherent continuous structures, spanning inward either in a straight (see e.g., NGC\,4941 in Fig.\,\ref{fig:Sample_extended_structures_perturbed_galaxies_residualvel}) or spiralling pattern (see  e.g., NGC\,5248 in Fig.\,\ref{fig:Sample_extended_structures_perturbed_galaxies_residualvel}), or as more discontinuous structures that nevertheless remain consistent with extended shocks. In some galaxies, these structures can be traced continuously over large scales, while in others they appear more fragmented, possibly due to strong local activity disrupting the propagation of the shocks. Even in such cases, they are still characterised by high residual velocities and/or large velocity differences, and often follow the continuous dust patterns that can be associated with loci of shock compression.} We classified \textcolor{black}{12 galaxies} in our sample as showing extended shocks: IC\,2051, NGC\,289, NGC\,1097, NGC\,1433, NGC\,1566, NGC\,3351, NGC\,3489, NGC\,4303, NGC\,4321, NGC\,4643, NGC\,5248 and \textcolor{black}{NGC\,7513.}

The second category is \textcolor{black}{\textbf{\textit{extended shocks and outflows}}. Although galaxies in this category exhibit extended shock signatures along their bar, within the innermost regions the kinematics are obscured by strong outflows from AGNs, preventing the tracing of shock signatures to the nucleus (see e.g., NGC\,613 in Fig.\,\ref{fig:Sample_extended_structures_perturbed_galaxies_residualvel})}. We classified 3 galaxies in our sample under this category: NGC\,613, NGC\,4941 and \textcolor{black}{NGC7177.}

The third category, defined as \textbf{\textit{other perturbations}}, includes galaxies that do not exhibit compelling signatures of extended shocks associated with inflow. However, the kinematic fields of these galaxies are strongly perturbed. In most of these galaxies, the perturbations within inner kpc can be associated with local shocks from SF and/or AGN .
We classified \textcolor{black}{4 galaxies} in our sample as having other perturbations: NGC\,1300, NGC\,3368, NGC\,3626 and NGC\,4457.

The final category is defined as \textbf{\textit{unperturbed}}. Galaxies under this category have relatively unperturbed kinematic fields and show no evidence of extended shocks. The residual velocity maps typically exhibit perturbations with low amplitudes. We classified 2 galaxies in our sample as unperturbed: NGC\,4237 and NGC\,4380. It is important to note that although both \textit{other perturbations} and \textit{unperturbed} categories include galaxies without shock signatures, galaxies within the former category have significantly perturbed kinematic fields, unlike the galaxies in the latter category. The kinematic category assigned to each galaxy in the sample we studied in this work is presented in the third column of \textcolor{black}{Table\,\ref{tab:galaxies_shock_classes}}. 

The grouping above indicates that \textcolor{black}{57\% (12 out of 21) }of the galaxies in our sample, including NGC\,1097 from Paper IV, exhibit signatures of extended shocks within their inner kpc region which can be traced to the innermost parsecs. In contrast, \textcolor{black}{29\% of the galaxies (6 out of 21) do not show such signatures,} with 2 of these galaxies displaying unperturbed kinematic fields. For the remaining \textcolor{black}{$\sim14$\% (3 out of 21)}, coherent structures with large deprojected residual magnitudes are observed along the bars, but AGN outflows obscure the innermost regions, preventing us from tracing these structures further. As a result, \textcolor{black}{the 57\% figure} may be a lower limit, as extended shocks could be reaching the nucleus in those 3 galaxies, but they are obscured by the dominating outflows. Therefore, this significant proportion -- at least \textcolor{black}{57\%} -- suggests that extended shocks could be the prevalent mechanism of gas inflow in the central regions of galaxies in our sample.

\subsection{Grouping of galaxies based on the nuclear emission type}
\label{sec:grouping-nuclear-emission}
We further categorised the galaxies in our sample based on the dominant emission type in their nuclear regions, to explore the correlation between the type of nuclear emission and the presence of extended shocks. In \textcolor{black}{Fig.\,\ref{fig:BPT_global}}, we present the \textit{\textbf{BPT diagram of the sample}}. Each point on the diagram represents a single galaxy, colour-coded according to the kinematic category assigned based on the signatures of extended shocks identified in each. The location of each point on the BPT diagram is determined by the ratio of emission line fluxes measured within a 150\,pc radius. The diagnostic lines based on relations from \citet{Kewley_01a}, \citet{Kauffmann_03} and \citet{Cid_Fernandes_10} separate the regions for SF, Composite, LINER, and Seyfert nuclear emission type as in \textcolor{black}{Sect.\,\ref{met:bpt}}. This classification is used for identifying the main ionisation source in the nucleus of each galaxy.

\begin{figure}
    \hspace{-0.4cm}
    \includegraphics[width=1.05\linewidth]{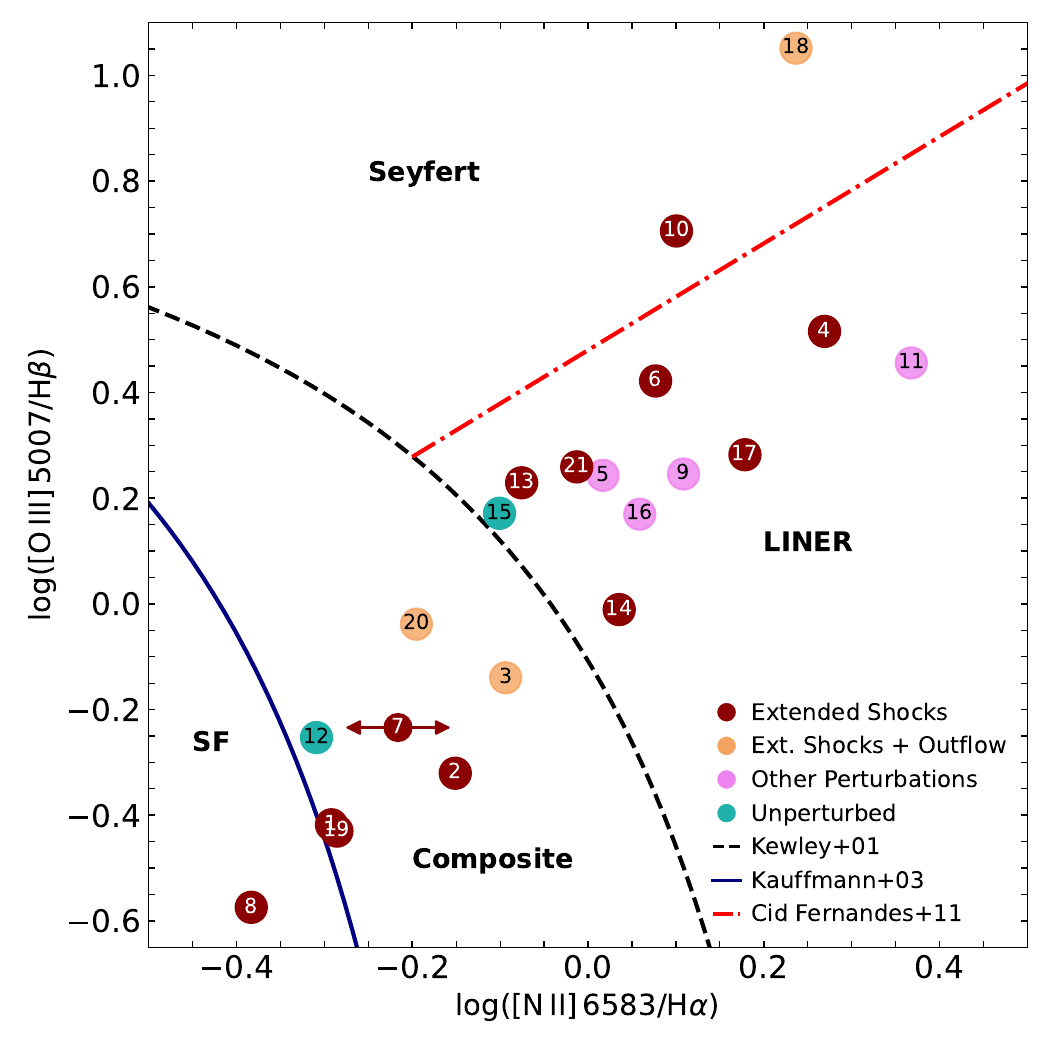}

    \caption{The BPT diagram of the galaxy sample analysed in this study, computed using integrated emission-line flux ratios within 150pc radius. The navy (solid) , black (dashed), and red (dotted dashed) lines show \citet{Kauffmann_03}, \citet{Kewley_01a}, and \citet{Cid_Fernandes_10} fits respectively. These lines determine the regions associated with different ionisation mechanisms: SF, Composite, LINER, and Seyfert. Each point in the diagram represents an individual galaxy, with the accompanying numbers listed in the index column of \textcolor{black}{Table}\,\ref{tab:galaxies_shock_classes} for identification. \textcolor{black}{The points are colour-coded as described in the legend according to the categories assigned based on the extended shock signatures.}}
    \label{fig:BPT_global}
\end{figure}

Consequently, based on the BPT diagram of the sample, we have assigned a nuclear emission type to each galaxy, which are presented in the \textcolor{black}{third} column of \textcolor{black}{Table\,\ref{tab:galaxies_shock_classes}}. We note that, because of strong broad emission from its AGN, NGC\,1566 had unreliable \ha and \nii flux measurements within its inner $\sim$1\arcsec~but the flux measurements of \oiii and \hb were not significantly affected. Therefore, to represent the uncertainty in the x-axis position of this galaxy on the BPT diagram of the sample, we have marked its location with two horizontal arrows, indicating its position could shift either to the left or right. While previous studies have classified this galaxy as Seyfert/LINER type (e.g., \citealp{Alloin_1985,Woo_2002}), we have decided not to assign a nuclear emission type obtained from other works to ensure consistency across the galaxies analysed in this study based on the BPT diagram of the sample. 

\textcolor{black}{In Fig.\,\ref{fig:Histogram_shocks}}, we present the distribution of galaxies in our sample grouped by kinematic category. The colour coding indicates the nuclear emission type assigned to each galaxy. We find no clear correlation between the nuclear emission types derived from the BPT diagram and the kinematic categories defined by the presence of extended shock signatures in the galaxies we studied. This lack of correlation is not surprising given the mismatch in characteristic timescales: \textcolor{black}{transporting gas from kpc scales to the central $\lesssim 10-100$\,pc typically requires $\sim 10^7-10^8$\,yr \citep[e.g.][]{Hopkins_Quataert_2010, Silva-Lima_2022}, whereas AGN activity is expected to be highly intermittent, with ``flickering'' on much shorter timescales of order \(\sim 10^{5}\,\mathrm{yr}\) \citep[e.g.][]{Schawinski_2015}. Such variability can therefore wash out any contemporaneous correspondence that might otherwise be expected. Additionally, the small sample size may further limit our sensitivity to subtle trends.}

\begin{figure}
\centering
   \includegraphics[width=1\linewidth]{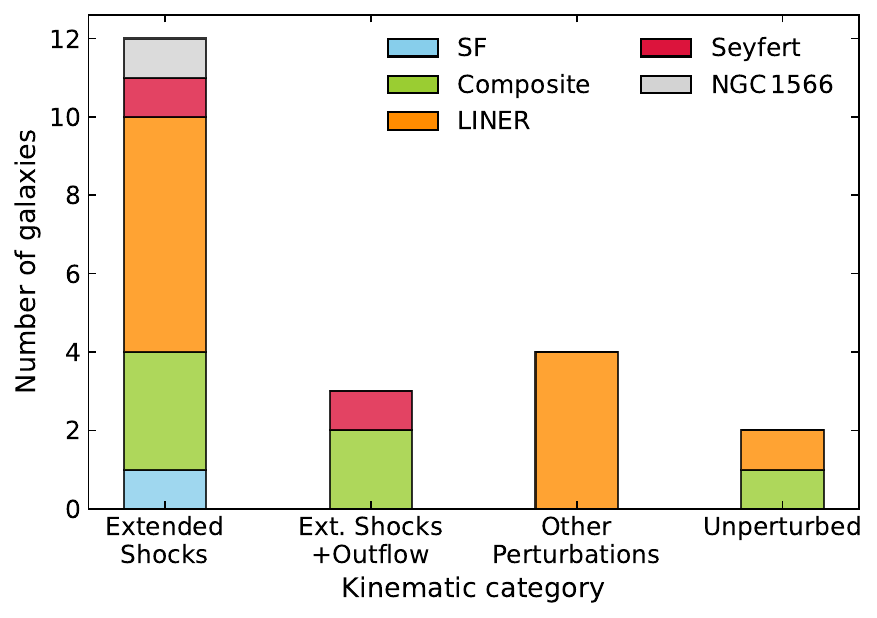}
    \caption{The distribution of galaxies stacked in each kinematic category, colour-coded based on their nuclear emission types, both classifications are defined in Sects.\,\ref{sec:grouping-shock-class} and \ref{sec:grouping-nuclear-emission}. Different colours represent SF in blue, composite emission in green, LINER in orange, and Seyfert in red. NGC\,1566, which has an uncertain nuclear emission type, is represented in grey.}
    \label{fig:Histogram_shocks}
\end{figure}

\section{Correlations between the observed quantities and other properties of the host galaxies}
\label{sec:discussion}

In Sect.\,\ref{sec:common-features-seen-across-galaxies}, we classified galaxies in our sample according to their gas kinematics in the central kpc region, as well as the type of nuclear emission deduced from the same data. We found no obvious correlations between classes in the two groupings. However, correlations are expected between other \textcolor{black}{properties of the host galaxies and their gas dynamics and nuclear activity. For example one expects from dynamical considerations that bars} in galaxies should stir gas, leading to shocks and disturbed kinematics, as well as to some nuclear activity, though whether this means nuclear star formation or feeding of the active nucleus remains being disputed. With the size of our sample, we can not only test those hypotheses, but we are also able to look for other correlations of the observed gas and shock parameters with e.g. the galaxy mass or the presence of other morphological features. Below we report on what correlations we found, their confidence level, their possible physical reasons, and their implications for gas flow in galaxies.

\subsection{The correlation between the kinematic category and the presence of large-scale bars}
\label{sec:correlation_between_kinematic_category_largescalebars}
One of the strongest observational findings of our analysis is that the two galaxies in our sample that have unperturbed kinematics, NGC\,4237 and NGC\,4380, are also the only two unbarred galaxies in our sample. Unlike the other galaxies we studied, these two galaxies exhibit very regular kinematic fields and they lack extended shock signatures. When attributed to their unbarred morphology, this observation confirms the significance of bars, causing perturbation in the disks, driving extended shocks and facilitating gas inflows to the nucleus. Although this conclusion is based on only two galaxies, statistically it appears significant, as with 2 unperturbed galaxies in a sample of 21, the likelihood of randomly selecting an unperturbed one is $2/21$, and if the presence of bar were unrelated to unperturbed gas kinematics, the likelihood of two unbarred galaxies displaying by chance unperturbed kinematics would be $(2/21)^2=0.9\%$.

Note that having only two unbarred galaxies in a sample of 21 is not representative of the population: according to \citet{Erwin_18}, in a sample of galaxies with stellar masses $\sim10^{10-11}M_{\odot}$, approximately 60\% are barred, so the remaining $\sim$40\% are unbarred. Taking this as a probability distribution, one might expect to find around 8 unbarred galaxies in a sample of 21, and the likelihood of having only two is 0.2\%. There are two sources of this discrepancy. First, in the original complete CBS sample of 53 galaxies, only 25\% are unbarred. This fraction is slightly reduced to 22\% when one limits the sample to 46 galaxies observed with MUSE. However, when the sample is further limited to galaxies with sufficient emission-line signal, this fraction drops to 12\%: only 20\% (2 out of 10) of unbarred galaxies in the subsample observed by MUSE have emission-line signal above our threshold, while among barred galaxies 47\% (17 out of 36) do.

Therefore, our analysis implies that in the CBS sample observed by MUSE, in which the fraction of barred galaxies is not far off the general population, 47\% of barred galaxies show gas emission in their central kpc sufficient for gas dynamics to be characterized, and in all cases this is a disturbed gas dynamics. On the other hand, in the same sample, only 20\% of unbarred galaxies show gas emission in their central kpc sufficient for gas dynamics to be characterized, and in all cases this dynamics is unperturbed.

\subsection{Correlations with the presence of inner bars and nuclear rings}
\label{sec:shocks_nuclearbars_nuclearrings}

Inner bars and nuclear rings play important roles in the transport of gas to the nucleus. While inner bars are seen as one of the primary mechanisms of inflow to the nucleus of galaxies \citep{Shlosman_1989,Shlosman_01,Martini_2003b,Namekata_09,de_Lorenzo-Caceres_2013,Li_2023}, nuclear rings are typically seen as barriers of gas transport to the nucleus, as the inflowing gas accumulates in them \citep{Piner_1995,Sheth_2005,Sormani_15}. 

In this section, we explore the correlation between the presence of extended shocks and the occurrence of inner bars and nuclear rings as identified within CBS. The CBS sample spans a diverse range of galaxies and is designed to consistently extract central stellar structures, including nuclear star clusters \citep{Ashok_2023}, inner bars, and nuclear rings. \textcolor{black}{Table\,\ref{tab:galaxies_shock_classes}} indicates whether the galaxies studied here host inner bars or rings, based on the CBS classification, which is carried out using a combination of isophotal ellipse fitting, unsharp masking, and visual inspection of optical and near-IR images \citep{Erwin_21,Erwin_24}. Unsharp masking is particularly valuable for confirming weak inner bars and for distinguishing them from nuclear spirals, discs, or rings, while optical imaging highlights the dust and star formation that reveal nuclear rings. Nuclear rings are further classified as star-forming, stellar, or dusty, depending on their colours, morphology, and associated emission features. While methodological details are discussed in \citet{Erwin_21,Erwin_24}, further refinements to the measurements of CBS galaxies will be presented in the forthcoming CBS publication.

\subsubsection{The correlation between the nuclear emission type and the presence of inner bars}

Amongst the 21 galaxies of our sample, 8 have inner bars, while 12 do not show an evidence of one. One galaxy, namely NGC\,1566, exhibits a potential presence of an inner bar. Because of this uncertainty, this galaxy is excluded from further statistical analysis. Therefore, the total number of galaxies considered for examining the correlation between the presence of extended shocks and inner bars is 20. \textcolor{black}{Figure\,\ref{fig:shocks_nuclearbar} }presents the fraction of galaxies in each kinematic category, separated according to the presence or absence of inner bars. Each galaxy is colour-coded according to the nuclear emission type derived from the BPT diagram shown in \textcolor{black}{Fig.\,\ref{fig:BPT_global}}.

Although the distribution of galaxies among kinematic categories does not seem to depend on whether the galaxy hosts an inner bar or not, \textcolor{black}{Fig.\,\ref{fig:shocks_nuclearbar}} reveals that all 8 galaxies with inner bars exhibit either LINER or Seyfert emission dominating their inner 150\,pc. Noticeably, no galaxies with inner bars have their innermost 150\,pc dominated by star formation or show composite emission in that region. On the other hand, the nuclear emission in the remaining 12 galaxies that do not have inner bars spans all categories: we classify 6 of those galaxies as composite, 5 as Seyfert/LINER and 1 as star-forming. Overall, there are 13 galaxies classified Seyfert/LINER in our sample of 20 galaxies. If we assume that the probability of showing a Seyfert/LINER emission by a galaxy from our sample is 13 out of 20, and that this probability is independent of whether the galaxy is hosting an inner bar, then the likelihood of all 8 galaxies in the subsample with an inner bar showing Seyfert/LINER emission is $(13/20)^8 \simeq 3\%$. In other words, with 97\% confidence we can say that the distributions of nuclear emission types for galaxies with and without inner bars differ: galaxies with inner bars are more likely to display Seyfert/LINER nuclear emission.

Different distributions of galaxies with and without inner bars are evident in the Global BPT diagram for our sample presented in \textcolor{black}{Fig.\,\ref{fig:BPT_global_innerbar}} (see Sect.\,\ref{sec:grouping-nuclear-emission} for an explanation of the diagnostic lines), colour–coded by the presence or absence of inner bars: galaxies hosting inner bars are found exclusively in the LINER or Seyfert regions of the diagram, with none occupying the star–forming or composite regions. By contrast, galaxies without inner bars span the full range of emission types. There is a clear trend in the diagram, where at low line ratios, only galaxies without inner bar are present, while galaxies with inner bar start appearing at intermediate line ratios, and eventually dominate at the highest line ratios. The BPT diagram therefore provides a clear visual confirmation of the trend that inner bars are associated with AGN-like nuclear emission.

Although the trend that we found does not imply that inner bars drive inflow to the galactic nuclei, it suggests that inner bars may inhibit star formation in the innermost regions, which in galaxies without inner bars can proceed unsuppressed.

\begin{figure}
  \hspace{-0.35cm}
     \includegraphics[width=0.5\textwidth]{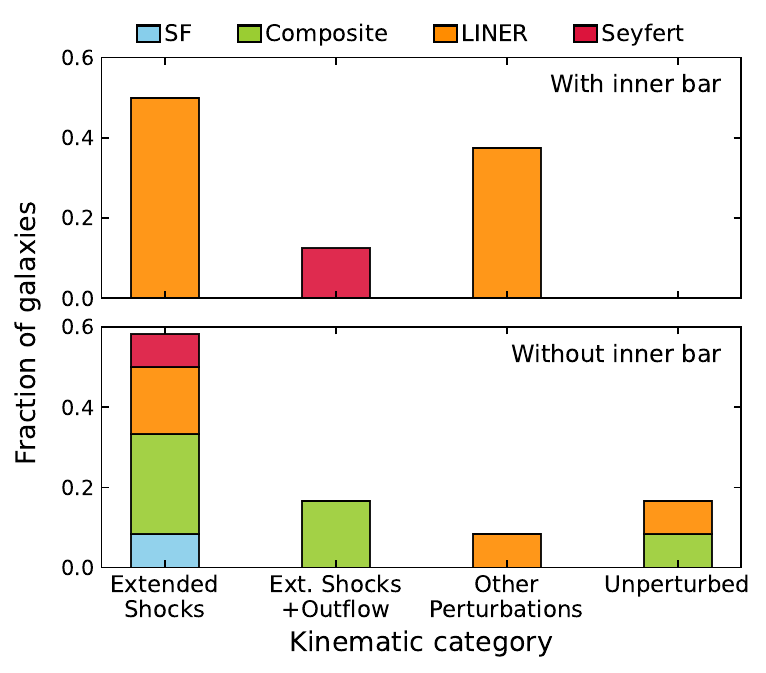}
    \vspace{-0.65cm}
    \caption{The fraction of galaxies in each kinematic category, grouped by the presence (top) and absence (bottom) of inner bars. The colours represent different nuclear emission types assigned to galaxies based on the BPT diagram of the sample: SF in blue, composite in green, LINER in orange, and Seyfert in red.}
    \label{fig:shocks_nuclearbar}
\end{figure}

\begin{figure}
    \hspace{-0.4cm}
    \includegraphics[width=1.05\linewidth]{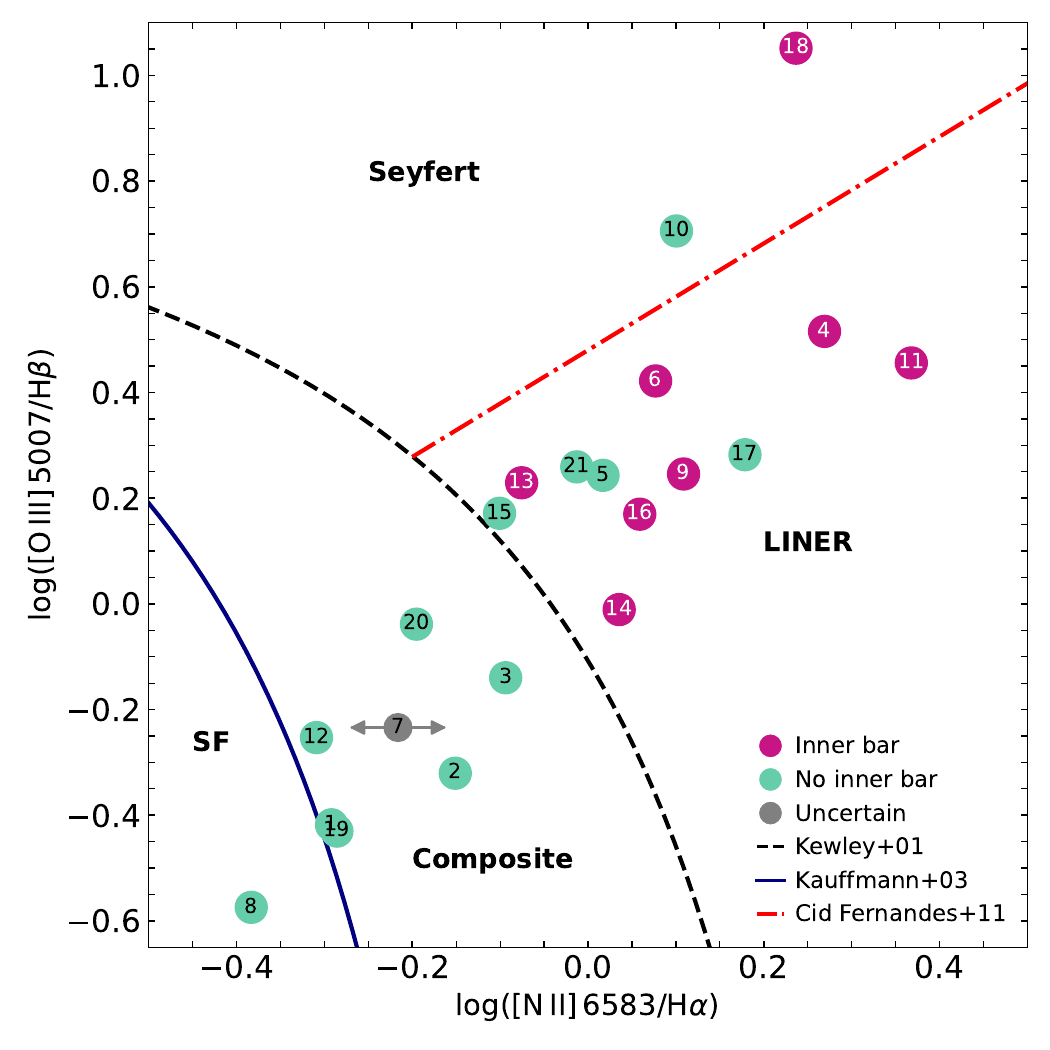}
    \caption{The BPT diagram for the galaxy sample analysed in this study, constructed using integrated emission-line flux ratios within a 150\,pc radius. The demarcation lines, identical to those in \textcolor{black}{Fig.\,\ref{fig:BPT_global}}, indicate the regions corresponding to different ionisation mechanisms: SF, Composite, LINER, and Seyfert. The numbers assigned to each galaxy correspond to the indexes shown in \textcolor{black}{Table\,\ref{tab:galaxies_shock_classes}}. Each point is colour-coded to indicate the presence (pink), absence (green), or uncertain (gray) classification of inner bars in the respective galaxies.   
    }

    \label{fig:BPT_global_innerbar}
\end{figure}

\subsubsection{The correlation between the kinematic category and the presence of nuclear rings}
Amongst 21 galaxies in our sample, 9 galaxies host nuclear rings, while another 9 galaxies do not. As for the remaining 3 galaxies, two of them are classified as potentially having a nuclear ring, one of them being NGC\,1566, which also has an uncertain nuclear emission type, and for one galaxy (marked with a ``?" in the nuclear ring column of Table\,\ref{tab:galaxies_shock_classes}), the presence of a nuclear ring has not been verified. Since these 3 galaxies do not provide the presence of nuclear rings confidently, we exclude them from our statistical analysis and description. Therefore, the total number of galaxies we consider when examining the correlation between the presence of nuclear rings and extended shocks is 18.

\textcolor{black}{Figure\,\ref{fig:shocks_nuclearring}} presents the fraction of galaxies in each kinematic category, grouped by the presence (top) and the absence (bottom) of nuclear rings within their inner kpc. As in \textcolor{black}{Fig.\,\ref{fig:shocks_nuclearbar}}, the colours represent galaxies with different nuclear emission types. While, contrary to what we found in the case of inner bars, we find no significant correlation between the presence of nuclear rings and the nuclear emission type, the distribution of galaxies among kinematic categories seen \textcolor{black}{Fig.\,\ref{fig:shocks_nuclearring}} appears to depend on whether the galaxy hosts a nuclear ring or not, again contrary to the case of inner bars seen in \textcolor{black}{Fig.\,\ref{fig:shocks_nuclearbar}}. \textcolor{black}{Approximately 78\% of galaxies} (7 out of 9) with nuclear rings are classified under in \textcolor{black}{\textit{extended shocks}} category. On the other hand, the fraction of galaxies {\it without} nuclear rings in the same category is low, \textcolor{black}{approximately 33\% (3 out of 9)}. Moreover, \textcolor{black}{$\sim 33\%$ (3 out of 9)} of galaxies without nuclear rings are in the \textit{other perturbations} category, while for galaxies {\it with} nuclear rings this fraction is only \textcolor{black}{11\%} (1 out of 9).

Although our sample is small, the observed trends support the concept of gas inflow along the bars contributing to the formation of the nuclear rings within the inner kpc of the galaxies. Moreover, strong disturbances caused by other perturbations, such as local SF and AGN activities, within the inner kpc of galaxies, which fall under the \textit{other perturbations} category, might inhibit the formation of nuclear rings in these galaxies.

\begin{figure}
  \hspace{-0.45cm}
      \includegraphics[width=0.5\textwidth]{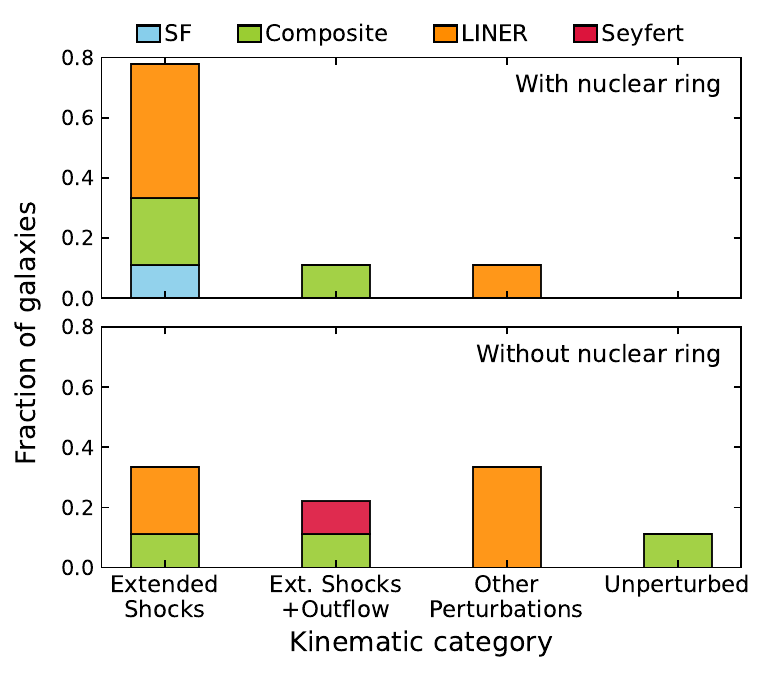}
    \vspace{-0.65cm}
    \caption{The fraction of galaxies in each kinematic category, grouped by the presence (top) and absence (bottom) of nuclear rings. The colours represent different nuclear emission types assigned to the galaxies based on the BPT diagram of the sample: SF in blue, composite in green, LINER in orange, and Seyfert in red.}
    \label{fig:shocks_nuclearring}

\end{figure}

\subsection{Relating the strength of extended shocks to the central gas velocity dispersion}
\label{sec:average_perturbation_against_shock_amplitudes}

As shown in diagnostic maps presented for each galaxy individually in Appendix\,\ref{sec:notes-on-galaxies}, the velocity dispersion of all emission line groups across the sample shows a large variety. While some galaxies, such as the ones with AGN outflows, show extremely high velocity dispersions, less perturbed systems like NGC\,5248, have smaller velocity dispersions across the MUSE FOV. Similar to AGN outflows, extended shocks with large amplitudes can potentially cause turbulence in the gas of galaxies, elevating velocity dispersions. Below we explore the correlation between the strength of extended shocks and the average \ha velocity dispersion within the inner kpc region of the galaxies in our sample. 

\subsubsection{Determination of shock strength and central gas velocity dispersion}
\label{sec:determining_averageHalpha_dispersion}

We quantify the strength of the shock as the deprojected residual velocity magnitude seen in the residual velocity maps and introduced in Sect.\,\ref{sec:determining_velocity_jumps}. In \textcolor{black}{Table\,\ref{tab:galaxies_shock_classes}}, four values of this magnitude are given: the largest and the smallest magnitude along the blueshifted and the redshifted structures in the residual velocity maps that we associate with the shocks. Thus we have two ranges of velocity magnitude, one for each Doppler shift, and below we analyze each range separately, with the values for the blueshifted one carrying the minus sign.

\begin{figure}
    \hspace{-0.4cm}
    \includegraphics[width=1.05\linewidth]{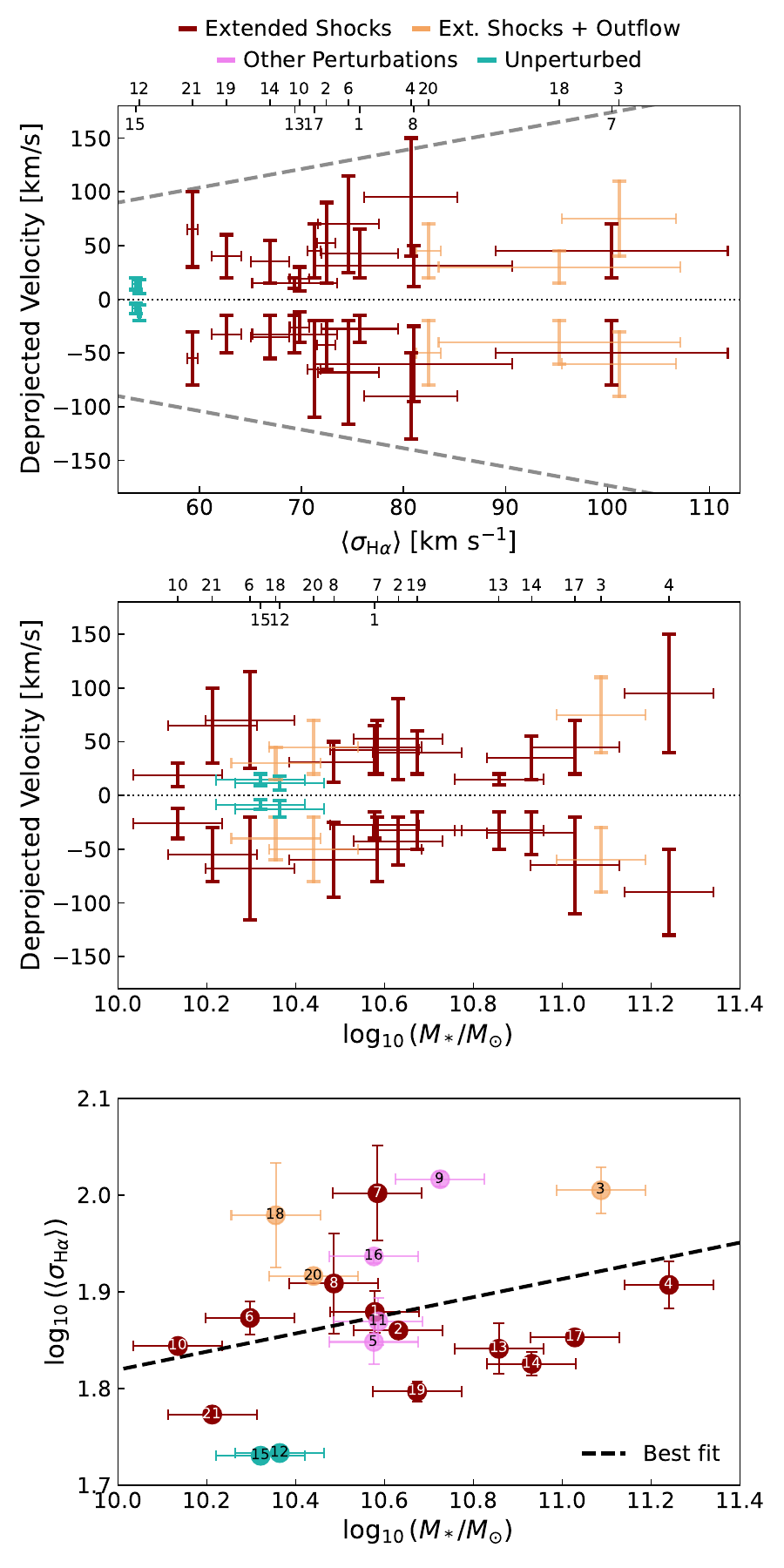}
    \vspace{-0.25cm}
    \caption{\textit{Top \& middle:} The range (minimum to maximum) of positive and negative deprojected residual velocity magnitudes measured along the extended kinematic structures (vertical bars), plotted against the average \ha velocity dispersion in the inner kpc of the galaxies (top) and stellar mass (middle). Gray dashed lines in the top panel mark deprojected velocity equal to $\sqrt{3}\sigma_{\rm H \alpha}$. \textit{Bottom:} Average \ha velocity dispersion plotted against stellar mass. Black dashed line shows unweighted least squares fit to the data points. In all panels, the index numbers of galaxies, specified in \textcolor{black}{Table}\,\ref{tab:galaxies_shock_classes} index column, are indicated either on the top axis or on the scatter points. \textcolor{black}{The markers are colour-coded as described in the legend according to the categories assigned based on the extended shock signatures.}}
    \label{fig:amplitudes_global_velocity_dispersion_mass}
\end{figure} 

In order to estimate the average \ha velocity dispersion within the inner kpc region for each galaxy, we select from the MUSE data spaxels within an elliptical region with the major axis aligned with the line of nodes, whose semi-major axis is 1\,kpc and semi-minor axis is $\cos{i}$\,kpc, where $i$ is the inclination of the galaxy given in Table\,\ref{tab:Disk_orientation_parameters}, together with its PA. We then apply the amplitude-over-noise (AON) threshold of 20 \citep{Sarzi_06} to exclude noisy spaxels within the defined elliptical region from calculating the average velocity dispersion. This threshold is determined by examining the spread of the \ha velocity dispersion measured in the individual spaxels as a function of the AON for each galaxy individually, as presented in Appendix\,\ref{app:suplementary_sample_plots}, \textcolor{black}{Fig.\,\ref{fig:AON_distributions}}. Points with low AON and high velocity dispersion indicate measurements from noisy spaxels, where the velocity dispersions are artificially high due to inaccurate Gaussian fitting, likely caused by low signal. 

After excluding the noisy spaxels, we compute the average \ha velocity dispersion, by taking an unweighted mean. This method of calculating the central velocity dispersion is not affected by the overall rotation, contrary to the line-width estimates for aperture-integrated spectra. The formal error on this quantity is very small ($\sim0.1$\,\kms) because of the large number of spaxels that we average over. However, it is not this random error, but a systematic error that dominates the uncertainty of the central velocity dispersion value. The true value should be the mean over the spaxels weighted by the gas mass in each spaxel, which is different from the unweighted mean. An AON-weighted mean, such that spaxels with higher AON contribute more strongly while lower-quality measurements carry less weight, or a mean weighted by the \ha\ flux in the spaxel are alternative ways of estimating the central velocity dispersion value, and although they are poor substitutes to weighting by mass, they give an indication of the systematic error associated with the derived central velocity dispersion value.

Therefore, for each galaxy in our sample, we calculated the unweighted, AON-weighted, and flux-weighted means of the velocity dispersion over selected spaxels which are presented in Table\,\ref{tab:gas_disturbance_weigthed}. For most galaxies all three means are within less than 25 \kms from each other, except for galaxies where AGN outflows interacting with the galaxy disc create bright regions of high velocity dispersion. In those galaxies, the flux-weighted mean can be twice higher than the other two. Therefore, we calculated the final average \ha velocity dispersion as the average of the unweighted and the AON-weighted means only, discarding the flux-weighted mean. Note that for galaxies under extended shocks and outflow category, velocity dispersions evaluated in this way will systematically differ from the ones derived from aperture-integrated spectra, which are intrinsically flux-weighted. The final average \ha velocity dispersion values for all galaxies are presented in \textcolor{black}{Table\,\ref{tab:galaxies_shock_classes}}. The errors given there are half the difference between the means, and therefore they indicate  uncertainty of the result against different reasonable weighting schemes.

\subsubsection{The role of extended shocks in driving gas turbulence in the central regions}
\label{sec:role_of_extended_shocks_driving_gas_turbulence}

In the top panel of \textcolor{black}{Fig.\,\ref{fig:amplitudes_global_velocity_dispersion_mass}, for each galaxy, in which we detected extended kinematic structures, the range of deprojected residual velocity magnitudes measured along these structures, marked by the vertical bars, is plotted as a function of its central gas (\ha) velocity dispersion.} The positive magnitudes give the values for the redshifted structures, and the negative magnitudes give them for the blueshifted ones. The \textit{unperturbed} category is included for comparison, as in galaxies in this category we also identified coherent kinematic structures for which we measured the velocity magnitudes, though these magnitudes are smaller than those expected in shocks. On the other hand, galaxies in the \textit{other perturbations} category are omitted as they lack structures associated with extended shocks. 

Generally, galaxies exhibiting larger deprojected residual velocity magnitudes also tend to display higher average velocity dispersions, especially for velocity dispersions up to $\sim$80\,\kms. Although this correlation is relatively weak, it suggests that extended shocks with large deprojected velocity magnitudes may induce turbulence in the interstellar medium, thereby increasing the overall level of kinematic disturbance and increasing the observed velocity dispersion. Along the dashed gray lines in this panel, the velocity is $\sqrt{3}$ times the observed one-dimensional velocity dispersion. Under the assumption of isotropic turbulent motions, this means that along these lines the energy in the turbulent motion is the same as the one associated with the deprojected residual velocity magnitude in the shocks. Thus this line allows us to relate the energy deposited to gas in shocks to the energy in turbulent motion. For points close to this line, the observed velocity dispersion could be attributed to disturbance caused by extended shocks.

For all galaxies in our sample, the bars indicating the ranges of velocity magnitudes in shocks lie in between the two dashed lines, which means that the energy deposited by shocks is smaller than that in the turbulent motions. Therefore the shocks cannot fully account for gas turbulent motions in the centres of galaxies, although they are likely to contribute to turbulence, as evidenced by the correlation between their strength and gas central velocity dispersion discussed above. This contribution can be significant, especially for NGC\,1097 (index 4), for which the bars in \textcolor{black}{Fig.\,\ref{fig:amplitudes_global_velocity_dispersion_mass}} reach close to the gray dashed line. However, this contribution is small compared to that from the AGN, as the orange-brown bars in \textcolor{black}{Fig.\,\ref{fig:amplitudes_global_velocity_dispersion_mass}}, representing the galaxies in \textit{extended shocks and outflow} category, indicate. For all three galaxies in this category, central gas velocity dispersions are higher than in the rest of the sample, and the bars, representing velocity magnitudes in extended shocks, are well within the dashed gray lines. This means that the turbulence in gas in the central kiloparsec of those galaxies is most likely mainly driven by AGN outflows. Also, for galaxies in our \textit{unperturbed} category (cyan bars in \textcolor{black}{Fig.\,\ref{fig:amplitudes_global_velocity_dispersion_mass}}), the velocity magnitudes are very small, around 10\,\kms (so that we do not classify the extended kinematic features there as shock structures), yet the observed central gas velocity dispersions there exceed 50\,\kms. Clearly something else drives central gas turbulence in those galaxies.

\subsection{\textcolor{black}{Dependence of the observed quantities on host galaxy mass}}
\label{sec:shocks_mass}

Following the CBS selection criteria, our sample consists of galaxies with masses differing by over one order of magnitude ($10^{10.1} - 10^{11.2}\,M_{\odot}$). Below we investigate how characteristics in our sample change with galaxy mass.
In the middle panel of Fig.\,\ref{fig:amplitudes_global_velocity_dispersion_mass}, we present the deprojected velocity magnitudes of extended shocks as a function of the stellar mass of the host galaxy listed in Table \ref{tab:targets-general}. There appears to be a positive trend, suggesting that stronger (higher-magnitude) shocks preferentially occur in galaxies with larger stellar masses, but this trend is weak. For example, a relatively low-mass galaxy such as NGC\,1433 ($10^{10.3}\,M_{\odot}$) may host shocks with deprojected velocity magnitudes up to 115\,\kms, almost as strong as those in a high-mass galaxy such as NGC\,1097 ($10^{11.24}\,M_{\odot}$), where shock velocity magnitudes reach 150\,\kms.

The Faber-Jackson relation \citep{Faber_Jackson_1976} links galaxy luminosity, and thus indirectly its mass, to its central stellar velocity dispersion. If there were a similar relation to the {\it gas} velocity dispersion studied above, the mass dependence could affect our results presented in \textcolor{black}{Sect.\,\ref{sec:role_of_extended_shocks_driving_gas_turbulence}}. The bottom panel of \textcolor{black}{Fig.\,\ref{fig:amplitudes_global_velocity_dispersion_mass}} shows the \ha velocity dispersion in the central kpc as a function of stellar mass for all galaxies in our sample, colour-coded by kinematic category. There is a very weak trend for more massive galaxies to display higher velocity dispersion, but the plot is dominated by scatter, and therefore such weak dependence of velocity dispersion on mass does not affect our conclusions in \textcolor{black}{Sect.\,\ref{sec:role_of_extended_shocks_driving_gas_turbulence}}. Nevertheless, the plot reveals a clear regularity in the distribution of galaxies across the different kinematic categories: \textit{unperturbed} galaxies lie below the average mass--dispersion relation marked by the dashed line, whereas galaxies in the \textit{extended shocks and outflows} category lie clearly above it.
This distribution supports our findings from Sect.\,\ref{sec:role_of_extended_shocks_driving_gas_turbulence}, even in the presence of the slight dependence of gas velocity dispersion on galaxy mass. Extended shocks do contribute to gas turbulence in the central regions, since galaxies without them show lower velocity dispersions than galaxies with them, but this contribution appears to be overtaken by AGN activity in galaxies with strong outflows.

Galaxies in the kinematic category \textit{other perturbations} exhibit a wide range of velocity dispersions, reflecting the diverse non-shock processes that disturb the gas, and they all have masses below $10^{10.8}\,M_{\odot}$, indicating that those perturbations are limited to lower-mass galaxies. This may be related to the fact that gas in higher-mass galaxies appears to move in a more orderly way under the stronger influence of the overall gravitational potential.

\section{Summary \& Conclusions}
\label{sec:conclusion}
Using MUSE/VLT data, we analysed ionised gas kinematic maps of 20 galaxies from the Composite Bulges Survey (CBS) and other ancillary surveys to identify signatures of extended shocks, similar to those found along bars, that can reach the innermost regions of galaxies, and to assess whether such shocks are the dominant mechanism driving inflow towards galactic nuclei (inner $\sim$100pc). This study builds on our pilot study on NGC\,1097 \citep{Kolcu_23}, where we developed a consistent methodology to highlight signatures of extended shocks in galaxy kinematic maps using both model-dependent and independent approaches.

Extended shocks driven by the stellar gravitational potential appear as coherent velocity jumps in gas kinematic maps, with large distinctive velocities. Their appearance can be distorted by inaccurate extraction of gas kinematics from the data, as well as by local perturbations in gas, such as stellar outflows or nuclear activity. Where line profiles show large excess of flux over a single Gaussian fit, we use multiple components in emission-line fitting to account for emission from multiple sources along the line of sight such as AGN outflows. Moreover, we show that removal of the global rotational flow by subtracting the circular velocity of a fitted flat disk can produce artefacts that obscure signatures of the extended shocks in the residual velocities if the fit is inaccurate. In order to exclude those artefacts, we examine residual velocity fields for disc models with a range of parameters. As an alternative, we examine the velocity differences of \ha and \nii emission lines. Moreover, we analyse the origin of ionisation through diagnostic line flux ratios and examine the dust morphology. Together with our kinematic study, these provide complementary evidence supporting our interpretation of extended shocks in the galaxies.

The individual analysis of the galaxies in our sample through various diagnostics revealed complex kinematic fields with unique structures in each galaxy. Finding common features among them allowed us to classify the galaxies in our sample into four kinematic categories based on the presence and characteristics of extended shock signatures identified through their diagnostic maps. \textcolor{black}{These categories include: (i) \textit{extended shocks}, where galaxies exhibit large-scale kinematic structures with high deprojected velocity magnitudes that are indicative of extended shocks.; (ii) \textit{extended shocks and outflow}, in which outflows from the AGN dominate the innermost regions of galaxies, preventing tracing the shock structures to the nucleus; (iii) \textit{other perturbations}, where the velocity field is clearly perturbed but it lacks extended shock signatures -- kinematics are likely dominated by other perturbations, such as local star formation or  AGN activities;} (iv) \textit{unperturbed}, where galaxies have relatively unperturbed kinematic maps with no evidence of shock activity.

The statistical analysis across the sample 
reveals the following.
\begin{enumerate}[label=\roman*., align=left, leftmargin=*]
    \item 
    \textcolor{black}{12 out of 21 galaxies in our sample ($\sim$57\%)} exhibit extended shock signatures within their inner kiloparsec. In \textcolor{black}{9} of these galaxies, the shock features can be traced down to the innermost 1--2 arcseconds (corresponding to $\sim$100\,pc, depending on distance), which is effectively MUSE resolution limit for finding extended structures, as they must stretch over the lengths of several times the spatial resolution in order to be identified. In the remaining three cases, extended shocks cannot be traced further in than $\sim$200--300\,pc. These findings suggest that shocks may represent a dominant mechanism driving gas inflow toward the nucleus. The \textcolor{black}{$\sim$57\%} fraction likely represents a lower limit, as galaxies classified under \textcolor{black}{\textit{extended shocks and outflow} ($\sim$14\% of the sample)} may also host bar-driven shocks extending into the nucleus, but their signatures are obscured by AGN outflows.

    \item Galaxies in the \textcolor{black}{\textit{extended shocks}} classes show deprojected residual velocity magnitudes of 50\,\kms or more across the shock signatures, consistent with models of bar driven shocks. 

    \item \textcolor{black}{Galaxies in the \textit{extended shocks} category have on average higher gas velocity dispersion in their inner kpc than galaxies classified as \textit{unperturbed}, but lower than galaxies in the \textit{extended shocks and outflow} category. Also, the deprojected residual velocity magnitudes in extended kinematic structures shows weak correlation with average gas velocity dispersion within the inner kpc.} This suggests that extended shocks contribute to turbulence in the host galaxy.
    
    \item We find no distinct correlation between the nuclear emission type and kinematic category. This may be attributed to the differing timescales of gas inflow and AGN activity, as well as the small sample size. 
        
\end{enumerate}

Further examining the correlation between extended shock signatures and the central stellar structures recovered in the CBS, we find that:

\begin{enumerate}[label=\roman*., align=left, leftmargin=*]
    \item The two unbarred galaxies in our sample are the only galaxies with unperturbed kinematic maps, showing no evidence of shock activity. Moreover, great majority (8 out of 10) of unbarred galaxies satisfying our selection criteria does not exhibit sufficient ionized emission to be included in our analysis. Our study, limited to the stellar mass range $10^{10.1} - 10^{11.2}\,M_{\odot}$, shows that while centres of barred galaxies usually show strong ionised emission and signatures of extended shocks, centres of unbarred ones show either no signatures of extended shocks, or ionised emission to weak to derive gas kinematics. 

    \item The nuclear emission in galaxies with inner bars is either of LINER- or of Seyfert-type, while in galaxies without inner bars it shows no preference, with star-forming or composite emission equally likely. This indicates that inner bars suppress SF activity in the innermost regions of galaxies. We found no statistical significance between the presence or absence of inner bars and any kinematic category.

    \item \textcolor{black}{Approximately 78\% of galaxies hosting nuclear rings are classified under \textit{extended shocks} vategory, compared to only 33\% of galaxies without nuclear rings.} 
\end{enumerate}

Methodology used in this work provides new means of finding coherent kinematic structures associated with extended shocks -- a mechanism which could be the predominant driver of inflows to the inner $\sim$100--150\,pc of galaxies. Although the sample size we presented is limited, this work offers a tailored strategy to characterise extended shocks over a broad range of galaxy types. Expanding this study to a larger sample size could strengthen the statistical significance of our findings by potentially uncovering stronger correlations between the presence of extended shocks and the characteristics of the galaxies such as the nuclear emission types, stellar masses, and distinct central stellar structures.

The main work for this paper was completed in close collaboration with \textit{\textbf{Dr. Peter Erwin}}, whose untimely passing is deeply mourned. Peter’s sharp insight, kindness, and dedication to understanding galaxies shaped both this work and those who worked alongside him. He provided the data and decompositions that form the foundation of this paper, but his influence went far beyond that -- through his careful feedback, patience, and genuine curiosity. His enthusiasm for astronomy and his generosity with his time and knowledge will always be remembered. This paper stands as a small reflection of his lasting impact on our work.

\section*{Acknowledgements}
We thank the anonymous referee for comments that gave us opportunity to clarify the presentation of this work. We thank Luis C. Ho, Aaron Barth and the Carnegie--Irvine Galaxy Survey group for granting us access to large-scale images of several galaxies used in this work, Ivan Baldry for his insights into improving the statistical analysis in our study, Ric Davies for guiding us through various AGN diagnostics, and Roberto P. Saglia for the nice discussions on the text. TK acknowledges the joint PhD studentship support from Liverpool John Moores University, the Faculty of Engineering and Technology and the Science and Technology Facilities Council. TK is also grateful for the financial support from The Leverhulme Thrust. WM is grateful to MPE for its hospitality, which created opportunity to discuss this work with the Infrared/Submillimeter Astronomy Group there. DAG is
supported by STFC grants ST/T000244/1 and ST/X001075/1.
AdLC  acknowledges financial support from the Spanish Ministry of Science and Innovation (MICINN) through RYC2022-035838-I and PID2021-128131NB-I00 (CoBEARD project).

\section{Data Availability}
MUSE data used in this work are accessed via the data archive of the European Southern Observatory at \url{http://archive.eso.org/wdb/wdb/adp/phase3$_$main/form}, under proposal IDs 0104.B-0404(A) and 109.22VU.001 (PI: Peter Erwin) provided by Composite Bulges Survey (CBS) collaboration; 0097.B-0640 (PI: Dimitri Gadotti) provided by the Time Inference with MUSE in Extragalactic Rings (TIMER) collaboration; 0100.B-0116(A) and 096.B-0309(A) (PI: Marcella Carollo) provided by the MUSE Atlas of Disks (MAD) collaboration.

\label{lastpage}

\bibliographystyle{mnras}
\bibliography{references} 

\appendix

\input{notes_on_galaxies}

\input{supplementary_appendix_figures}


\end{document}

%% file: notes_on_galaxies.tex
\section{Individual analysis of sample galaxies}

\label{sec:notes-on-galaxies}

In the following subsections, we provide details on the analysis and specific results for each galaxy in our sample. The main results for each galaxy are visualised in multi-panel Figures \ref{fig:ic2051_4by4}--\ref{fig:NGC7513}, one for each galaxy, each following the same arrangement. Here we describe the contents of the panels, and we introduce their respective names (in italics) that we use in the remainder of this section. The top left panel shows a {\it large-scale image} of the galaxy, with the MUSE FOV marked with a red box.
These images are taken from either the Carnege-Irvine Galaxy Survey (\citealp{Ho_2011}; ``CGS'' in the lower-left corner of the image)\footnote{\url{https://cgs.obs.carnegiescience.edu/CGS/database_tables/sample1.html}} or from DR16 of SDSS (\citealp{Ahumada_20}; ``SDSS'' in the lower-left corner)\footnote{\url{https://skyserver.sdss.org/dr16/en/home.aspx}}. Immediately below this (leftmost panel in the second row) is a {\it colour image} created by dividing the 4650–9300 \AA\ wavelength range of MUSE data cubes into three segments: $\sim$ 4650--6200\AA, 6200--7750\AA, 7750--9300\AA. These segments were assigned to blue, green, and red channels and combined to produce the colour image. The cyan box overplotted on the colour image marks the inner 40\arcsec~$\times$ 40\arcsec~region which was further analysed. All other panels in the figure show this same 40\arcsec~$\times$ 40\arcsec~~field of view. Further below, in the leftmost panel of the third row, we show an {\it unsharp mask image} generated from the MUSE data following the procedure described in \textcolor{black}{Sect.\,\ref{method:dustmorphology}}.

The remaining three panels of the first row show the fluxes of \ha, \nii and \oiii emission lines (from left to right) in units $10^{-20}$ ergs$^{-1}$ cm$^{-2}$. For brevity, these are referred to as 
\textit{\ha}, \textit{\nii} and \textit{\oiii flux maps}.  White spaxels in the flux maps indicate flux below the detection limits. The three panels below them (the second row) show the LOS velocities for the three emission line groups defined in Section 3.2: the Balmer lines, the low-ionisation lines and the high-ionisation lines. Because of the lines dominating in each group, these are referred to as \textit{\ha}, \textit{\nii} and \textit{\oiii velocity maps}. In the corresponding panels of the third row, we plot the {\it dispersion maps} showing the velocity dispersion for the same line groups. Note that spaxels of low flux have inaccurate velocity and dispersion estimates, which results in areas of spaxels with close-to-random distribution of the kinematic values which should be disregarded.

Finally, in the fourth row, the leftmost two panels connected together show information from emission-line ratios: the {\it BPT map} to the left, where individual spaxels are color-coded by their location in the {\it BPT diagram} immediate to the right (see \textcolor{black}{Sect.\,\ref{met:bpt}} for the details on the construction of the BPT maps). The right two panels in the fourth row show two different maps created by \textit{subtracting} emission-line velocities. The first one, the {\it residual velocity map}, shows the observed Ha velocity minus the best-fitting circular-rotation model (see \textcolor{black}{Sect.\,\ref{met:residual_velocities}}). The second, the {\it velocity difference map} is created by subtracting the \ha velocity map from the \nii velocity map (see \textcolor{black}{Sect.\,\ref{met:velocity_differences}}).

The gray contours in the flux, velocity and dispersion maps, as well as in all the maps in the bottom row of panels, are isophotes of the H$\alpha$ emission. The cross mark is the centre of the galaxy. 

\subsection{IC2051}
\label{sec:IC2051_notes}
\begin{figure*}
    \centering
    \includegraphics[width=1\textwidth]{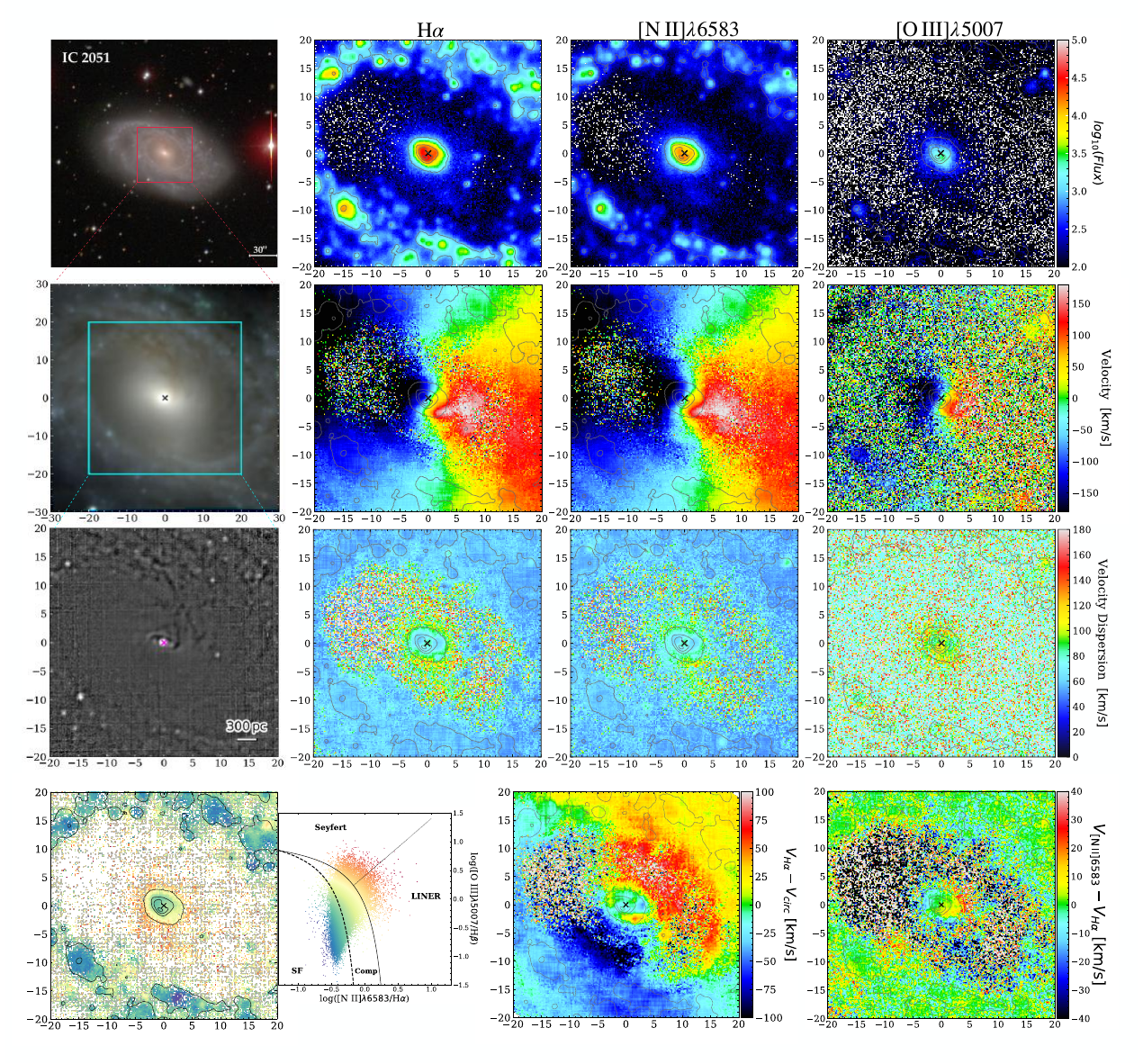}
    \caption{Images and diagnostic maps of IC\,2051. Panels show the same as in Fig.\,\ref{fig:NGC4303}}.
    \label{fig:ic2051_4by4}
\end{figure*}

The images and the diagnostic maps of IC\,2051 are presented in Fig.\,\ref{fig:ic2051_4by4}. As shown in the large-scale and colour images, within the MUSE FOV, the galaxy has a luminous core, a bar situated within the $\sim$11\arcsec~radius, a star-forming inner ring at approximately 12\arcsec~radius which encompasses the bar, and an outer ring encircling the inner one. The HST observations \citep{potw1950a} of the galaxy reveal a small dusty nuclear ring within the $\sim$2\arcsec~radius. The MUSE FOV covers the entire bar of the galaxy and a portion of the inner ring surrounding it. Starting from the inner edge of the inner ring, $\sim$11\arcsec~north from the centre, a dust lane stretches towards the south along the bar. After reaching $\sim$3\arcsec~west of the centre, the dust lane curves towards the centre connecting to the nuclear ring; a counterpart of this dust lane is seen to the north of the centre. These distinctive dust lane features are more prominent in the unsharp mask image of the galaxy, and resemble two curving dust lanes often seen in bars. 

As shown in the flux maps, the strongest emission in all three line groups is concentrated within the radius of $\sim$3\arcsec. Particularly, the \ha emission forms a crescent of radius $\sim$1--2\arcsec~southeast of the centre. Outside the $\sim$3\arcsec~radius, around the bar, the emission is strongly depleted. The inner ring has relatively high \ha and \nii emission, featuring clumps of enhancements dispersed throughout due to local star formation; the \oiii emission within the inner ring is relatively weak.

The velocity maps of all emission line groups show signs of non-circular motions throughout the entire MUSE FOV. The zero velocity line forms an ``S" shape, characteristic of the flow in bars. The velocity dispersion maps of \ha and \nii exhibit peak values at the centre, reaching roughly 100\kms. Past $\sim$1\arcsec~from the centre, the velocity dispersion drops and reaches the lowest values, approximately 50\kms, in two lobes at $\sim$2\arcsec~east and west of the centre. These two lobes extend beyond the crescent seen in \ha and \nii emission. Outside the two lobes, the velocity dispersion increases yet we can not reliably determine the velocity dispersion values beyond this region due to poor signal. The random motions of the gas within the inner ring are mild, with velocity dispersion values for \ha and \nii staying around 50--60\kms. 

Within the $\sim$3\arcsec~radius, the BPT map of the galaxy exhibits composite emission with contribution from SF, SF being the strongest within the $\sim$2\arcsec~radius, particularly within the crescent in \ha emission. This BPT pattern together with the dust and low velocity dispersion is consistent with the small nuclear ring of the galaxy. The inner ring encompassing the bar has prominent star formation activity throughout.

The poor signal within the bar region makes IC\,2051 a challenging target when fitting a circularly rotating disk to its velocity field. Results from \texttt{Kinemetry}, when run with free parameters and a fixed centre, show a roughly stable radial distribution of \textit{i} converging on  $\sim$41\degr, but the distribution of PA of the LON is fluctuating. In an attempt to obtain more consistent parameter distributions, we binned the velocity map by aggregating the data into 2$\times$2, 3$\times$3, and 4$\times$4 spaxel blocks. While this technique decreases the resolution of the velocity map, it increases the signal within each bin. However, even after applying binning, the results did not change. Therefore, we ran \texttt{Kinemetry} with different \textit{i} and PA of the LON values, trying to minimise the artefacts arising from their incorrect values in the residual velocity maps. A model fitted with a fixed centre, fixed \textit{i} at 41\degr and the PA of the LON at 60\degr effectively showed the minimum artefacts within the $\sim$3\arcsec~radius. Therefore, we fixed the parameters to these values for further analysis. However, these values differed from the values obtained in CBS from large-scale fits to outer isophotes (see Table\,\ref{tab:Disk_orientation_parameters}). The resultant residual velocity map is presented in Fig.\,\ref{fig:ic2051_4by4}. Since we aim to trace the inflow to the innermost regions, we focused on minimising the artefacts within the $\sim$3\arcsec~radius. Beyond this region, the artefacts of incorrectly fitted PA of the LON dominate the residual velocity map, observed as $m$=1 harmonic terms. 

The best fit of the rotating disk exposes two structures within the $\sim$4--5\arcsec~radius, exhibiting amplitudes of $\sim$30--40\kms in the residual velocity map, one with positive and one with negative velocity residuals. The corresponding minimum and maximum deprojected residual velocity amplitudes measured along these structures are $-15$ and $-40$\kms for negative and $20$ and $65$\kms for positive velocities. The feature exhibiting negative velocity residuals overlaps with the dust lane on the west, coinciding with its curve and extension towards the centre. In the velocity difference map, there is a coherent structure with positive values, with amplitude of $\sim$10--15\kms, extending from $\sim$4\arcsec~southwest of the centre towards $\sim$4\arcsec~east. One could argue that there is a counterpart of this feature in the northeast, exhibiting negative velocities around similar velocity amplitudes. The largest amplitudes of these two structures are co-spatial with the largest amplitudes of the features seen in the residual velocity map. 

\textcolor{black}{The coherent kinematic features in IC\,2051 extend over several hundred parsecs, which are outlined in Fig.\,\ref{fig:Sample_extended_structures_perturbed_galaxies_residualvel}, and exhibit large deprojected residual velocity magnitudes of up to $\sim$65\kms. We therefore attribute these features to coherent extended shocks in the gas. A summary of the properties of the extended structures, including their radial extents and velocity magnitudes, is provided in Table \ref{tab:galaxies_shock_classes}.} While a part of the gas shock structure seen in the residual velocity map coincides with the dust lanes, the structures seen in the velocity difference map are $\sim$1\arcsec~downstream from the dust lanes.

\subsection{NGC\,289}
\label{sec:NGC289_notes}

\begin{figure*}
    \centering
    \includegraphics[width=1\textwidth]{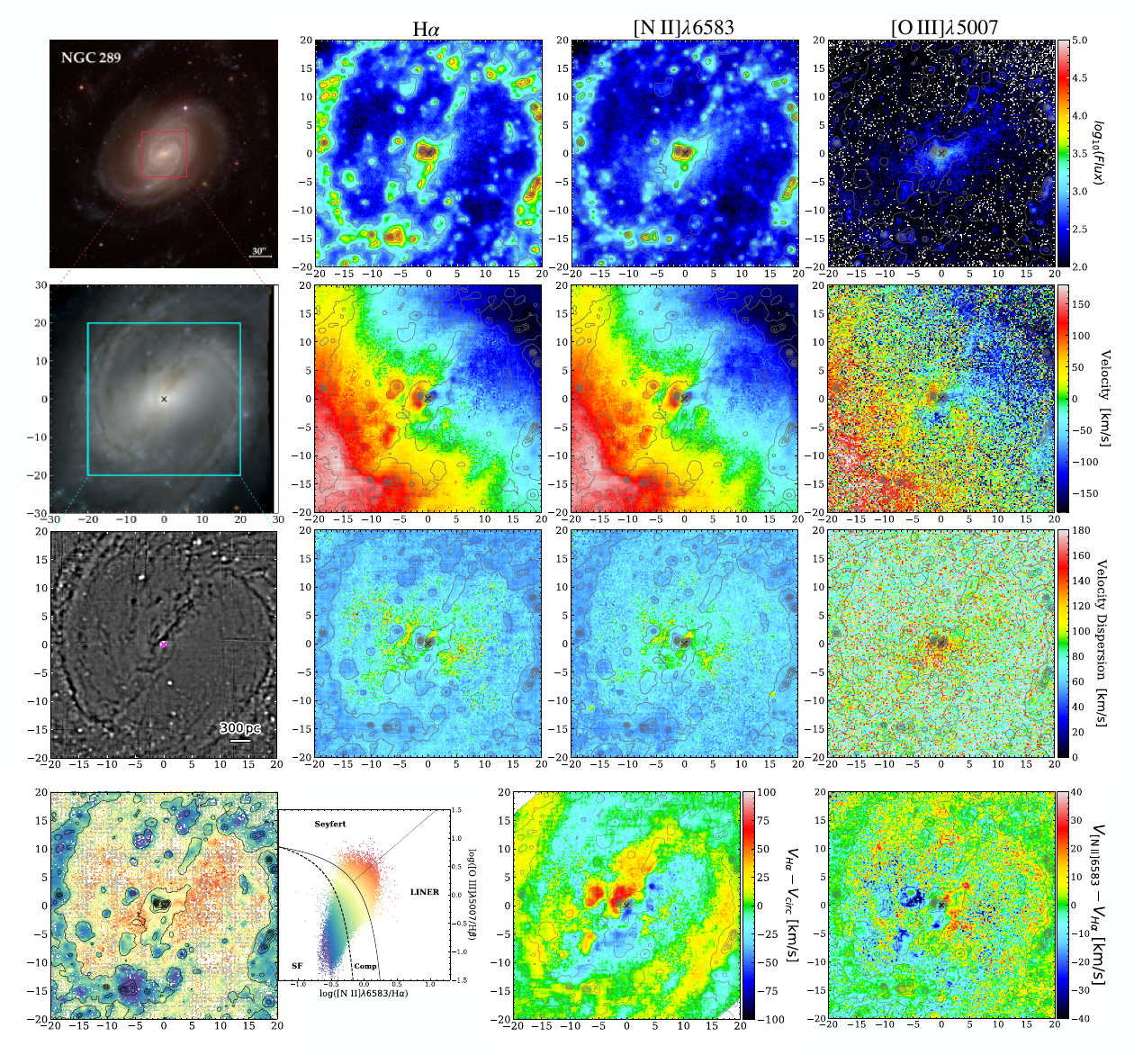}
    \caption{Images and diagnostics maps of NGC\,289. Panels show the same as in Fig.\,\ref{fig:NGC4303}}
    \label{fig:NGC289}
\end{figure*}

The images and diagnostic maps of NGC\,289 are presented in Fig.\,\ref{fig:NGC289}. As shown in large scale and colour images, NGC\,289 has a luminous core, a bar located within the $\sim$15\arcsec~radius and two star-forming spiral arms surrounding the bar. Both spiral arms extend from the bar region several tens of kpcs towards the outer parts of the galaxy, where they both appear disturbed; this disturbance is associated with the interacting companion of the galaxy \citep{ARP_1981,Bendo_Joseph_2004}. The MUSE FOV covers the entire bar region, together with a portion of the spiral arms. The dust lanes along the spiral arms and the bar are notably seen in the unsharp mask image of the galaxy. 

The \ha and \nii flux maps show two central emission peaks. One at the nucleus and the other $\sim$2\arcsec~east of the centre. These regions coincide with the region of a small nuclear disc with $\sim1\arcsec$ radius that was found in \citet{deSaFreitas_23}. Within the bar region, there is clumpy \ha emission likely originating from local SF, as well as extended emission elongated with the bar, close to the bar's major axis and on its leading side, likely related to shocks in the bar, which upon close inspection can be attributed to a set of aligned emission clumps. These structures are also seen, though relatively weaker, in \nii emission. The spiral arms show prominent \ha and \nii emission throughout their extent. Compared to \ha and \nii, \oiii emission is notably weaker and is mostly confined within the $\sim$4\arcsec~radius of the galaxy, its elongation at the larger radii not correlated with the orientation of the bar.

The \ha and \nii velocity maps show signs of non-circular motions, particularly along the zero velocity line where several distinct features are seen. Approximately 1.5--2\arcsec~towards the northwest and 2--3\arcsec~southeast from the centre, \ha and \nii velocities reach a local maximum of around 120\kms. These two regions are not co-spatial with the central peaks in the emission. However, throughout the region where emission is notably high, within the $\sim$3--4 radius, the \ha and \nii velocities also show an increase, with values $\sim$90--100\kms. Approximately 6\arcsec~northeast from the centre, a velocity peak with \ha and \nii velocities, $\sim$80--100\kms is seen; this coincides with a clumpy structure in \ha emission, which may indicate a young stellar cluster moving with respect to the ambient medium.

The centre of the galaxy exhibits \ha and \nii velocity dispersion values around 100--120\kms, spanning over a region from the centre $\sim$4\arcsec~towards the north and southeast, almost manifesting an ``S" shape. From the centre $\sim$2\arcsec~towards the east, velocity dispersion reaches a minimum, exhibiting values around 50\kms. This region coincides with a region with high \ha velocity. Moreover, $\sim$6\arcsec~towards the northeast from the centre, another velocity dispersion minimum is seen which coincides with the \ha clump of velocity higher than its surroundings. Several other features with low \ha and \nii velocity dispersion are seen along the bar, coinciding with the two regions with elevated \ha flux. Throughout the extent of the spiral arms, the velocity dispersion of both \ha and \nii remains around 50--60\kms.  

The disk orientation parameters, PA of the LON = 125\degr and \textit{i}=46\degr obtained from \texttt{Kinemetry} agree with the values from large-scale studies (\citealp{Walsh_1997}: PA of the LON = 130\degr and \textit{i}=46\degr). Thus we fit the rotating disk by fixing the parameters to these values. The residual velocity map of the galaxy reveals several distinct coherent structures. Outside the bar region, velocity residuals take a spiral pattern exhibiting amplitudes of $\sim$20--30\kms. Within the bar region, the velocity residuals form a straight feature with amplitudes of $\sim$50\kms, one with positive and one with negative velocities. The deprojected residual velocity amplitudes across these structures are large, up to 70--90\kms. The innermost part of the structure with positive velocities, at $\sim$2\arcsec~northeast from the centre, coincides with a region of high \ha and \nii velocities. The spiral and straight structure with positive velocity residuals meet at the northwest end of the bar. The straight structure with negative velocities extending along the bar passes the centre by $\sim$7\arcsec~towards the northwest end of the bar. Approximately 5\arcsec~northeast from the centre, high positive velocity residuals with amplitudes of $\sim$50\kms are seen in the \ha clump.

The velocity difference map is generally noisy. Along the bar, two structures with velocity difference amplitudes around $\sim$15--20\kms are visible, one stretching from the centre $\sim$5 arcseconds towards the south, and the other stretching from the centre $\sim$3\arcsec~towards the northwest. Approximately 4--5\arcsec~northeast of the centre, surrounding the clumpy region exhibiting high \ha flux, the velocity differences reach $\sim$30\kms likely indicating a spatial overlap of two regions with different physical conditions.

The BPT map of the galaxy reveals composite emission in the centre. Approximately 2\arcsec~east of the centre, coinciding with the off-centre \ha clump, contribution from SF is stronger than the central clump. Along the bar, the emission is of a composite nature with contribution from SF. Within the bar region, several distinct star-forming clumps are present, including the region at $\sim$6\arcsec~northeast from the centre, where we observed high \ha flux. Outside the star-forming clumps, the ionisation is dominated by the LINER region. Furthermore, the spiral arms predominantly show active SF throughout their entire extent.

Although rather discontinuous, coherent spiral and straight structures seen in the residual velocity map \textcolor{black}{extend across several hundreds of parsecs, exhibiting large deprojected velocity magnitudes of $\sim$60--90 \kms (see outlined features in  Fig.\,\ref{fig:Sample_extended_structures_perturbed_galaxies_residualvel}, also a summary of the properties of the extended structures, including their radial extents and velocity magnitudes, is provided in Table \ref{tab:galaxies_shock_classes}). Moreover, the alignment of these kinematic features with the dust lanes are compelling evidence of the presence of extended shocks in gas.} Additionally, we hypothesise that a portion of the straight feature with negative velocities, which passes the centre by $\sim$7\arcsec~towards the northwest end of the bar, is the overshot gas after being funnelled in towards the centre from the southeast end of the bar. Instead of falling further inwards, the gas continues to flow towards the northwest bar end, where it collides with the gas present there.


\subsection{NGC\,613}

\label{sec:NGC613_notes}

\begin{figure*}
    \centering
    \includegraphics[width=1\textwidth]{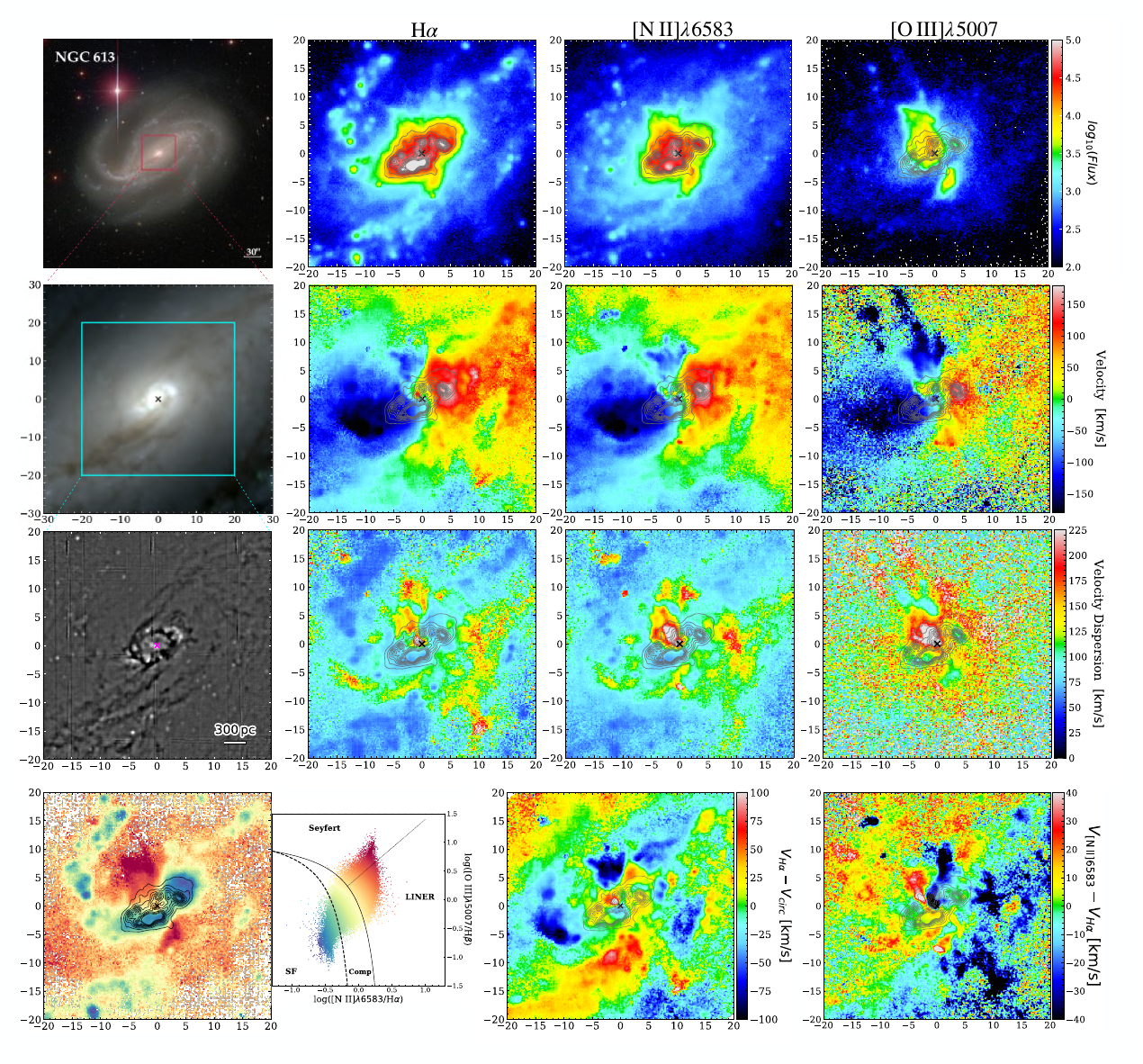}
    \caption{Images and diagnostics maps of NGC\,613. Panels show the same as in Fig.\,\ref{fig:NGC4303}}
    \label{fig:NGC613}
\end{figure*}

The images and diagnostic maps of NGC\,613 are presented in Fig.\,\ref{fig:NGC613}. As shown in the large scale and colour image, NGC\,613 has a luminous core, a star-forming nuclear ring of $\sim$6\arcsec~radius, a bar and star-forming spiral arms surrounding the bar region. The MUSE FOV encompasses the core, the nuclear ring and nearly the entire bar. Along the bar, two prominent dust lanes are seen. These dust lanes are prominent in the unsharp mask image of the galaxy, where they curve to follow the nuclear ring over a full circle.

As shown in the flux maps of the galaxy, the \ha and \nii emission is mostly confined within the $\sim$5--6\arcsec~radius. The \ha emission has several peaks in the nuclear ring and one $\sim$0.5\arcsec~southeast of the nucleus. The main peak of the \nii emission is $\sim$0.5\arcsec~southeast of the nucleus, the peaks in the nuclear ring correspond to the ones in \ha emission, but they are rather weaker than the nuclear peak. \oiii emission, rather strong and extended, is confined within a region stretching from the central $\sim$8--9\arcsec~towards the northeast, indicating possibly an off-plane outflow. Some \ha and more \nii emission appear to extend from the nuclear ring towards that direction as well. Moreover, there is enhanced emission across all line groups in an elongated region at $\sim$3--6\arcsec~southwest of the centre. While \ha and \nii emission in this region appears joined with the emission from the nuclear ring, \oiii emission appears notably separated. The separation overlaps with the region where the dust lane to the southeast connects to the nuclear ring. Clumpy \ha emission is seen along the bar lanes, but also towards the northeast of the centre, where it appears to group into a spiral arm offset tangentially by roughly 90\degr from the dust lanes.

The velocity fields of all emission line groups are strongly perturbed, showing signs of non-circular motions. Outside the nuclear ring, there is a region with large negative velocities across all line groups, covering a PA between 0 and 45\degr and extending radially out to $\sim$10\arcsec~in \ha and \nii and $\sim$15\arcsec~in \oiii, with amplitudes around $\sim$80--100\kms for \ha and \nii, and $\sim$150\kms for \oiii. This region appears separate from the main region with blueshifted velocities along the LON and is likely associated with the elevated \oiii emission and with the outflow.

Within a region extending from the centre towards $\sim$5--6\arcsec~northeast, the velocity dispersion across all emission line groups reaches a peak, with values exceeding 150\kms for \ha and \nii, and 200\kms for \oiii. This region coincides with the regions of negative velocities across all line groups. The velocity dispersion values of \ha and \nii in the nuclear ring are around $\sim$70\kms. Outside the nuclear ring, the velocity dispersion across all line groups is high, around $\sim$100--140\kms but in the spiral towards the northeast with clumpy \ha emission, \ha and \nii velocity dispersion values reach a minimum of $\sim$50--70\kms.

The disk orientation parameters, the PA of the LON=115\degr and \textit{i}=37\degr, obtained from \texttt{Kinemetry} agree well with the parameters estimated from large-scale fits; the PA of the LON differing by 5\degr from the value estimated from \citet{Bajaja_1995} (see Table\,\ref{tab:Disk_orientation_parameters}). The residual velocity map of the galaxy reveals several strong localised extremes within the nuclear ring, likely originating from the complex interaction between the inflow, SF and the AGN outflows. Throughout the region to the northeast of the nucleus, associated with AGN outflows, strong negative residual velocities with amplitudes exceeding 50\kms are seen. Coinciding with the dust lanes along the bar, two coherent structures with amplitudes $\sim$30--50\kms are seen, one exhibiting negative and one exhibiting positive velocity residuals. The corresponding deprojected residual velocity amplitudes across these coherent structures along the bar range from 30--40\kms to 90--110\kms. While the structure with positive velocities can be traced towards its connection to the nuclear ring, the structure with negative velocities is dominated by the strong AGN outflows likely located in front of it. To the east of the nuclear ring, there is a coherent region where the amplitudes of the residual velocity exceed 50\kms. This region is not co-spatial with the SF spiral structure with clumpy \ha emission seen in the BPT map; only their innermost parts overlap.

The velocity difference map of the galaxy exposes two coherent structures with amplitudes of $\sim$40--50\kms. One of these structures, characterised by negative velocity differences, extends from the central $\sim$10--11\arcsec~towards the north. The other structure, characterised by positive velocity differences, is situated $\sim$3\arcsec~east of the centre and extends $\sim$6\arcsec~towards the northeast. These two structures nearly encircle the area exhibiting coherent negative velocity residuals, which are associated with the AGN outflows. This suggests the presence of two distinct regions of gas with differing physical conditions along these structures, where there is a transition from emission dominated by the disc to that dominated by outflow. Moreover, there is a faint coherent continuous structure with amplitudes of $\sim$10--15\kms upstream from the coherent structure with positive residual velocities along the dust lane to the southeast. 

The core of NGC\,613 is classified as a LINER type \citep{Audibert_19}. In the BPT map, LINER emission is seen at the very core as well as in the northeast region, where velocity dispersion across all line groups peaks, and in the southeast region with enhanced emission across all line groups. The latter two regions resemble a bi-symmetric structure, co-spatial with pronounced \oiii emission. This confirms that these features are likely originating from AGN outflows; consistent with similar emission-line diagnostics. The interpretation of outflows in this galaxy is provided by \citet{Silva-Lima_2025}. However, the emission from a high energy mechanism occurs where the innermost part of the dust lane connects to the nuclear ring, which may indicate the inflowing gas is irradiated by the AGN, except for the northeast region, where the nuclear ring is dominated by SF. Furthermore, the clumpy \ha emission in the northeast appears as a coherent spiral in SF, winding from the northeast of the centre and connecting to the nuclear ring at the east. This spiral was not identified clearly in the emission or characterised by the dust lanes, but it is notably revealed in the BPT map.

\textcolor{black}{Various aspects of our methodology provide evidence for extended shocks along the dust lanes in the bar, connecting to the nuclear ring. As presented in Table\,\ref{tab:galaxies_shock_classes}, these structures span galactocentric radii from several hundred parsecs to a few kiloparsecs, with deprojected residual velocity magnitudes of approximately 90--110\kms.} However, the presence of strong AGN outflows within the nuclear ring prevents identifying the signatures of extended shocks to the innermost regions. \textcolor{black}{Therefore, based on the grouping described in Sect.\,\ref{sec:grouping-shock-class}, this galaxy is classified under \textit{extended shocks and outflows} kinematic category.}

\subsection{NGC\,1300}
\label{sec:NGC1300_notes}

\begin{figure*}
    \centering
    \includegraphics[width=1\textwidth]{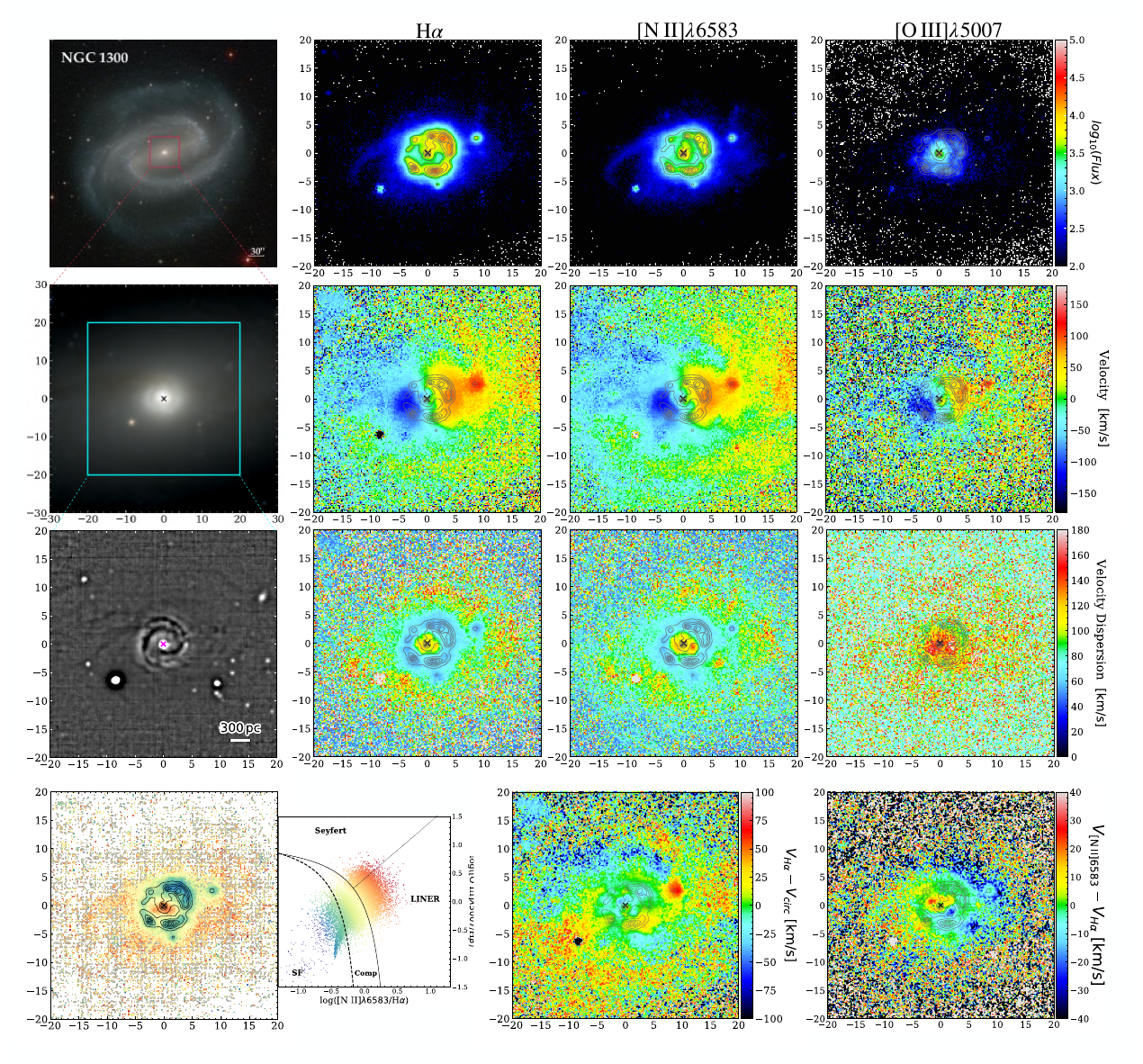}
    \caption{Images and diagnostics maps of NGC\,1300. Panels show the same as in Fig.\,\ref{fig:NGC4303}}
    \label{fig:NGC1300}
\end{figure*}

The images and the diagnostic maps of NGC\,1300 are presented in Fig.\,\ref{fig:NGC1300}. As can be seen in large scale and colour images, the galaxy has a luminous core encircled by a nuclear ring or disk and a prominent classical bar \citep{Lindblad_1996,Lindblad_1997}. It also has two star-forming spiral arms surrounding the bar region. The MUSE FOV encompasses the core, the nuclear ring, and approximately two-thirds of the bar. Along the bar, we see two dust lanes, one stretching from the east towards the centre, connecting to the nuclear ring from the north, and the other as the weak counterpart of the former, stretching from the west towards the centre, connecting to the nuclear ring from the south. Several other faint dust lanes downstream from the main dust lane to the east are seen, yet these do not join to the nuclear ring, and instead they continue towards the northwest and join the dust lane to the west. Within the nuclear ring, we notice two dust spirals curving towards the core of the galaxy. They appear to connect with the dust lanes in the bar; at the inner end of the western lane, south of the nucleus, the spiral lane looks doubled. The unsharp mask image of the galaxy delineates these faint dust structures. 

As shown in the flux maps, \ha and \nii emission is high in the nucleus, drops at $\sim$2--3\arcsec~from the centre, and then is high in the nuclear ring of $\sim$5\arcsec~radius. In comparison with \ha and \nii, the \oiii emission is relatively weak and is confined within the $\sim$3\arcsec~radius. Outside the nuclear ring, within the bar region, the emission across all line groups is weak, yet several round regions with enhanced \ha and \nii emission are seen, likely originating from local SF. 

The velocity maps of \ha and \nii present signs of non-circular motions. Outside the nuclear ring, the zero velocity line has a strong indent and inside the nuclear ring, it shows a strong wiggle due to several local velocity peaks. The velocity dispersion of all line groups shows a peak within the inner $\sim$3\arcsec~region, with values $\sim$110--140\kms. The values decrease and remain around $\sim$50--60\kms within the nuclear ring. Outside the nuclear ring, within the round region where enhanced \ha and \nii emission is seen, at $\sim$9\arcsec~northwest of the centre, there is low \ha and \nii velocity dispersion around $\sim$60\kms. Moreover, there is a relative increase in velocity dispersion outside the nuclear ring but we can not determine the values reliably due to poor signal within the bar region. 

Because of the poor signal within the bar region, the disk orientation parameters of the galaxy are not well constrained with \texttt{Kinemetry}; the estimated distributions of \textit{i} and PA of the LON fluctuate as a function of the radius of the fitted ellipse. Therefore, we ran \texttt{Kinemetry} with different \textit{i} and PA of the LON values, to fit various rotating disks to the \ha velocity field and, we studied the corresponding residual velocity maps to identify parameter values that minimise the artefacts. A rotating disk fitted with a fixed centre, PA of the LON at 107\degr and \textit{i} at 35\degr minimised the artefacts seen in residual velocity maps. Consequently, these values are selected as the final orientation parameters for the disk. When compared to the disk orientation parameters estimated from the large-scale fits in \citet{Lindblad_1997} (\textit{i}=35\degr and PA of the LON =87\degr), the settled value for \textit{i} aligns well with theirs, but there is a notable 20\degr difference in the PA of LON. However, this is anticipated and can be attributed to the large-scale fits incorporating contributions from the entire bar and the asymmetric spiral arms of the galaxy \citep{Elmegreen_1996,Martini_03}.

In the residual velocity map of the galaxy, inwards from the nuclear ring, three clumpy regions with negative velocity residuals of amplitudes $\sim$20--30\kms, situated $\sim$1\arcsec~towards the northeast, southeast and southwest from the centre are exposed. The northern region exhibits the strongest amplitudes. While these three regions resemble $m$=3 harmonic terms, raised from wrongly fitted \textit{i} in the disk model, they also coincide with the local velocity peaks seen in the original velocity maps, and therefore they are real and driven by the data. None of these regions overlap with the dust spirals inside the nuclear ring. External to the nuclear ring, we notice a coherent structure with negative velocities, with amplitudes around $\sim$30--40\kms, winding from $\sim$15\arcsec~east of the centre towards the northwest, coinciding with one of the faint dust lanes downstream from the main dust lane. Where this coherent structure ends, $\sim$10\arcsec~northwest of the centre, a round region with high positive residual velocities, with amplitudes exceeding 50\kms, is seen. This region coincides with enhanced \ha and \nii emission and low velocity dispersion.

Inwards from the nuclear ring, the velocity difference map reveals two peaks, one with positive and one with negative values, with amplitudes of $\sim$15--25\kms. Both peaks are located at a similar distance, $\sim$4\arcsec~from the centre, one towards the northeast and the other towards the southwest. These two peaks also appear to have an extension of elevated velocity difference amplitudes in the clockwise direction. Considering their similar distances from the centre, and the possible extended natures, these two kinematic structures can be counterparts of one another. Outside the nuclear ring, where velocity residuals were showing an increase, the velocity differences are also coherent, exhibiting amplitudes around $\sim$20--30\kms. This trend persists throughout the dust lane. However, once reaching the round region, where enhanced \ha and \nii emission and low velocity dispersion are seen, the amplitudes are $\sim$0\kms;  around this region, the velocity differences remain consistent with those observed in the dust lane.

The core of NGC\,1300 is classified as a weak low-mass AGN \citep{Atkinson_05,Hughes_05}. As shown in the BPT map of the galaxy, the nuclear ring exhibits composite emission with contribution from SF, with the most significant SF occurring in the northwest. Inwards from the nuclear ring, ionisation is predominantly driven by high-energy mechanisms. Outside the nuclear ring, two clumps, each showing contribution from SF, are observed -- one  $\sim$10\arcsec~northwest of the centre and the other $\sim$6\arcsec~south of the centre. These structures coincide with round regions displaying enhanced \ha and \nii emission. Considering high residual velocities co-spatial with the structure at the northwest, and elevated velocity differences surrounding it, we hypothesise that this structure can be a young star cluster moving with respect to the surrounding medium of different physical conditions.

Although NGC\,1300 shows a clear coherent spiral structure in dust that can be traced down to $\sim$1\arcsec~from the nucleus, this structure does not appear to have a coherent kinematic counterpart. The kinematic structures in the residual velocity and velocity difference maps are likely the result of local shocks and are not associated with extended shocks.

\subsection{NGC\,1433}
\label{sec:NGC1433_notes}

The images and diagnostic maps of NGC\,1433 are shown in Fig.\,\ref{fig:NGC1433}. The large-scale image of the galaxy reveals a luminous core, a prominent nuclear disk, an outer bar, an inner ring surrounding the outer bar region, outside of which there are two or possibly four spiral arms. The MUSE FOV, shown in the colour image of the galaxy, covers the core, the nuclear disk and roughly half of the bar. There remains a subject of ongoing debate about whether NGC\,1433 hosts an inner bar. While several photometric \citep{Erwin_04, Buta_2015} and kinematic \citep{Bittner_19} studies present evidence in favour of an inner bar, contrasting kinematic studies challenge its existence \citep{deLorenzo_2019, Bittner_21}. Along the outer bar, two prominent dust lanes are seen which curve around the nuclear disk. Moreover, a faint dust lane is seen following the same pattern as the dust lane to the southwest, yet it extends further beyond and connects to the dust lane to the northeast. Further inwards, there are several dust spiral arms within the nuclear disk, curving towards the centre. These dust features are more clearly seen in the unsharp mask image of the galaxy, where they appear as a continuation of the straight dust lanes along the outer bar.

\newcounter{tempfig}
\setcounter{tempfig}{\value{figure}}

\renewcommand{\thefigure}{\thesubsection a}

\stepcounter{figure}
\begin{figure*}
    \centering
    \includegraphics[width=1\textwidth]{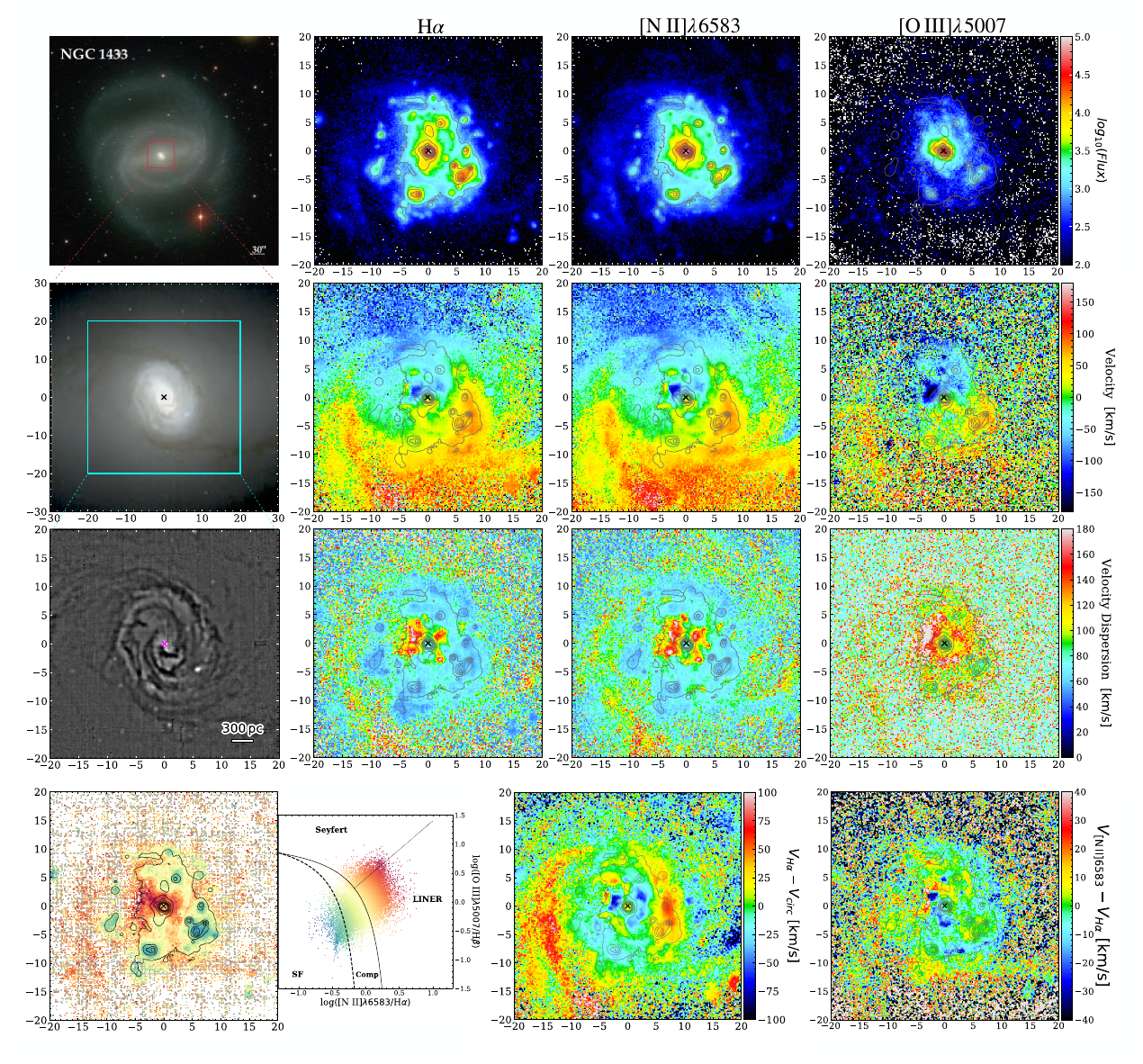}
    \caption{Images and diagnostics maps of NGC\,1433. Panels show the same as in Fig.\,\ref{fig:NGC4303}}
    \label{fig:NGC1433}
\end{figure*}

The \nii flux, velocity and velocity dispersion maps of NGC\,1433 referred to in the paragraph below show the measurements obtained from single Gaussian fits, presented in Appendix\,\ref{app:suplementary_sample_plots}, first row of Fig.\,\ref{fig:NGC1433_NGC4303_NGC4321_R1}. \ha and \oiii flux, velocity and velocity dispersion measurements can be represented with the maps shown in the main diagnostic maps presented in Fig.\,\ref{fig:NGC1433}. These measurements are obtained from fits including a secondary component to \nii, yet \ha and \oiii are not affected by this correction. More details on this are provided in the following paragraph. 

As shown in the flux maps, presented in Fig.\,\ref{fig:NGC1433} and in the first row of Fig.\,\ref{fig:NGC1433_NGC4303_NGC4321_R1}, emission across all line groups is confined within the nuclear disk. The strongest emission comes from the core, confined within $\sim$3--5\arcsec~radius for \ha and \nii emissions, and within $\sim$3\arcsec~radius for \oiii. There is no nuclear ring in \ha or \nii emission. Within the nuclear disk, distinct clumpy regions with enhanced \ha and \nii emissions are seen. One of the most prominent of these regions is seen as a single blob at $\sim$9\arcsec~southeast of the centre. Another notable region with enhanced emission is characterised by three maxima and is situated $\sim$6--8\arcsec~southwest of the centre. In comparison to \ha emission, the \nii emission from these regions is weaker. These regions also exhibit elevated \oiii emission. The velocity maps of all emission line groups exhibit signs of non-circular motions, with indents in the zero-velocity-line characteristic to strong streaming motion in the spiral arms. Outer parts of the nuclear disc, at $\sim$8\arcsec, have the smallest velocity dispersion with $\sim$60\kms. Throughout the nuclear disk, where enhanced clumpy \ha and \nii emissions are seen, the velocity dispersion of \ha and \nii remains around $\sim$60--70\kms. Closer to the nucleus, there is an extended region ($\>$5\arcsec~across) of high velocity dispersion to the east and two small ones ($\sim$2\arcsec~in size) to the west where high dispersion values exceed $\sim$150\kms. These regions also exhibit anomalous \ha and \nii velocities. Due to poor \oiii emission, the velocity dispersion values within the nuclear disk are not reliably estimated, but in the regions with high \ha and \nii velocity dispersion, values are also elevated for \oiii.

The high velocity dispersion in the nucleus prompted us to examine the Gaussian fits in the corresponding spaxels. As we explained in Sect.\,\ref{sec:excess_masking}, if there are multiple emission sources along the LOS, which can appear as excess emission in the spectra, the standard single Gaussian fitting approach will produce a poor and broad fit in the spectra of spaxels that are affected from the excess emission. Such poor fits correspondingly can shift the recorded velocity of main lines, and can artificially increase the velocity dispersions. Correspondingly, the artificial measurements can significantly influence the residual velocity and velocity difference maps we construct to search for the signatures of extended shocks. Upon checking the spectra, we noticed that spectra in several spaxels within the inner $\sim$3\arcsec~radius are poorly fitted due to an excess emission in the blue wing of the \nii line. Therefore, to address this excess emission, we followed the same procedures introduced in Paper IV, and allowed for fitting a secondary component to \nii. 

Further, we compare the kinematic measurements obtained from single Gaussian fits and fits including a secondary component to determine regions that are the most affected by the excess emission. In general, \ha and \oiii are not affected throughout the entire FOV. However, within a localised region extending from $\sim$3\arcsec~east of the centre $\sim$4\arcsec~towards the north, the introduction of the secondary component shifts the derived velocity of \nii significantly, by $\sim$15--20\kms. The velocity dispersion of \nii in this region remains high. This region is presented in Fig.\,\ref{fig:1433_mask}, after certain mask conditions are applied which we explain below. 

\renewcommand{\thefigure}{\thesubsection b}
\stepcounter{tempfig}

\begin{figure}
    \centering
    \includegraphics[width=1\columnwidth]{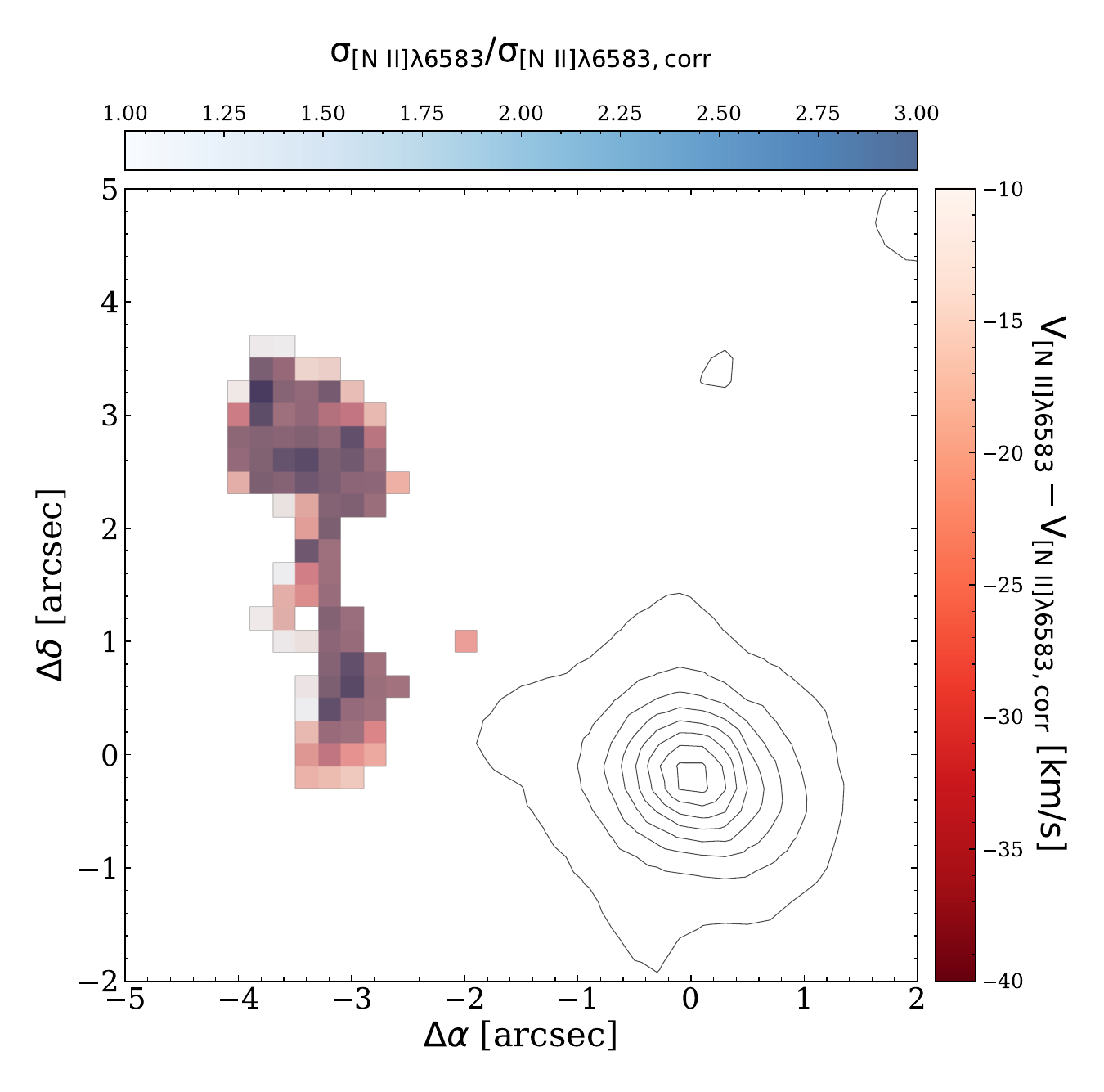}
    \caption[Mask of spaxels in NGC\,1433]{The mask of spaxels for which the fits including secondary component were used in further analysis of NGC\,1433. The red and blue colour maps show the velocity shift of \nii and the ratio of \nii velocity dispersion, respectively, before and after including the secondary component in the fitting.}
    \label{fig:1433_mask}
\end{figure}

\addtocounter{figure}{-2} 
\renewcommand{\thefigure}{\arabic{figure}}

Therefore, by establishing the criteria explained further, we determine the region where the fits to spaxels are the most affected by the excess emission, to correct the \nii kinematic measurements in the regions, by using the values for the main component from the fits that include a secondary component there. The criteria are as follows:

\begin{enumerate}[label=\roman*., align=left, leftmargin=*]
    \item $\rm V_{[N\,II]} - V_{\rm [N\,II], corr} \leq -10$ \kms
    \item $\rm \sigma_{[N\,II]} / \sigma_{\rm [N\,II], corr} \geq 1.2$
    \item $\rm \sigma_{[N\,II]w} \leq 300$\kms
    \item $-4$\arcsec$\leq \Delta\alpha \leq-1$\arcsec~and $-1$\arcsec$\leq \Delta\delta \leq4$\arcsec
\end{enumerate}
where \textit{corr} indicates the values extracted from the fits including the secondary component. The third condition, the velocity dispersion limit of the secondary \nii component ([N\,II]w) ensures that the secondary component is real and is not fitted to noise. The spatial cuts are applied to select the region where the change in velocity and velocity dispersion ratio is consistently large. For further analysis, the flux, velocity, and velocity dispersion values for all emission line groups within the mask region are replaced with those from the corresponding main components in fits that include the secondary component. Although \ha and \oiii measurements are not affected, we continue to use measurements from fits with a secondary component to maintain consistency. The corrected flux, velocity and velocity dispersions maps of all line groups are presented in Fig.\,\ref{fig:NGC1433}.

When run with a fixed centre the distributions of the \textit{i} and PA of the LON obtained from \texttt{Kinemetry} show strong variation with radius. Therefore, we adopted initial values PA of the LON=19.7\degr and \textit{i}=29\degr from the large-scale study \citealp{Lang_20}, and by adjusting these values with small increments, we ran \texttt{Kinemetry} with different fixed PA of the LON and \textit{i} values to identify the values that raise the minimum level of artefacts in the residual velocity map. Despite these adjustments, we did not observe a significant change in the residual velocity map. Therefore, we selected the initial parameter values as the final \textit{i} and PA of the LON. 

In the residual velocity map of the galaxy, there are several pronounced coherent structures within the nuclear disk. One of these structures exhibiting positive velocities with amplitudes of $\sim$30--50\kms, forms an arch $\sim$8\arcsec~west of the centre, extending for over 15\arcsec. The corresponding deprojected residual velocity amplitudes across this structure reach up to 115\kms. A possible counterpart to this structure, exhibiting negative velocities, extends from $\sim$10\arcsec~south of the centre towards the northeast, exhibiting deprojected residual velocity amplitudes around 70--80\kms. Both structures align well with the outer curvature of the nuclear disk, and the dust lanes. Another coherent structure with negative velocity residuals is seen, with residual velocity amplitudes around 40--50\kms, forming an arch $\sim$4\arcsec~west of the centre extending almost 10\arcsec. Across the extent of this structure, the deprojected residual velocity amplitudes are high, reaching up to 115\kms. The largest amplitude of this structure coincides with one of the peaks of \ha and \nii velocity and velocity dispersion. Moreover, throughout the extent of this structure, the \nii velocity dispersion is $\sim$30--40$\%$ larger than that of \ha. Between $\sim$2\arcsec~and $\sim$5\arcsec~to the northeast of the nucleus there are clumpy regions of high negative and positive velocity residuals -- they are likely related to the AGN outflows, explained further below. In addition, throughout the nuclear ring, several other regions exhibit high residual velocity amplitudes. However, these regions rather have an amorphous structure and are likely originating from local SF. Outwards from the nuclear ring, extending from the southeast of the centre towards the northeast for $\sim$20\arcsec, a coherent structure forms an arch, exhibiting velocity residual amplitudes around $\sim$30\kms, with corresponding deprojected residual velocity amplitudes around 80--90\kms. This structure coincides with the faint dust lane seen in the same region, likely indicative of overshot gas.

In the velocity difference map of the galaxy, regions of high positive and negative values are seen to the south and southeast of the nucleus, where the velocity dispersion is high. In addition, several clumpy regions with enhanced velocity differences, up to $\sim$20\kms are seen throughout the nuclear disk, that do not particularly coincide with the structures seen in the residual velocity map.

The classification of the galaxy's core remains the subject of a debate \citep{Veron-Cetty_1986,Sosa-Brito_01}. Nevertheless, the contribution from the AGN to the ionisation is evident in the BPT map of the galaxy. Up to $\sim$4\arcsec~radius, in two regions exhibiting a distinct horizontal biconical shape, characteristic to AGN outflows, the ionisation is primarily influenced by a high-energy mechanism. Particularly along east-west direction, in the regions where the velocity dispersion is high. A part of this region coincides with where we corrected for excess emission in the fitting, indicating the excess emission seen in this region is likely originating from AGN outflows. Furthermore, throughout the nuclear disk, there is composite emission with a contribution from SF. The regions with the strongest SF, to the southeast and southwest, are co-spatial with the most prominent \ha and \nii enhancements seen in emission. These coincide with the two coherent structures with high positive and negative residual velocity amplitudes, situated at the curvature of the nuclear disk. These alignments are indicative of SF being triggered by enhanced gas density at the leading side of the bar and provide evidence of large-scale extended shocks propagating within those regions.

\textcolor{black}{Within the nuclear disk of NGC\,1433, the residual velocity map exhibits multiple elongated structures, some confined to galactocentric radii of a few hundred parsecs and others extending out to a few kiloparsecs (see also features outlined in Fig.\,\ref{fig:Sample_extended_structures_perturbed_galaxies_residualvel}). These features show large deprojected residual velocity amplitudes, reaching $\sim$100\kms\, and coinciding with dust spirals.} These features together resemble a nuclear spiral shock, but they lack the continuity expected from such a shock signature. Nevertheless, the alignment of the dust structures with features seen in residual velocity, together with enhanced velocity differences and the elevated SF activity occurring in these features can all be associated with extended shocks in gas. We hypothesise that strong SF activity within the nuclear disk and potentially the AGN activity, interrupt the propagation of the shock, leaving portions of it visible in the residual velocity map. \textcolor{black}{Therefore, based on the grouping described in Sect.\,\ref{sec:grouping-shock-class}, this galaxy is classified under \textit{extended shocks} kinematic category.}

\subsection{NGC\,1566}
\label{sec:NGC1566_notes}
\renewcommand{\thefigure}{\thesubsection}
\begin{figure*}
    \centering
    \includegraphics[width=1\textwidth]{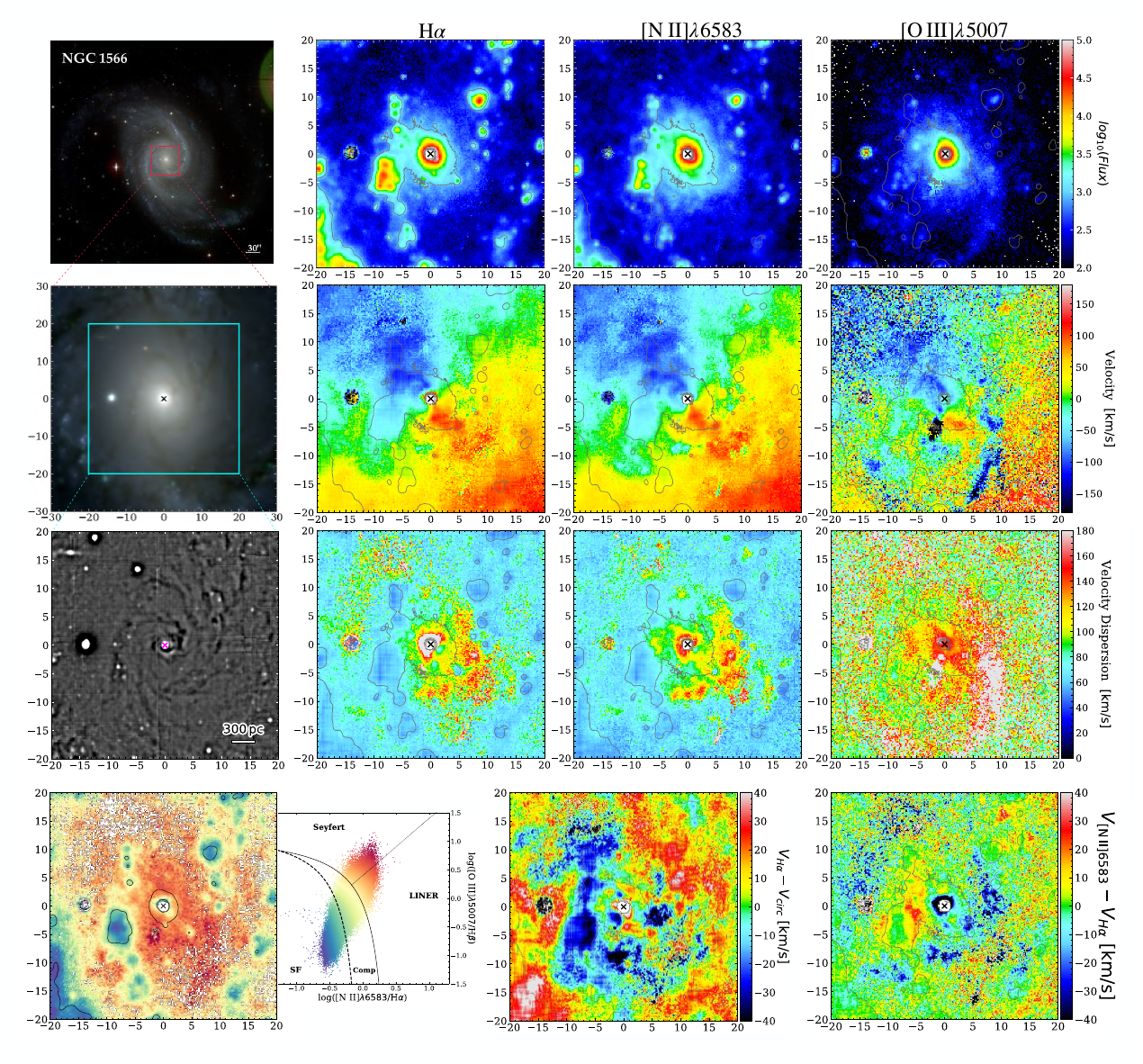}
    \caption{Images and diagnostics maps of NGC\,1566. Panels show the same as in Fig.\,\ref{fig:NGC4303}}
    \label{fig:NGC1566}
\end{figure*}

The images and the diagnostic maps of NGC\,1566 are presented in Fig.\,\ref{fig:NGC1566}. As shown in the large-scale and colour images, NGC\,1566 has a bright core, which is classified as Seyfert 1.2 type based on its optical broad line ratio, yet also exhibiting optical narrow line ratio typical of Seyfert 2 cores \citep{Alloin_1985,Kriss_1991,Smajic_2015}. The galaxy also has a weak bar \citep{deVaucouleurs_1991, Sheth_10}, and two or possibly four star-forming spiral arms encircling the bar region. The MUSE FOV encompasses the core and the entire bar. In the bar region, originating in the north and south, there are dust lanes winding in clockwise direction. These dust features are more prominent in the unsharp mask image. Particularly, within the inner 15\arcsec, we notice the multiplicity of spiral arms in clockwise direction. 

As presented in the flux maps, the emission across all line groups is concentrated within $\sim$3\arcsec~radius. In the bar region, two strong lanes of emission are seen with clumpy \ha and \nii emission; some of these regions align as a straight line, likely indicating enhanced SF within the bar lanes. The \ha emission in these regions is stronger than \nii emission. We do not discern a nuclear ring in emission. \oiii emission beyond the inner $\sim$3\arcsec~region significantly weakens. However $\sim$8\arcsec~from the centre towards south--southeast enhanced \oiii emission is seen; since there is no counterpart to this structure seen in any other emission, it is likely originating from AGN outflows.

The velocity maps reveal significant non-circular motions of no clear pattern across all emission line groups. In the \oiii velocity map, $\sim$4--8\arcsec~from the centre, at PA $\sim$160\degr, a region of high negative velocities (up to $\sim$300--400\kms) occurs at the location of enhanced \oiii emission, confirming its origin being AGN outflows. There is also a blueshifted quasi-linear structure $\sim$10\arcsec~to the southwest of the nucleus, co-spatial with a filament seen in \oiii, but not in \ha or \nii, inconsistent with the spiral pattern, possibly related to the AGN outflows.

Within the $\sim$2\arcsec~radius, an ambiguous round structure with high velocity across all line groups is seen. Upon closer examination of the spectra, within $\sim$1\arcsec~radius of this region, we noticed strong broad emission primarily in the \ha and [N\,II] wavelength ranges, which can be attributed to the AGN. The emission from AGN was so strong that it even caused saturation in the innermost couple of spaxels. As a result, the Gaussians are not well fitted to the \ha and [N\,II] lines, indicating that the kinematic measurements derived from them are unreliable within the inner 1\arcsec. Broad emission is also seen in \oiii, but the Gaussian fits to \oiii are still reliably fitted. Although the broad emission is associated with the characteristic of the AGN, there can also be a potential contribution from a small nuclear ring in that region, which has a size smaller than 100\,pc, standing as one of the smallest nuclear rings ever found \citep{deSaFreitas_23}. Nevertheless, the erroneous Gaussian fits due to broad emission in the spectra indicate that the measurements within the $\sim$1\arcsec~radius are not reliable. The complexity of the broad emission extends beyond the standard approach of including a secondary component in the fitting process and requires careful attention to be addressed. Given that addressing this issue is beyond the scope of this work, we excluded the measurements within the 1\arcsec~radius in our further analysis. In Fig.\,\ref{fig:NGC1566}, the excluded spaxels appear as an empty white region with no measurements.   

As shown in the velocity dispersion maps, within the $\sim$3--4\arcsec~radius, beyond the excluded region, the velocity dispersion across all line groups is high, exceeding 150\kms. In the bar region, particularly to the southwest, the velocity dispersions remain high, with values $\sim$100--130\kms which can be related to AGN features explained above. Along the dust lanes in the bar region, the velocity dispersion of \ha and \nii decreases and remains around $\sim$50--70\kms. Where strong blueshifted velocities are seen, to the southwest of the centre, the velocity dispersion of \oiii is high, with values $\sim$170--200\kms.

The disk orientation parameters estimated with \texttt{Kinemetry}, \textit{i} = 40\degr and position angle (PA) of the Line of Nodes (LON) = 37\degr, are consistent with the parameters estimated from large-scale fits from \citet{Lang_20}, differing by approximately $\pm$5\degr (see Table\,\ref{tab:Disk_orientation_parameters}). Therefore, we settle on these parameters in the final disk model. The residual velocity map of the galaxy reveals coherent continuous structures with velocities around $\sim$30\kms. The structure with negative velocities is seen extending from the south of the centre towards the northeast. While a part of this structure curves towards the centre in the northeast, another part continues extending further north by another $\sim$5--6\arcsec. These structures likely reflect the inflow along the two dust lanes seen extending from the south towards the north. A counterpart, exhibiting positive residual velocities, extends from the north of the centre towards the west, coinciding with the dust lane to the north. Throughout the extent of these coherent structures, the deprojected residual velocity amplitudes are around 70--80\kms. These large amplitudes further confirm the shock nature of the observed signatures in residual velocity maps.

In the velocity difference map, two extended regions are seen with amplitudes of $\sim$15--25\kms; one extending from the southeast towards north with positive velocity differences and the other extending from the northwest towards west with negative velocity differences. These two structures appear upstream from the two lanes seen in emission, and the coherent positive and negative residuals, by $\sim$1\arcsec.

As shown in the BPT map of the galaxy, the majority of the emission is of LINER/Seyfert type. The star-forming regions are rather in clumps grouped along several strings that largely correspond to the two lanes seen in emission. The region with the strongest SF coincides with the strongest \ha enhancement seen in emission, and with a portion of the coherent structure with negative residual velocities. Moreover, in the region where the coherent structure with negative residual velocities continues further north, several aligned round regions with enhanced SF are seen.

Even though the presence of AGN outflows and the possible contribution from the small nuclear ring to the emission prevents us from resolving the propagation of shocks into the very core, there is compelling evidence of extended shocks within the inner kpc, along the bar lanes and curving towards the nucleus, reaching the innermost regions, down to the inner $\sim$2\arcsec~region. As shown in Fig.\,\ref{fig:Sample_extended_structures_perturbed_galaxies_residualvel} and Table\,\ref{tab:galaxies_shock_classes}, these structures extend across several kpcs and exhibit large deprojected residual velocity magnitudes of $\sim$70--80\kms.

Even though the presence of AGN-driven outflows and possible emission from the small nuclear ring prevents us \textcolor{black}{from resolving whether the shocks propagate into the very core, the residual velocity map provides compelling evidence for structures associated with extended shocks in the inner regions of the galaxy. Coherent elongated structures are observed along the bar dust lanes, curving towards the nucleus and reaching down to the inner $\sim$2\arcsec~region. As outlined in Fig.\,\ref{fig:Sample_extended_structures_perturbed_galaxies_residualvel}, these structures extend over several kiloparsecs and exhibit large deprojected residual velocity magnitudes of $\sim$70--80\kms. Their large spatial extent, coherent morphology, and high residual velocities therefore support their interpretation as extended shocks. Therefore, based on the grouping described in Sect.\,\ref{sec:grouping-shock-class}, this galaxy is classified under \textit{extended shocks} kinematic category.}


\subsection{NGC\,3351}
\label{sec:NGC3351_notes}

\begin{figure*}
    \centering
    \includegraphics[width=1\textwidth]{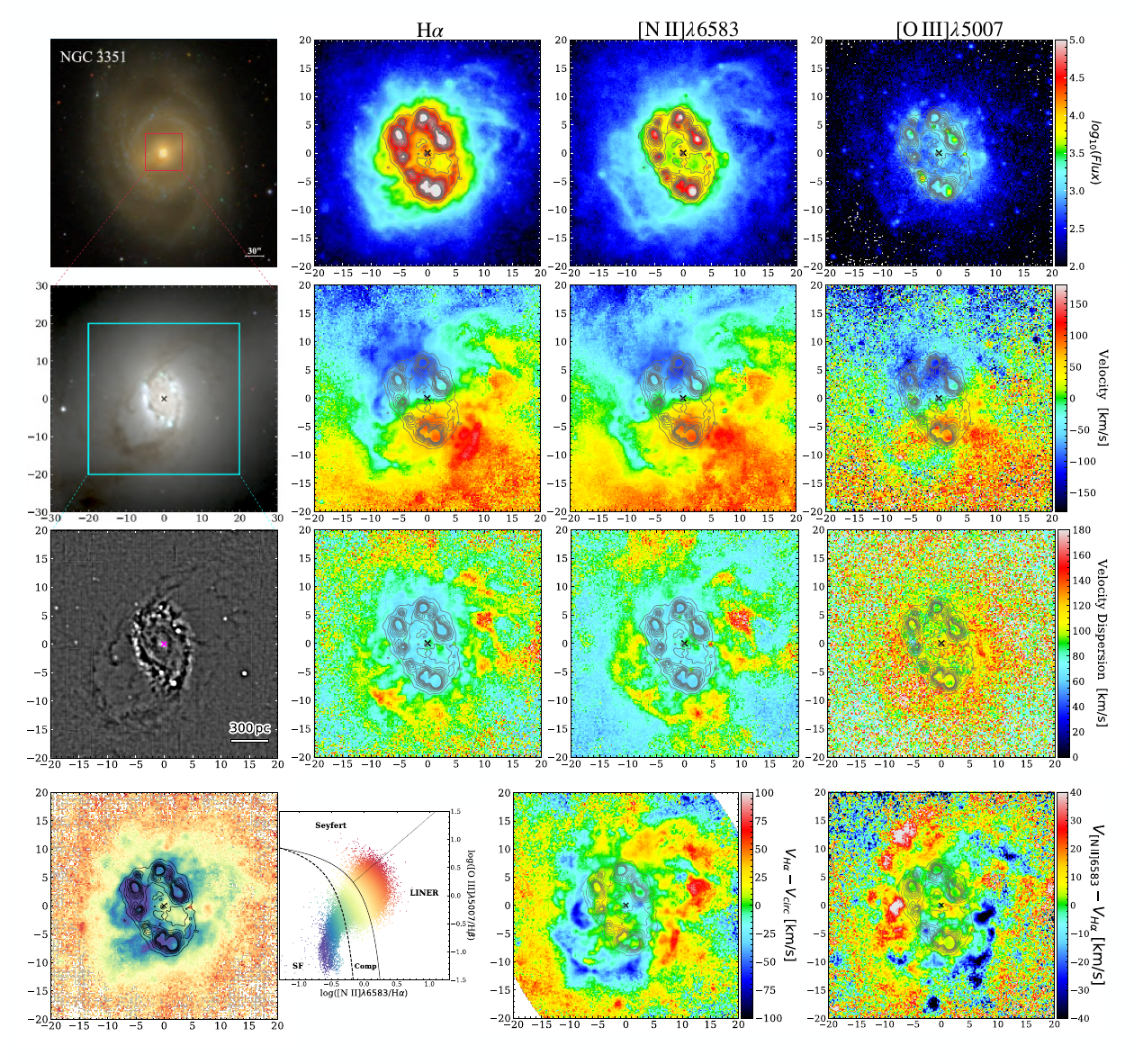}
    \caption{Images and diagnostics maps of NGC\,3351. Panels show the same as in Fig.\,\ref{fig:NGC4303}}
    \label{fig:NGC3351}
\end{figure*}

NGC\,3351 is a well studied galaxy \citep{Mazzalay_14,Smirnova_19,Gadotti_19, Leaman_19,Calzetti_21}, and is one of the members of the Leo Group. The images and diagnostic maps of this galaxy are presented in Fig.\,\ref{fig:NGC3351}. As shown in large-scale and colour images, NGC\,3351 has a luminous core, a star-forming nuclear ring within the $\sim$10\arcsec~radius, a prominent bar, an inner ring surrounding the bar region and two or more spiral arms surrounding the inner ring. The MUSE FOV encompasses the core, the nuclear ring and almost the entire bar of the galaxy. Along the bar, two faint dust lanes are seen; one of these extends from the northwest towards the centre and the other extends from the southeast towards the centre. These two dust lanes continue to curve within the nuclear ring over a full circle. Furthermore, within the bar region, an additional and more prominent dust lane is seen, manifesting an ``L" shape. More specifically, this dust lane appears just outside the nuclear ring, where it is initially almost parallel to the southeastern lane and at $\sim$10\arcsec~towards south from the centre becoming orthogonal to it. To the east and the west of the nuclear ring, several faint dust lanes are also seen. All dust structures are more prominently seen in the unsharp mask image of the galaxy.

As presented in the flux maps, the emission across all line groups is the strongest in the nuclear ring. Inwards from the nuclear ring the emission decreases, though it remains relatively high. In comparison to \ha and \nii, the \oiii emission is notably weak. The velocity maps of \ha and \nii show strong distortions from circular motion. Outside the nuclear ring, the zero velocity line has a strong indentation at $\sim$8-10\arcsec~northwest and southeast from the centre. The indent is less pronounced in the OIII velocity field, but there it extends further, all the way to the nuclear ring. As shown in the velocity dispersion maps, the velocity dispersion of \ha and \nii is the lowest in the nuclear ring, reaching around $\sim$50--60\kms inside the nuclear ring the velocity dispersion values increase up to 90\kms in the region $\sim$3\arcsec~southwest of the nucleus. Outside the nuclear ring, the velocity dispersion values increase up to well above 120\kms. The velocity dispersion values for \oiii can not be reliably estimated due to poor signal, although even the noisy values suggest the same trends as in \ha and \nii. 

After fitting the rotating disk with \texttt{Kinemetry}, \textit{i} and PA of the LON converged to 46\degr and 23\degr respectively. While the \textit{i} aligns well with measurements from a large-scale study of \citet{Erwin_08} (\textit{i}$=$46\degr, PA of the LON$=$10\degr), the PA of the LON slightly deviates from them. The residual velocity map of galaxy exposes two coherent structures with amplitudes of $\sim$50\kms, one exhibiting negative and one exhibiting positive velocities. The feature with positive velocities can be traced from the right edge of the frame at $\Delta\delta$$\sim$8\arcsec~towards the northern tip of the nuclear ring, coinciding with one of the main faint dust lanes along the bar. The feature with negative velocities coincides with the L-shaped dust lane. Along the dust lanes within the nuclear ring, there are two more coherent structures, one with positive and one with negative velocities with amplitudes up to $\sim$50\kms, corresponding to negative and positive deprojected velocities up to $-$95\kms and 50\kms respectively.

Furthermore, several other relatively extended structures with high residual velocity amplitudes are exposed to the east and west of the nuclear ring; one exhibiting negative and two exhibiting positive velocities. The structure with negative velocities has velocity amplitudes $\sim$40\kms, and extends $\sim$6\arcsec~northeast from the centre directly towards the south, coinciding with the faint dust lane observed to the east of the nuclear ring. The two other features with positive velocities, have velocity amplitudes above $\sim$50\kms, each having roughly $\sim$5\arcsec~extent that curves towards the west of the nuclear ring. This appearance of the residual velocity map challenges the textbook understanding of shocks in bars, as in NGC\,3351 outside the nuclear ring their kinematic signatures extend along the bar only over a short distance, and then take a right-angle turn, past which they coincide with dust lanes \textit{perpendicular} to the bar. 

The velocity difference map of the galaxy exposes two coherent structures with amplitudes of $\sim$30-40\kms, one at the west of the nuclear ring exhibiting negative velocity, and the other at the east of the nuclear ring exhibiting positive velocity. These two structures are among the structures with highest velocity difference amplitudes we observed in other galaxies we studied in this work; not related to AGN. They are consistent with the textbook picture of shocks in bars, they occur downstream from the dust lanes associated with those shocks, but their morphology shows little or no consistency with that of the velocity residuals. Inwards from the nuclear ring, near locations of the peak residual velocities, the velocity differences also show an increase, reaching amplitudes of $\sim$15--20\kms.  

The BPT map shows that the emission from the nuclear ring is strongly dominated by SF. Outside the nuclear ring, there is composite emission with a contribution from SF. Particularly, to the west of the nuclear ring, there is composite emission with a contribution from LINER emission. Inwards from the nuclear ring, towards the southwest of the nucleus, there is composite emission.

NGC\,3351 exhibits clear signatures of extended shocks. \textcolor{black}{As outlined in Fig.\,\ref{fig:Sample_extended_structures_perturbed_galaxies_residualvel}, these structures extend across several hundreds of parsecs, have elevated deprojected residual velocity magnitudes reaching $\sim$50--95\,{$\rm kms^{-1}$}}, and they are cospatial with the coherent morphology of dust structures, especially inside the nuclear ring. Along with elevated velocity differences, there is dominating composite emission inwards from the nuclear ring. Therefore, even though these structures appear rather discontinuous, and are not continuously extending from outer regions towards the nucleus, we still associate them with extended shocks, dominating the inner kpc of the galaxy.

\subsection{NGC\,3368}
\label{sec:NGC3368_notes}

\begin{figure*}
    \centering
    \includegraphics[width=1\textwidth]{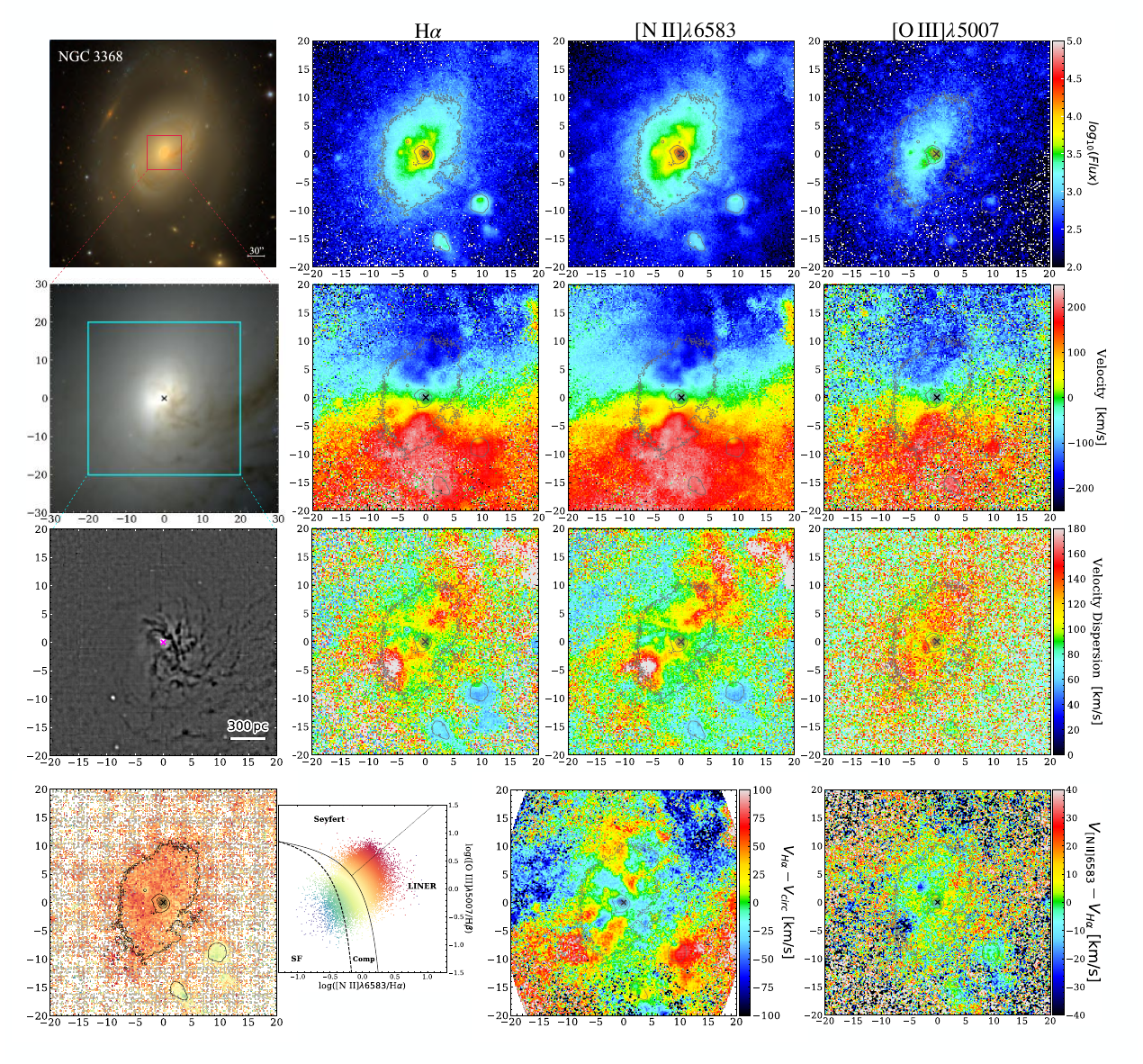}
    \caption{Images and diagnostics maps of NGC\,3368. Panels show the same as in Fig.\,\ref{fig:NGC4303}}
    \label{fig:NGC3368}
\end{figure*}

The images and the diagnostic maps of NGC\,3368 are presented in Fig.\,\ref{fig:NGC3368}. NGC\,3368 belongs to the Leo Group together with the galaxy NGC\,3351. As shown in the large-scale and colour images, the galaxy hosts a luminous dusty core and two bars -- an outer bar with a length of 75\arcsec~and an inner bar with a length of 5\arcsec~\citep{Erwin_04,Erwin_15}; in between the two bars, there is a nuclear disk \citep{Mollenhoff_2001,Erwin_04,Erwin_15}. The outer bar is surrounded by spiral arms. The MUSE FOV covers the core with roughly two-thirds of the outer bar. The dust structures around the core of the galaxy are more clearly seen in the unsharp mask image.

As presented in the flux maps, the strongest emission across all line groups is concentrated within the $\sim$3\arcsec~radius, mostly to the east of the centre. The velocity field of all emission line groups does not exhibit strong perturbations; the zero velocity line remains relatively straight. Two structures, showing local velocity maxima, at PA 100\degr and $-$80\degr, extend radially by $\sim$8\arcsec. Another linear structure, showing velocity maxima, also at a PA $-$80\degr, is located slightly outside the frame, roughly at $\sim$20\arcsec~from the centre. This structure has positive velocities, counter-rotating with respect to the surroundings.

As velocity dispersion maps indicate, the overall velocity dispersion of \ha and \nii is high, with values around $\sim$100--150\kms. The regions with the highest dispersion, where values exceed 150\kms, form a biconical structure extending from the centre towards the northwest and southeast. This structure is more prominently seen in the \nii velocity dispersion map. Another region with velocity dispersions exceeding 150\kms appears towards the edge of the frame, where the gas is counter-rotating, as seen in the velocity map. The velocity dispersion values of \oiii are overall noisy due to poor signal within the FOV, yet the values still suggest an increase along the biconical structure seen in \ha and \nii fields. 

The disk orientation parameters estimated with Kinemetry, \textit{i}=46\degr and PA of the line of nodes (LON)=172\degr, agree with estimations from large-scale studies (e.g., \citealp{Erwin_08}: \textit{i}=50\degr and PA of the LON=172\degr). Upon analysing the residual velocity map of galaxy, within the $\sim$5\arcsec~radius towards the northeast, there are extended structures that show similar winding with dust spiral arms which may be related to the kinematics. Moreover, two structures, one with positive and one with negative velocities extend along the biconical structure seen in velocity dispersion, the highest amplitudes of velocity residuals roughly coincide with the highest velocity dispersion. The velocity difference map is noisy and does not exhibit coherent structures. However, where the highest velocity dispersion values are seen along the biconical structure, the velocity differences also show an increase. The region to the southeast exhibits negative velocity differences and the region to the northwest exhibits positive velocity differences, both with amplitudes around $\sim$20--30\kms. The biconical morphology of the observed structure with extreme velocity dispersions, along with structures seen in velocity residual and velocity difference maps confirm that this structure outflows from the AGN. The counter-rotating portion of the gas is a consequence of the outflowing gas, moving in different conditions with respect to its surroundings.

Furthermore, the BPT map of the galaxy reveals that the ionisation throughout the entire FOV is predominantly coming from the LINER emission with weak contribution from composite emission. 

Although the galaxy is dominated by an AGN, which could potentially prevent identifying extended shocks near the nucleus, we have resolved structures within the innermost regions that are not obscured by AGN outflows. Generally, there is no compelling evidence of extended shocks within the inner kpc of NGC\,3368. \textcolor{black}{Therefore, based on the grouping described in Sect.\,\ref{sec:grouping-shock-class}, this galaxy is classified under \textit{other perturbations} kinematic category.}

\subsection{NGC\,3489}
\label{sec:NGC3489_notes}

\begin{figure*}
    \centering
    \includegraphics[width=1\textwidth]{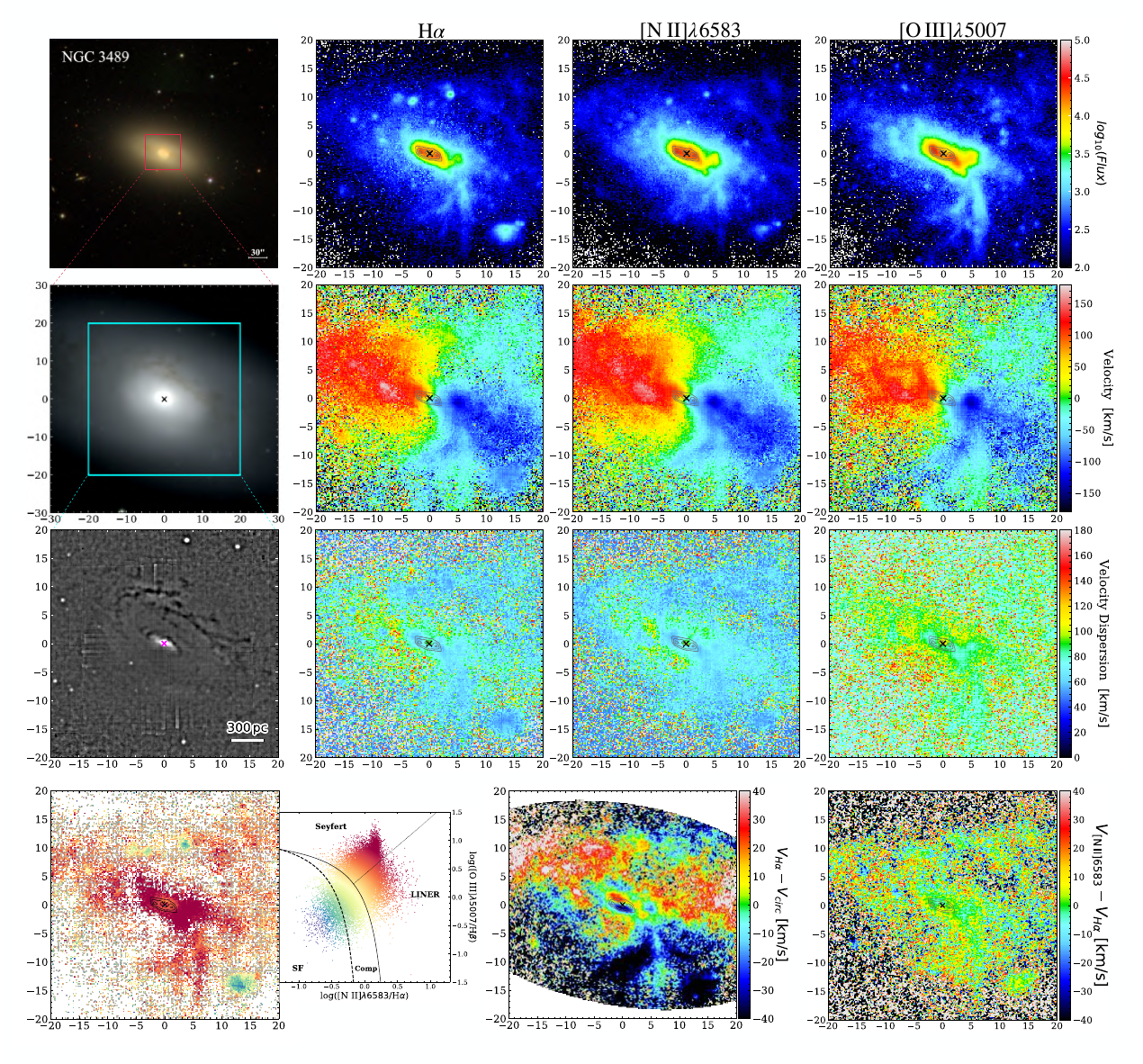}
    \caption{Images and diagnostics maps of NGC\,3489. Panels show the same as in Fig.\,\ref{fig:NGC4303}}
    \label{fig:NGC3489}
\end{figure*}

The images and the diagnostic maps of NGC\,3489 are shown in Fig.\,\ref{fig:NGC3489}. As presented in  large-scale and colour images, NGC\,3489 hosts a bright core, and a bar with $\sim$20\arcsec~in size \citep{Erwin_03}, embedded in a dusty disk. The spectral classification of the galaxy's core falls within the Seyfert region \citep{Ho_1997, Sarzi_06}. The MUSE FOV covers the core, the bar and the disk. The unsharp mask image of the galaxy outlines a prominent dust spiral, originating from the west of the frame and circling the bar region in the counter-clockwise direction. Moreover, to the northeast and southwest, two very faint straight dust lanes are seen along the bar; these two lanes appear to connect a dusty ring of semi-major axis $\sim$3--4\arcsec.

As presented in the flux maps, the strongest emission across all line groups is mostly confined within the inner $\sim$5\arcsec. Within the radius of approximately $\sim$2--3\arcsec, the strongest emission exhibits several off-centred and elongated peaks, indicative of a small nuclear ring. A similar ring structure of size $\sim$3\arcsec~is reported in former works \citep{Erwin_03,Nowak_2010}. Within the same region, \oiii emission is exceptionally strong, exhibiting quasi-radial extensions to the southwest, that may be related to shocks in the bar.

The velocity fields across all line groups reveal a wiggly zero velocity line, where it forms an ``S" shape within the $\sim$4\arcsec~radius, associated with the characteristic flow in bars. There is also a peculiar "blue arm", with negative velocities extending to the southeast, strongest in \oiii. As demonstrated in the the velocity dispersion maps, dispersion across all line groups is high in the nucleus, with values around $\sim$100--120\kms. Within the nuclear ring, the velocity dispersion decreases; values remain consistent towards the southwest end of the frame. 
To the northeast and southwest of the centre, we discern two local velocity dispersion minima with values $\sim$50--60\kms; these structures coincide with the off-centred peaks in emission. Throughout the dust spiral surrounding the bar region, the velocity dispersion values of \ha and \nii remain around $\sim$50--60\kms, values for \oiii cannot be estimated reliably due to poor signal in those regions, yet the dispersion is generally low.

After fitting the rotating disk to the \ha velocity field, the \textit{i} and the PA of the LON converged to 60\degr and 80\degr respectively. While \textit{i} agrees well with the measurements estimated from large-scale fits, the PA of LON differs by 10\degr (e.g., \citealp{Erwin_08}: PA of the LON=71\degr, \textit{i}=58\degr). However, when employing the large-scale PA of the LON in the disk model, artefacts of $m$=1 harmonic terms appear in the residual velocity field. These artefacts were not observed when using values deduced by \texttt{Kinemetry} in the model. Consequently, we chose to use the \texttt{Kinemetry} inferred parameters in the final disk model.

The residual velocity map of the galaxy exposes two structures located $\sim$2\arcsec~towards the north and south of the centre; one with negative and one with positive velocities, both exhibiting amplitudes $\sim$40\kms. The largest deprojected residual velocity amplitudes observed across these structures are around 30--40\kms. Both structures have an additional $\sim$2\arcsec~extension towards northeast and southwest with weakened residual velocity amplitudes, $\sim$10--20\kms -- the deprojected residual velocity amplitudes in these regions are around 30\kms. These extensions coincide with the faint dust lanes along the bar and they are consistent with shocks in gas. Moreover, strong residuals of radial extend, with values around $\sim$40--50\kms or more, are also seen where there is quasi-radial emission.

The velocity difference map appears rather noisy, yet it does reveal very faint negative and positive velocities, with values around $\sim$5--10\kms, within the $\sim$2\arcsec~radius, coinciding with the regions with high residual velocity amplitudes. The aligning coherent structures of elevated velocity residuals and velocity differences, together with the aligning dust morphology, can be associated with extended shocks in gas, extending from the outer regions towards the core.

The BPT map of the galaxy reveals strong AGN dominance in the radiation from a Seyfert core. In the nuclear ring, there is a slightly reduced AGN contribution, coinciding with the region with low velocity dispersion, suggestive of the nuclear ring being fed by the inflowing gas. In literature, the galaxy is described as having a rejuvenated nucleus with a young central stellar population based on luminosity-weighted ages \citep{Annibali_2007}, and this is further associated with the potential gas inflow within the central region towards the AGN \citep{Annibali_2010,Marino_11}.  Therefore, the interpretation of the coherent structures being extended shocks propagating towards the nucleus appears to align with these findings. 

Generally, the overall emission in NGC\,3489 seems to be of an AGN nature, but the gas flow is not affected by the AGN. \textcolor{black}{The galaxy exhibits coherent structures extending over several hundred parsecs, which are outlined in Fig.\,\ref{fig:Sample_extended_structures_perturbed_galaxies_residualvel}, with deprojected residual velocity magnitudes reaching up to $\sim$40\kms. We therefore associate these features with signatures of extended shocks.}

\subsection{NGC\,3626}
\label{sec:NGC3626_notes}

NGC\,3626 is one of the most complex galaxies in our sample. It is classified as an unbarred galaxy in \citet{deVaucouleurs_1991}, but several photometric studies show evidence of a double barred morphology \citep{Mollenhoff_2001,Gutierrez_11}; the inner bar is within $\sim$2--2.5\arcsec~radius and outer bar is within $\sim$20\arcsec~radius. Examining the stellar velocity dispersion and higher order moments of NGC\,3626, which are presented in Fig.\,\ref{fig:NGC3626_stellar}, we identify the velocity dispersion signatures, $\sigma$-hollows, characteristic to double-barred galaxies (see \citealp{deLorenzo_08}), agreeing well with the proposed orientation and size of the inner bar. Therefore, for further analysis, we settle on using double barred morphology for NGC\,3626. The gas in the galaxy is counter-rotating with respect to the stars \citep{Ciri_1995, Garcia-Burillo_1998}, which we confirm after comparing the stellar and gas velocity maps.

\setcounter{tempfig}{\value{figure}}
\renewcommand{\thefigure}{\thesubsection a}
\begin{figure*}
    \centering
    \includegraphics[width=1\textwidth]{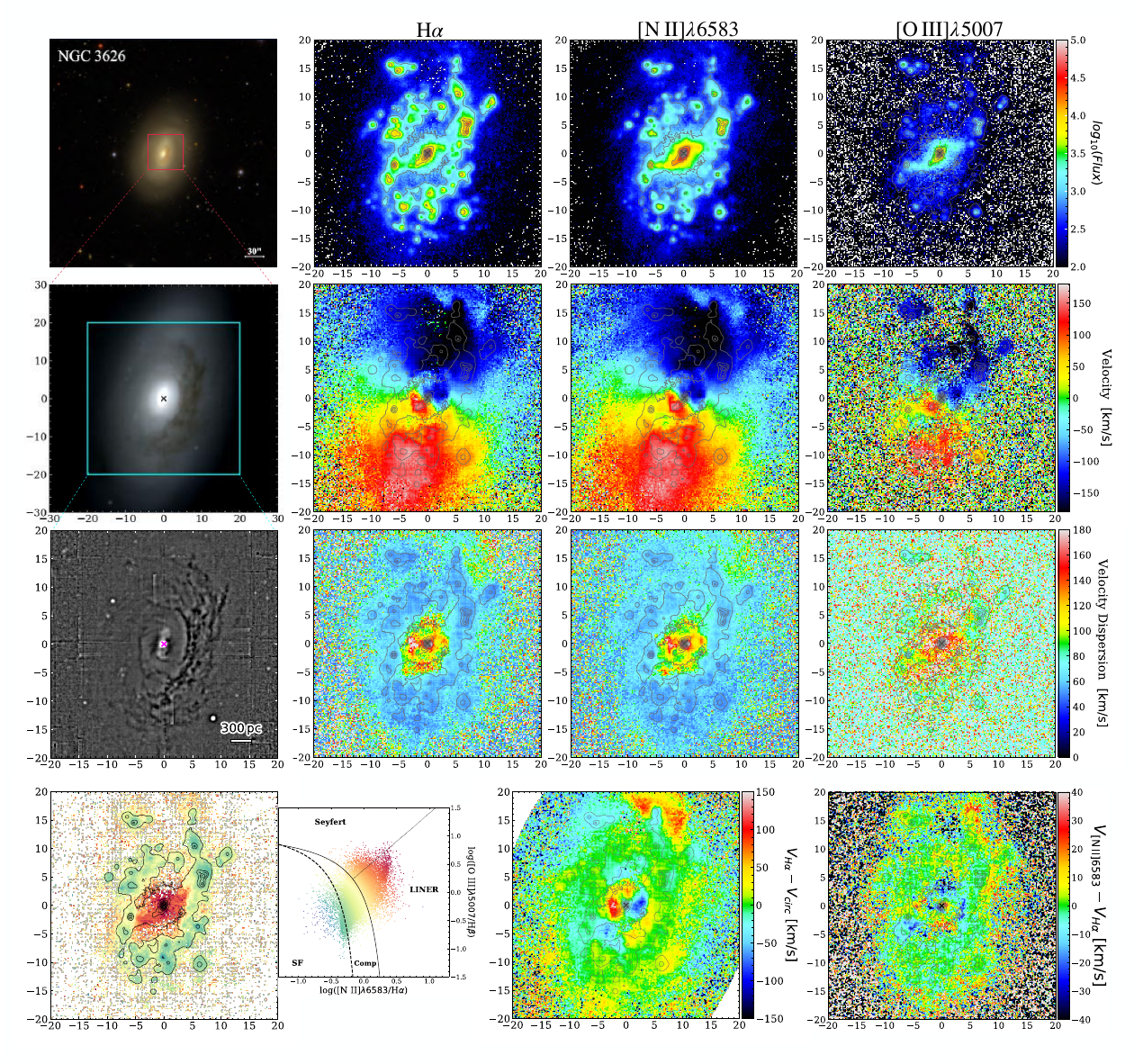}
    \caption{Images and diagnostics maps of NGC\,3626. Panels show the same as in Fig.\,\ref{fig:NGC4303}}
    \label{fig:NGC3626}
\end{figure*}

\renewcommand{\thefigure}{\thesubsection b}
\stepcounter{tempfig}
\begin{figure}
    \centering
        \includegraphics[width=0.5\textwidth]{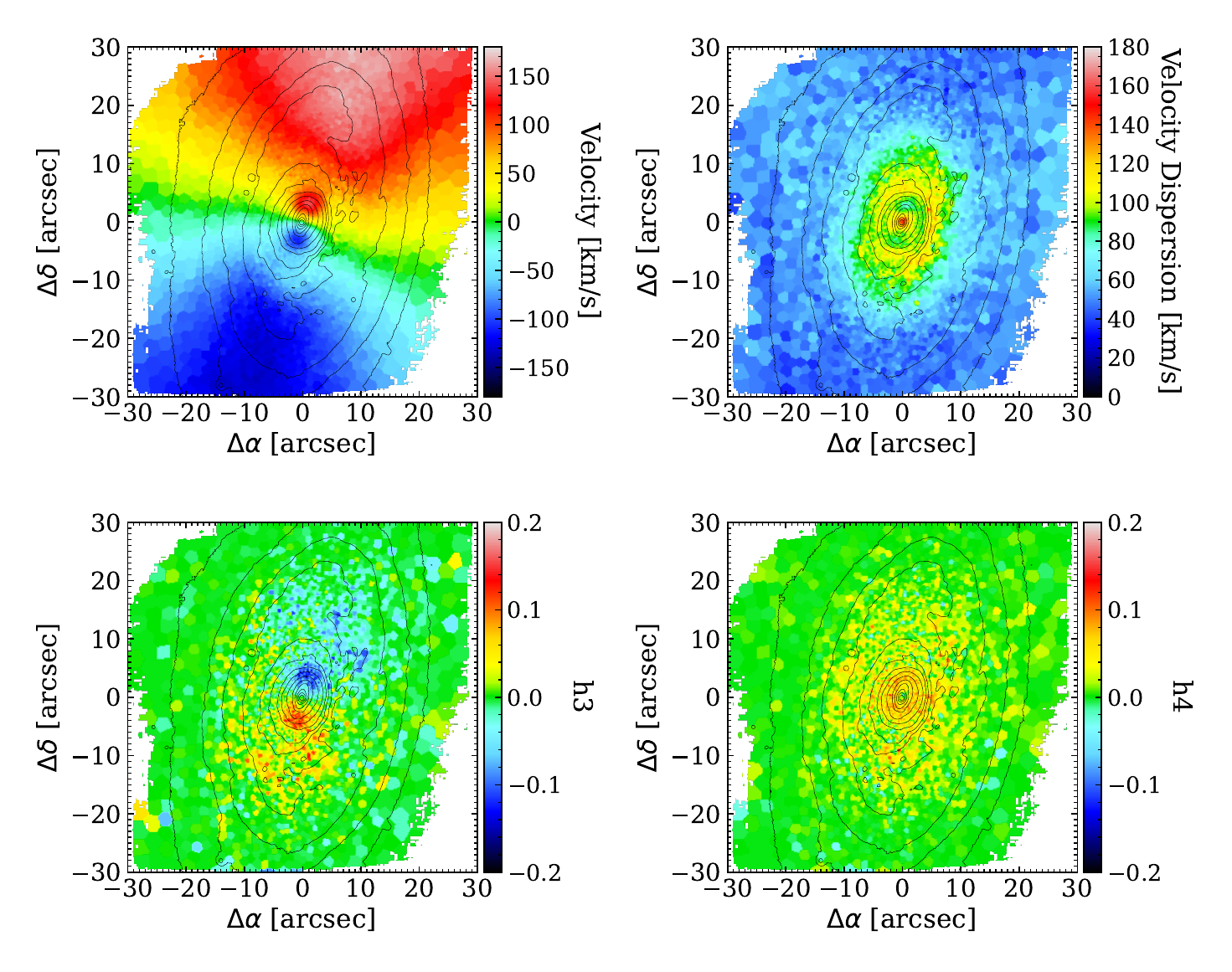}
        \caption{NGC\,3626. Panels show the same as in Fig.\,\ref{fig:NGC4303_stellar_conscsie}}
        \label{fig:NGC3626_stellar}
\end{figure}
\addtocounter{figure}{-1} 
\renewcommand{\thefigure}{\arabic{figure}}

The standard images and diagnostic maps we use for the rest of the discussion of NGC\,3626 are presented in Fig.\,\ref{fig:NGC3626}. In the large-scale and colour images, it can be seen that NGC\,3626 hosts a luminous core and an inner bar embedded in a dusty disk; the outer bar is not well defined in the colour image. The MUSE FOV covers more than half of the entire galaxy. As notably seen in the unsharp mask image, there are prominent dust lanes surrounding the outer bar region within the inner 20\arcsec. 

As shown in the flux maps across all line groups, the emission within the $\sim$4\arcsec~radius is relatively strong and extended towards the northwest and southeast direction, manifesting an ``S" shape. There is also strong and clumpy emission forming a boxy ring structure of $\sim$15\arcsec~radius, coinciding with some of the dust lanes. Extending from the ends of the inner bar towards northwest and southeast, two lanes of emission connect to the boxy ring structure. These two emission lanes resemble the characteristic emission along the bar if gas were to rotate in the same direction as stars, but the gas is counter-rotating in that region. This indicates that the gas in NGC\,3626 likely has two different origins.

As shown in the velocity maps, the velocity across all emission line groups shows peculiar structures. Overall the velocity maps follow the typical spider pattern, yet inwards from the boxy ring, there is a notable change in the PA of the zero velocity line. Within the $\sim$5\arcsec~radius, the PA of the LON is around 130\degr, while beyond this region, the PA of the LON is almost perpendicular to the zero velocity line, measuring $\sim$160\degr. This shift in the PAs suggests that the gas is moving in two different planes, indicating likely two different origins. At $\sim$3\arcsec~towards the northwest and southeast from the centre, two local velocity peaks are seen. 

Throughout the boxy ring, the velocity dispersion of \ha and \nii remains around 50\kms. Inwards from the boxy ring, the velocity dispersions are generally high, with the highest values being in the core. Almost perpendicular to the inner bar, a region extending from the centre towards $\sim$4\arcsec~east and west, shows consistently high velocity dispersions, with values around $\sim$130\kms; in certain spaxels values even exceed $\sim$150\kms. Additionally, situated $\sim$1-2\arcsec~beyond the endpoints of the inner bar, the velocity dispersions are also high, with values around 130\kms. This observation further confirms the presence of two kinematically distinct regions inwards from the boxy ring, supporting the argument for gas having two different origins there. The velocity dispersion of \oiii is not reliably estimated due to poor signal, but the values generally show a similar trend to \ha and \nii velocity dispersion values. 

While the distributions of PA of the LON and \textit{i} obtained from Kinemetry strongly vary inwards from the boxy ring, the distributions estimated outwards from the boxy ring converge around \textit{i}=53\degr and the PA of the LON=163\degr, which agree well with the values estimated from large-scale photometric fits in CBS (see Table\,\ref{tab:Disk_orientation_parameters}). This indicates that outwards from the boxy ring, the gas is moving within the galaxy plane, while inward from the boxy ring, the gas is moving out of the galaxy plane and can not be described by a planar motion. Consequently, the velocity residual map of the galaxy reveals distinct artefacts of wrongly fitted PA of the LON, $m$=1 harmonic terms, inwards from the boxy ring. Despite attempting various parameter adjustments for fitting the rotating disk better inwards from the boxy ring, these artefacts persisted and were not minimised. Although several extended structures are present throughout the boxy ring, they do not exhibit a turning behaviour towards the inner regions.

The velocity difference map is noisy and does not show prominent extended structures. However, only within $\sim$1--2\arcsec~radius, we notice a blue and a red structure with residual velocity amplitudes $\sim$20\kms, possibly counterparts of one another, though these structures are rather amorphous. Moreover, from the centre $\sim$4\arcsec~towards the north and south, there are also two regions with elevated velocity differences, one region exhibiting negative and the other region exhibiting positive velocities, with amplitudes $\sim$15--20\kms.

As shown in the BPT map, inwards from the boxy ring, the ionisation is dominated by the AGN, where there is mostly LINER emission. Throughout the boxy ring, there is composite emission with weak contribution from star formation.

The complex nature of kinematically distinct gas regions within NGC\,3626, prevents us from accurately modelling the disk inwards from the boxy nuclear ring; thus the residual velocity map of the galaxy was dominated by the artefacts. Although we identify emission along the inner bar and a possible flow along the inner bar from the velocity difference map, these structures are rather weak to be interpreted as extended gas shocks. Therefore the galaxy is not exhibiting signatures of extended shocks \textcolor{black}{and based on the grouping explained in Sect.\,\ref{sec:grouping-shock-class}, it is classified under \textit{other perturbations} category.}

\subsection{NGC\,4237}
\label{sec:NGC4237_notes}

\begin{figure*}
    \centering
    \includegraphics[width=1\textwidth]{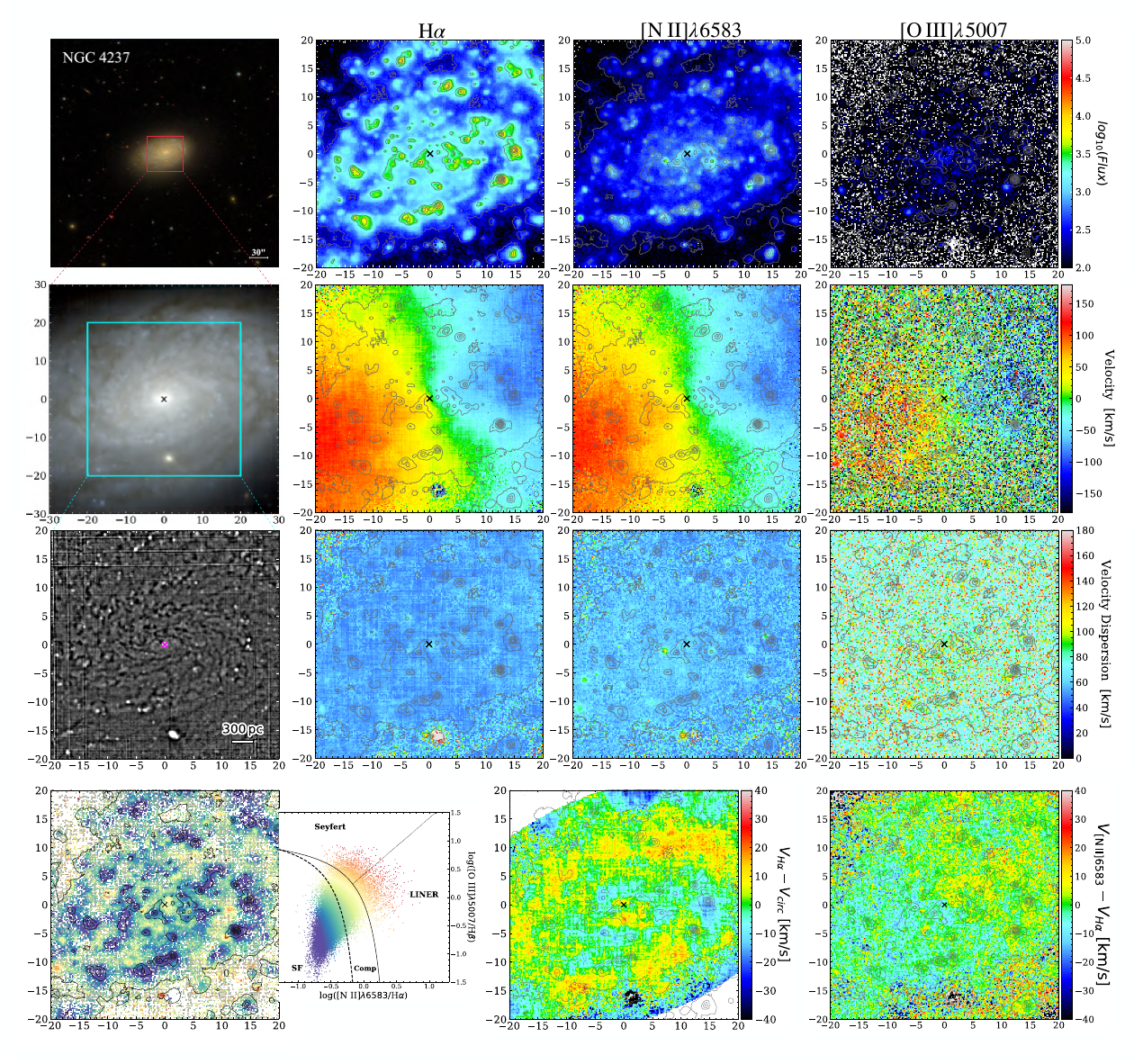}
    \caption{Images and diagnostics maps of NGC\,4237. Panels show the same as in Fig.\,\ref{fig:NGC4303}}
    \label{fig:NGC4237}
\end{figure*}

The diagnostic images and maps of NGC\,4237 are presented in Fig.\,\ref{fig:NGC4237}. NGC\,4237 is a member of the Virgo Galaxy Cluster \citep{Tully_1988_NGC}. As shown in its large scale image the galaxy is unbarred and has prominent star formation and dusty spiral arms encircling its bright core. The MUSE FOV covers almost the entire galaxy. The dust structures in spiral arms are enunciated in the unsharp mask image.

As shown in the flux maps, there is clumpy emission across all line groups forming a broken spiral structure. The \ha emission, likely originating from SF, is significantly stronger compared to \nii and \oiii. The velocity maps of \ha and \nii are relatively unperturbed, exhibiting motion close to circular. The \ha and \nii velocity dispersions remain around $\sim$50\kms across the whole FOV. 

The PA of the LON=107\degr and \textit{i}=53\degr obtained from \texttt{Kinemetry} agree well with the same parameters estimated in large-scale studies, \citep{Chemin_06}, as shown in Table\,\ref{tab:Disk_orientation_parameters}. Residual velocity map of the galaxy exposes weak quasi-spiral structures, with amplitudes of $\sim$5--10\kms, corresponding to deprojected residual velocity amplitudes of $\sim$20\kms or lower. Due to their low velocity amplitudes, we do not interpret these structures as extended shocks. Across the entire FOV, the velocity difference map is noisy. In this map, we observe a weak $m$=1 signature of amplitude $\sim$5--10\kms mirroring the velocity distribution, which for a presumably single-component gaseous medium may indicate a slight offset between the derived \ha and \nii velocity fields that is proportional to the velocity magnitude. This further confirms the absence of extended shocks in the field. 

The BPT map of the galaxy exhibits predominant SF in multiple spiral arms, together with some composite emission. Several round regions show the emission of the LINER type, coinciding with the regions with enhanced \nii velocity dispersion, which may indicate local shocks.

NGC\,4237 has relatively unperturbed kinematic fields and does not exhibit signatures of extended shocks.

\subsection{NGC\,4303}
\label{sec:NGC4303_notes}

The diagnostic images and maps of NGC\,4303 are presented in Fig.\,\ref{fig:NGC4303}. As the large-scale and colour images show, the galaxy has a luminous core surrounded by a star-forming nuclear ring of $\sim$3\arcsec~radius \citep{Colina_wada_00}. This galaxy also has a nested bar system; an inner bar of $\sim$2\arcsec~radius and an outer bar of $\sim$20\arcsec~radius \citep{Laine_02, Schinnerer_Maciejewski_02}, the outer bar is encircled by spiral arms. The MUSE FOV, covers the core, the nuclear ring and two bars. As enunciated in the unsharp mask image, two prominent dust lanes along the outer bar are connecting to the nuclear ring; inwards from the nuclear ring, the dust lanes along the inner bar are also visible.

The \nii flux, velocity and velocity dispersion maps of NGC\,4303 referred to in paragraph below show the measurements obtained from single Gaussian fits, presented in Appendix\,\ref{app:suplementary_sample_plots}, second row of Fig.\,\ref{fig:NGC1433_NGC4303_NGC4321_R1}. \ha and \oiii flux, velocity and velocity dispersion measurements can be represented with the main diagnostic maps presented in Fig.\,\ref{fig:NGC4303}. These measurements are obtained from fits including secondary component to \nii, yet \ha and \oiii are not affected by this correction, more details on this are provided in further paragraphs. 

\newcounter{tempfig4}
\setcounter{tempfig4}{\value{figure}}


As shown in Figs.\,\ref{fig:NGC4303} and \ref{fig:NGC1433_NGC4303_NGC4321_R1} (second row), the strongest emission across all line groups is confined within the nuclear ring and the core. Outside the nuclear ring, there is clumpy \ha and \nii emission; \ha being stronger. The velocity field of all emission line groups reveals perturbations superimposed on the circular rotation, clearly manifested in the rugged zero velocity line. Also, $\sim$5--10\arcsec~west to northwest of the nucleus, where the velocities are anticipated to be redshifted, several regions with blueshifted velocities are seen, most notably in \nii and \oiii. In the \oiii velocity map, there is also a region with strong blueshifted velocities, extending from the centre $\sim$7\arcsec~towards the south, suggesting a contribution from outflows. Inwards from the nuclear ring, the velocity dispersion values are up to $\sim$100\kms in \ha and up to $\sim$150\kms in \nii in two peaks $\sim$2\arcsec~offset from the centre. In the nuclear ring and to the east of it, there is low \ha and \nii velocity dispersion, with values around $\sim60$\kms. Outside the nuclear ring, both \ha and \nii velocity dispersion values are generally around 60--80\kms, but to the northwest and southeast of the centre, the values are high, around 100--130\kms. Particularly, in a small region $\sim$7\arcsec~to the northwest of the centre, the \nii velocity dispersions reach their highest values, exceeding 200\kms. The \oiii velocity dispersions, exhibit values over 200\kms in a ring of $\sim$1--4\arcsec~around the centre, and $\sim$6\arcsec~extension to the south. The region extending to the south coincides with the strong blueshifted \oiii velocities. 

The high velocity dispersion of \oiii prompted us to search for an excess emission in the spectra, and the Gaussian fits. If there is an excess emission in the spectra, fitting the excess and the main emission line with single Gaussians is not sufficient. The fitted Gaussian will be broad and will artificially increase the velocity dispersion and alter the velocity of the main emission line. Upon closer examination of the spectra where velocity dispersion of \oiii is high, we noticed a weak asymmetry in the \oiii line profile. To ensure this asymmetry does not artificially influence the measurements, following the same concept introduced in Paper IV, we allowed for fitting a secondary component to \oiii. If the emission is dominated by two components along the LOS, coming from both the excess emission sources and the disk, fitting two components will allow separating them. After allowing to fit a secondary component to \oiii, in the region where highly blueshifted \oiii velocities and high \oiii velocity dispersions are seen, we did not notice a significant \oiii velocity shift. The velocity dispersion of \oiii in the same region was reduced by 30--40\%; despite this reduction, the values remained high, confirming that the disturbance of the \oiii is real. The extreme kinematics of \oiii likely indicate a contribution from AGN outflows. Since kinematics are not significantly affected by the excess emission, to reduce the number of free parameters of the fit we only use single Gaussians to fit \oiii.

Not only in \oiii but high velocity dispersions are also seen in \nii. Thus, to ensure these values are real, following the same concept and procedure in Paper IV, we allowed for fitting a secondary component to the \nii within the entire FOV. We note that an excess emission to the blue wing of \nii can affect the Gaussian fits to both \ha and \nii, causing artificially high velocity and velocity dispersions for both, as seen in our pilot study with NGC\,1097. This issue can consequently affect the residual velocity map of NGC\,4303, which we construct by fitting a disk model to the \ha velocity field. Comparisons between measurements obtained from single Gaussian fits and fits including the secondary component, did not show a significant velocity shift or velocity dispersion change for \ha, indicating that the residual velocity map is not affected by an excess emission. However, the introduction of the secondary component led to a \nii velocity shift of $\sim$5--10\kms and a \nii velocity dispersion reduction of $\sim$40\% in the regions to the east and west of the centre, where the velocity dispersion of \nii was originally high, but also to south of the centre, which coincide with highly blueshifted \oiii velocities and high \oiii velocity dispersions. This spatial coincidence likely indicates that \nii is also affected by the AGN outflows. Such differences in \nii measurements are large enough to recognise the contribution from the AGN outflows but have a negligible effect on the velocity difference map, which incorporate \nii velocity field together with \ha. Therefore, to the east, west and south of the centre, we do not apply masking and the measurements are kept with those obtained from the single Gaussian fits.

However, the most prominent \nii velocity shifts, $\sim$40--60\kms, occur outside the nuclear ring, $\sim$7\arcsec~northwest of the centre, where the highest \nii velocity dispersions are originally seen; see Fig.\,\ref{fig:NGC4303_mask}. Including the secondary component to the fitting reduced the velocity dispersions there by $\sim$60--70\%. This region is outlined by a pink box in the second panel of Fig.\,\ref{fig:NGC4303_mask}, also presented in the third panel of the same figure after certain mask conditions are applied, which is explained below in more detail. Additionally, this region has anomalously blueshifted \nii velocities, while \ha velocities are at similar values to the surrounding. 

Because of such large changes in \nii velocity and velocity dispersion after introducing the secondary component to the fitting, we had a closer look at the spectra of spaxels in that region. The \ha and \nii wavelength region of spectra exhibited features much more complex than those seen in NGC\,1097. To demonstrate this, in the first panel of Fig.\,\ref{fig:NGC4303_mask}, we present two spectra: (a) from a spaxel where the fits are unaffected by the excess emission, (b) from a spaxel where the fits to [N\,II] components are significantly affected from the excess emission, (c) from the same spaxel in (b), after the excess emission is fitted with the secondary component. In all three sub-panels, there are two emission line components associated with components of [N\,II] doublet and of \ha; unlike what is seen in the spectra of NGC\,1097, NGC1433 and NGC4321 where the excess emission only appears as an asymmetry to the main emission line.
In the spaxel marked with an ``x" in the second panel (left) of the figure, fitting single Gaussians to each emission line recovers well the parameters of the main component. However, in the spaxel marked with an ``+" in the second panel (left) of the figure, the fit to the \ha line incorporates the main component only, and the fit to \nii line incorporates both the main and secondary components which artificially increases the \nii velocity dispersion and blueshifts the \nii velocities for those spaxels. Since \ha is fitted well, the measurements are not affected by the second component of \nii. Moreover, since [N\,II]$\lambda$6548 and \nii fits are kinematically bound, the wrongly fitted single Gaussian to \nii also affects Gaussian fits to [N\,II]$\lambda$6548. As shown in sub-panel (c), including the secondary component effectively separates the two \nii components, correcting the Gaussian fits. Consequently, the \nii velocity dispersions are reduced and the blueshifted velocities are brought back to values similar to the surrounding, as shown in the region outlined with a pink box in the second panel (right) of Fig.\,\ref{fig:NGC4303_mask}.

\newcounter{tempfig5}
\setcounter{tempfig5}{\value{figure}}

\renewcommand{\thefigure}{\thesubsection}
\begin{figure}
    \centering
    \includegraphics[width=0.42\textwidth]{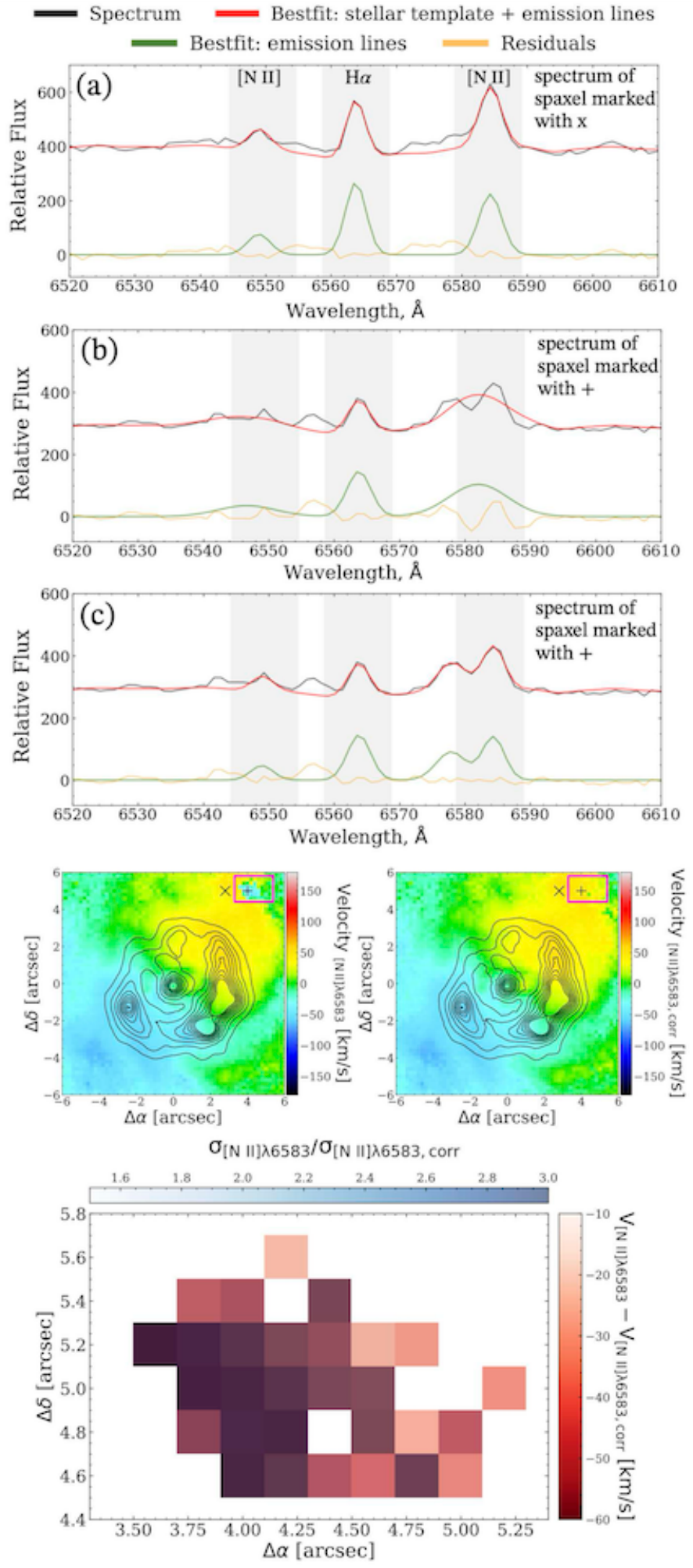}
    \caption{Spectra and masked region of NGC\,4303. \textit{first:} Spectrum of spaxels: marked with an ``x" in the second panel, excess emission does not affect the fits (a); marked with a ``+" in the second panel, excess emission affects the fits (b); same spaxel as in (b), after the excess emission is fitted with a secondary component (c). The lines represent the MUSE spectrum (black), the best fit of the stellar template with emission lines (red), the best fit of the emission lines (green) and the difference between the MUSE spectrum and the combined best fit with stellar template with emission line templates (orange). \textit{second:} velocity map of \nii when emission lines are fitted with single Gaussians (left), and with the secondary component included (right). The pink box represents the region where \nii measurements are the most affected by the excess emission. Black contours represent the \ha emission from the nuclear ring. \textit{third:} region enclosed by the magenta box in the second panel, showing the mask of spaxels for which the fits including secondary component were used in further analysis. The red and blue colour maps show the velocity shift of \nii and the ratio of \nii velocity dispersion, respectively, before and after including the secondary component in the fitting.}
    \label{fig:NGC4303_mask}
\end{figure} 

\addtocounter{figure}{-1} 
\renewcommand{\thefigure}{\arabic{figure}}

Since such large \nii velocity shifts, with $\sim$40--60\kms, would affect the velocity difference map significantly, we proceed to mask the artificial measurements in the affected region, and then use the measurements of the main line component of fits including the secondary component. The spaxels that are most affected by the excess emission are presented in the third panel of Fig.\,\ref{fig:NGC4303_mask}. To correct the measurements in these spaxels, we have followed the masking concept explained in Sect.\,\ref{sec:excess_masking}; ensuring there is a significant velocity shift and velocity dispersion change for \nii between measurements from the two fitting approaches. Since \ha is not affected by the secondary component, we do not include any conditions regarding \ha measurements, unlike in the case with NGC\,1097 presented in Paper IV. The conditions applied to construct the mask are as follows:

\begin{enumerate}[label=\roman*., align=left, leftmargin=*]
    \item $\rm V_{[N\,II]} - V_{\rm [N\,II], corr} \leq -10$ \kms
    \item $\rm \sigma_{[N\,II]} / \sigma_{\rm [N\,II], corr} \geq 1.5$
    \item $\rm \sigma_{[N\,II]w} \leq 300$\kms
    \item $3$\arcsec$\leq \Delta\alpha \leq6$\arcsec~and $3.5$\arcsec$\leq \Delta\delta \leq6.5$\arcsec
\end{enumerate}
where \textit{corr} indicates the values extracted from the fits including the secondary component. The third condition, the velocity dispersion limit of the secondary \nii component ([N\,II]w), ensures that the secondary component is real and is not fitted to noise. The spatial cuts are applied to select the region where the change in velocity and velocity dispersion ratio is consistently large. For further analysis, the flux, velocity, and velocity dispersion values for all emission line groups within the mask region are replaced with those from the corresponding main components in fits that include the secondary component. Although \ha and \oiii measurements are not affected, we continue to use measurements from fits with a secondary component to maintain consistency. The resultant flux, velocity and velocity dispersion maps of all emission line groups are presented in Fig.\,\ref{fig:NGC4303}.

The PA of LON returned from \texttt{Kinemetry} varies from 110\degr, within the outer regions, to 140\degr within regions inside the nuclear ring. Since the flow is more circular inside the nuclear ring, we explored around 140\degr, and converged using 135\degr as the PA of the LON since this value minimised the artefacts seen in the residual velocity map; 135\degr also agrees with values obtained from large-scale fits within a couple of degrees (\citealp{Chemin_06}: PA of LON=132\degr). \textit{i} inferred from \texttt{Kinemetry} converged around 42\degr, which differs by 18\degr from the inclination obtained from large-scale fits (\citealp{Chemin_06}: \textit{i}=23.5\degr). Upon several checks, the residual velocity map resultant from subtracting a disk with \textit{i}=23.5\degr, reveals weak m=3 harmonic terms, while such artefacts are less emphasised when the disk model is constructed with \textit{i}=42\degr. Thus for further analysis, we settle on using the \textit{i} and PA of the LON inferred from \texttt{Kinemetry}. As shown in Fig.\,\ref{fig:NGC4303}, the residual velocity map reveals a coherent structure exhibiting negative velocities, with amplitudes of $\sim$20\kms to the west of the nucleus. This structure extends from the north towards the south and connects to the nuclear ring at the southwest, roughly coinciding with the northwestern dust lane external to the nuclear ring. The largest deprojected residual velocity amplitudes observed across this structure are roughly $\sim$50\kms. After crossing the nuclear ring, the same coherent structure extends towards the nucleus from the west of the centre, roughly along the major axis of the inner bar. Although several extended structures exhibit enhanced positive residual velocities throughout the FOV, exhibiting deprojected residual velocity amplitudes up to $\sim$20\kms, particularly in the nuclear ring, we do not see a clear counterpart to the structure with negative residual velocities.

Outside the nuclear ring, roughly coinciding with the coherent structures in the velocity residuals there, a coherent structure with negative velocity differences of $\sim$15--20\kms is seen, extending inwards from the nuclear ring and approaching to the nucleus from the east of the centre. This structure is located $\sim$0.2--0.5\arcsec~downstream of a dust lane, which appears to be related to the inner bar; thus this kinematic structure can be indicative of an inflow along the inner bar. Moreover, two regions $\sim$1--2\arcsec~northwest and southeast from the centre, show elevated velocity differences of amplitudes $\sim$15\kms, one with negative and one with positive velocities, supporting the argument of potential inflow seen along the inner bar. 

The BPT map of the galaxy shows predominant SF in the nuclear ring and outside of it where clumpy \ha emission is seen. In the remaining regions, the ionisation is dominated by composite and LINER emission. Inwards from the nuclear ring, the emission is of LINER kind, agreeing with the classification from former works (e.g., \citealp{Ho_1997}). 
 
Both the residual velocity and velocity-difference maps reveal consistent signatures of extended shocks along the outer and inner bars of NGC\,4303. \textcolor{black}{These coherent structures extend over several kiloparsecs and exhibit large deprojected residual velocity magnitudes of up to $\sim$50\kms, which can be seen as the outlined features in Fig.\,\ref{fig:Sample_extended_structures_perturbed_galaxies_residualvel}. Their coherent morphology, large spatial extent, and high residual velocities therefore support their association with extended shocks.} This consistency across different methodologies indicates that signatures of extended shocks can be reliably detected using \textcolor{black}{complementary} approaches. \textcolor{black}{These findings reinforce the importance of searching for extended shocks using both model-dependent and model-independent methods.}

\subsection{NGC\,4321}

\label{sec:NGC4321_notes}

NGC\,4321, also known as M100, is a member of the Virgo Cluster and has been extensively studied in former works \citep{Knapen_1995a,Knapen_1995b,Knapen_2000,Ryder_2001,Allard_05,Garcia-Burillo-05, Haan_08, Pohlen_2010, Munoz_22,Liu_2023}. The images and diagnostic maps of the galaxy are presented in Fig.\,\ref{fig:NGC4321}. As shown in the large-scale and colour images, the galaxy has a bright core and a nuclear ring within the $\sim$8--10\arcsec~radius. The galaxy is double barred, hosting an inner bar within $\sim$5\arcsec~radius and an outer bar within $\sim$26\arcsec~radius \citep{Knapen_1995a,Knapen_1995b,Erwin_04}. There are several star-forming spiral arms surrounding the outer bar. Both large-scale and MUSE FOV images, and the unsharp mask image reveal the curved dust lanes along the outer bar, joining the nuclear ring and continuing towards the nucleus. 
\newcounter{tempfig6}
\setcounter{tempfig6}{\value{figure}}

\renewcommand{\thefigure}{\thesubsection a}
\begin{figure*}
    \centering
    \includegraphics[width=1\textwidth]{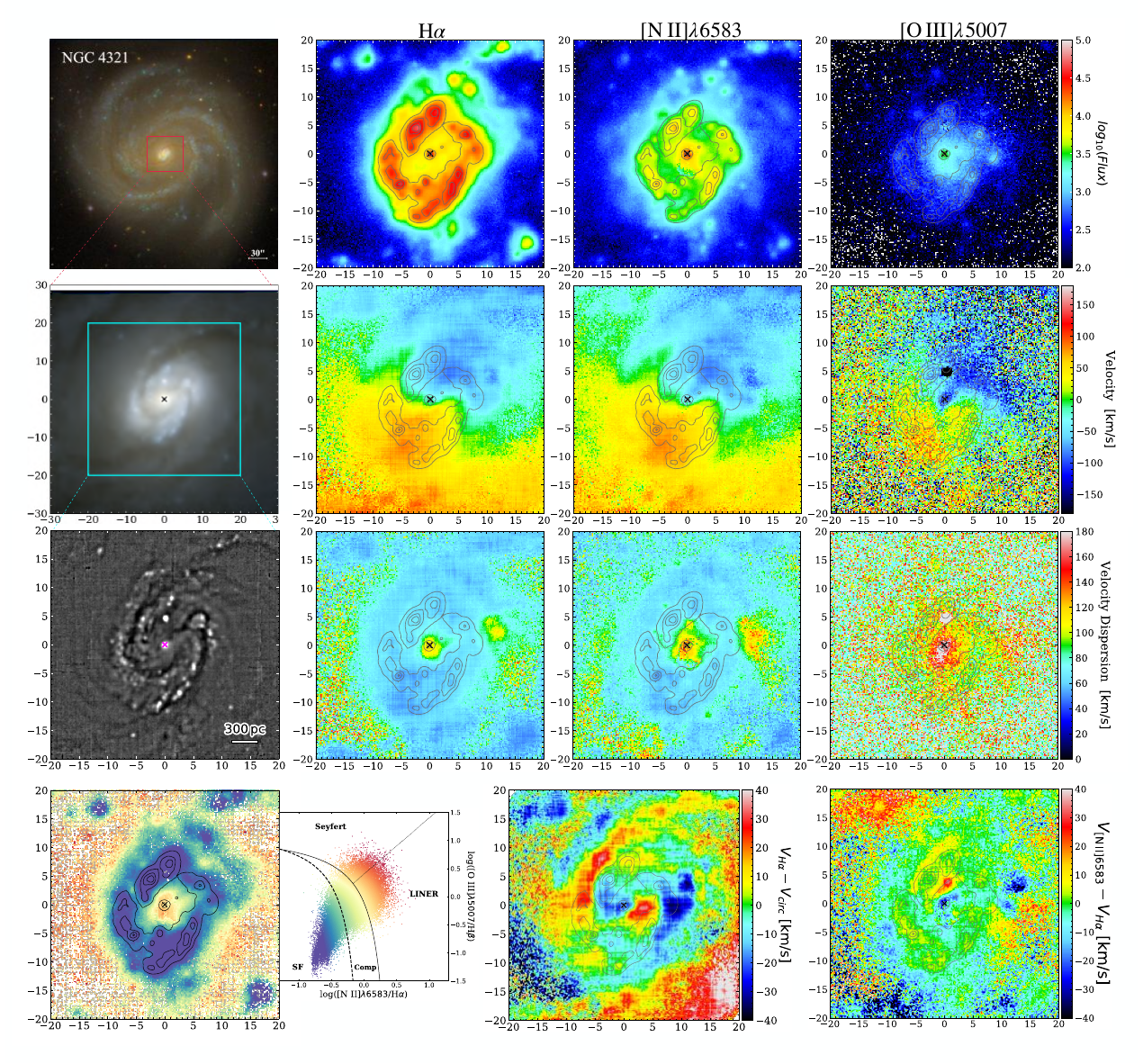}
    \caption{Images and diagnostics maps of NGC\,4321. Panels show the same as in Fig.\,\ref{fig:NGC4303}}
    \label{fig:NGC4321}
\end{figure*}

The \nii flux, velocity and velocity dispersion maps of NGC\,4321 referred to in paragraph below show the measurements obtained from single Gaussian fits, presented in Appendix\,\ref{app:suplementary_sample_plots}, third row of Fig.\,\ref{fig:NGC1433_NGC4303_NGC4321_R1}. \ha and \oiii flux, velocity and velocity dispersion measurements can be represented with the main diagnostic maps presented in Fig.\,\ref{fig:NGC4321}. These measurements are obtained from fits including secondary component to \nii, yet \ha and \oiii are not affected by this correction, more details on this are provided in the following paragraphs.

As presented in Figs.\,\ref{fig:NGC4321} and \ref{fig:NGC1433_NGC4303_NGC4321_R1} (third row), the strongest \ha and \nii emission is in the nuclear ring and the core. \ha emission is significantly stronger than \nii emission. Outside the nuclear ring, clumpy \ha and \nii emission is seen. \oiii emission is weak throughout the FOV, yet the most pronounced emission is confined within the core. Across all emission line groups, velocity maps reveal perturbations from circular rotation with the zero velocity line having strong indents outside the nuclear ring at $\sim$5--7\arcsec~radius. There is also an indent in the zero velocity line, $\sim$3\arcsec~southeast of the nucleus, evident in \nii and \oiii, but absent in \ha. In the nucleus, the velocity dispersion of all emission line groups is high. Moreover, in a region extending from the central $\sim$4\arcsec~towards the south, which partially overlaps with the indent in the zero velocity line, the \nii velocity dispersions are up to $\sim$170\kms and \oiii velocity dispersions are up to $\sim$200\kms; while \ha velocity dispersions are not elevated there. Throughout the nuclear ring and the curved dust lanes, the velocity dispersion of \ha and \nii is low, around $\sim$60--70\kms. Outside the nuclear ring, to the west and southeast of the centre, several regions exhibit enhanced \ha and \nii velocity dispersions with values ranging from 90\kms to 120\kms. 

Examining the spectra in the regions south of the nucleus, where the highest \nii velocity dispersions are seen, we noticed an excess to the blue wing of \nii, similar to that seen in NGC\,1097 and NGC\,1433. This indicates that the high velocity dispersion in those spaxels is artificial and caused by poorly fitted Gaussians. Thus, to account for the excess, we allowed for fitting a secondary component to \nii, following the approach introduced in Paper IV. The secondary component fitted to the excess will allow separating the main emission coming from the disk and an excess source.

Next, to identify the regions that are the most affected by the excess emission, we compared the kinematic measurements of main emission lines obtained from single-Gaussian fits and the fits with the secondary component. The \ha and \oiii measurements throughout the FOV are generally unaffected. However, extending from the centre $\sim$4\arcsec~towards the south, we identified a localised region with large \nii velocity shifts, up to $\sim$45\kms, between the two kinematic measurements explained above; this region is shown in the top-left panel of Fig.\,\ref{fig:NGC4321_mask}, after the masking criteria explained in further are applied. 

\newcounter{tempfig7}
\setcounter{tempfig7}{\value{figure}}

\renewcommand{\thefigure}{\thesubsection b}

\begin{figure}
    \hspace{-0.4cm}
    \includegraphics[width=1.1\columnwidth]{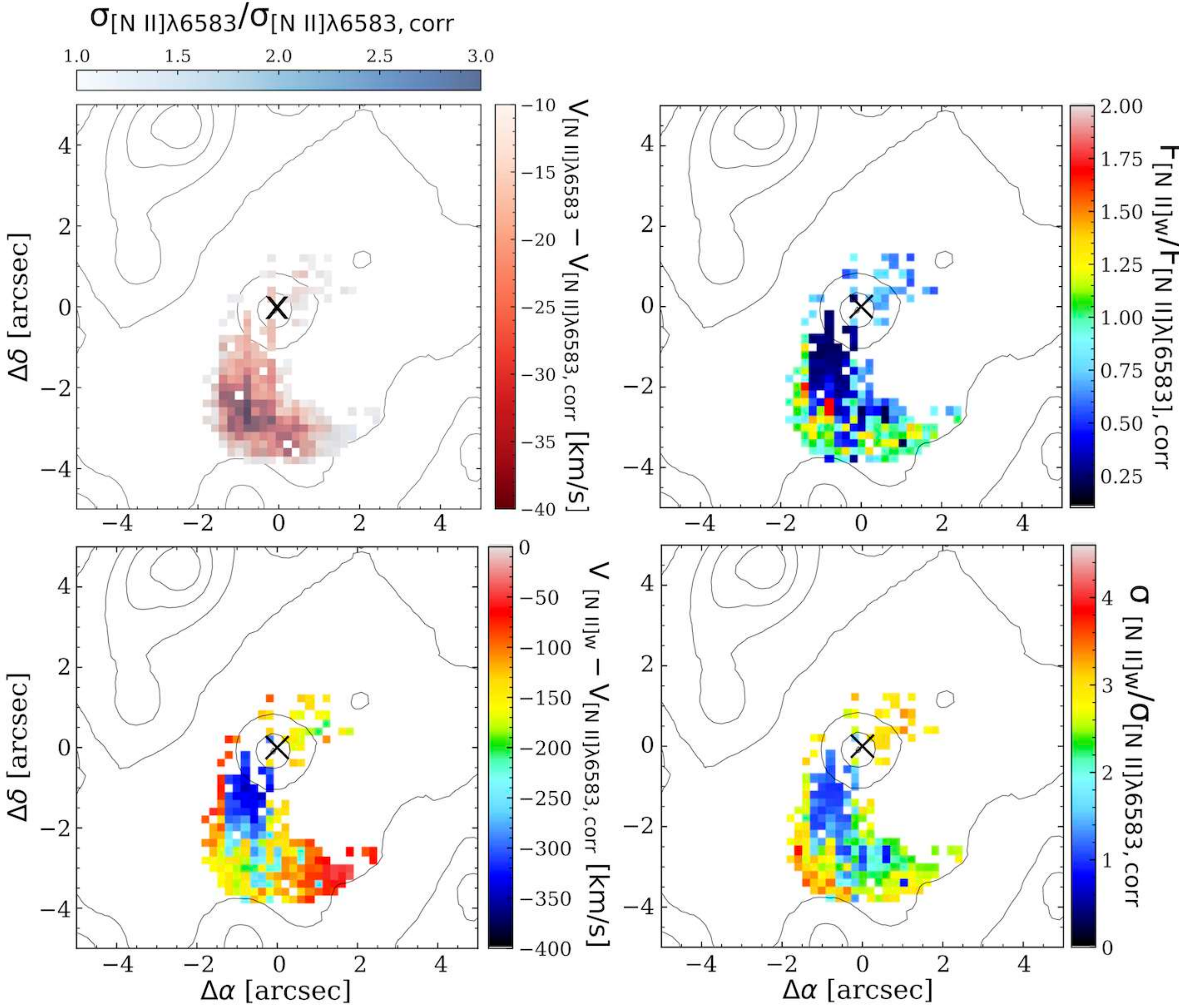}
    \caption{Mask of spaxels for which the fits including secondary component were used in further analysis of NGC\,4321 (top left). The red and blue colour maps show the velocity shift of \nii and the ratio of \nii velocity dispersion, respectively, before and after including the secondary component in the fitting. Comparison of the secondary and main component: their flux ratio (top right) velocity difference (bottom left) and velocity dispersion ratio (bottom right).}
    \label{fig:NGC4321_mask}
\end{figure}

\addtocounter{figure}{-1} 
\renewcommand{\thefigure}{\arabic{figure}}

To correct the kinematic measurements of \nii, which are the most affected by the excess emission in spectra, we must define the region where including the secondary component alters the kinematic measurements of the \nii significantly. To define this region, we used the following criteria:
\begin{enumerate}[label=\roman*., align=left, leftmargin=*]
    \item $\rm V_{[N\,II]} - V_{\rm [N\,II], corr} \leq -10$ \kms
    \item $\rm \sigma_{[N\,II]} / \sigma_{\rm [N\,II], corr} \geq 1.2$
    \item $\rm \sigma_{[N\,II]w} \leq 300$\kms
    \item $-2.5$\arcsec$\leq \Delta\alpha \leq$2.5\arcsec~and $-4$\arcsec$\leq \Delta\delta \leq$2\arcsec
\end{enumerate}
where \textit{corr} indicates the values extracted from the fits including the secondary component. The third condition, the velocity dispersion limit of the secondary \nii component ([N\,II]w) ensures that the secondary component is real and is not fitted to noise. The spatial cuts limit, defined in criterion iv., is applied to select the region where the change in velocity and velocity dispersion ratio is consistently large. The above criteria ensure that within the region, where the derived flux, velocity and velocity dispersion measurements are significantly affected by the presence of the excess \nii emission, those measurements are replaced by the ones from the fit that include the secondary component. Although, for NGC\,4321 only \nii measurements are affected by the excess emission, we also replaced the \ha and \oiii measurements obtained from single Gaussian fits with those obtained from the fits with the secondary component, to maintain consistency across different line groups. After these corrections, \ha and \nii velocities and velocity dispersion are in much better agreement than before the correction. The velocity dispersion of \nii within the masked region is reduced by roughly $\sim$30$\%$. The resultant flux, velocity and velocity dispersion maps of all emission line groups are presented in the Fig.\,\ref{fig:NGC4321}; \oiii and \ha maps are not affected.

As shown in the bottom panels of Fig.\,\ref{fig:NGC4321_mask}, the secondary component appears blueshifted by $\sim$300\kms with respect to the main component, and the blueshift decreases away from the galaxy centre. The velocity dispersion of the secondary component is larger than that of the main component by up to 20\%, and this ratio increases away from the nucleus. The coherent morphology of the region where excess is dominating, and extreme kinematics of the secondary component within this region suggest the presence of AGN outflows.

The disk orientation parameters inferred from \texttt{Kinemetry}, PA of the LON=155\degr and \textit{i}=36\degr, are within only a couple of degrees from parameters estimated from large-scale fits (e.g., \citealp{Haan_08}: PA of the LON=152\degr, \textit{i}=32\degr). The residual velocity map of galaxy exposes two strong coherent structures, one with positive and one with negative velocities, with amplitudes of $\sim$25--30\kms. The structure with positive velocities extends from the northwest and joins the nuclear ring from the east, and the structure with negative velocities extends from the southeast and joins the nuclear ring from the southwest. After joining the nuclear ring both structures continue extending along the nuclear ring. These coherent structures are downstream from the dust lanes by $\sim$0.5--1\arcsec~and are consistent with being their counterparts.  The largest deprojected residual velocity amplitudes across these structures are around 55\kms. Outside the nuclear ring, to the northwest, a part of the coherent structure with negative velocities continues extending straight, towards the north rather than following the ring; this region coincides with the region with enhanced \nii and \ha velocity dispersion, likely indicating the overshooting of the postshock gas there. Inwards from the nuclear ring, two regions with enhanced residual velocities with amplitudes of $\sim$20--30\kms are seen; one exhibiting negative and one positive velocities. These regions can be a continuation of the structures extending along the nuclear ring and likely indicate the continuity of inflow to the innermost region. 

The velocity difference map of the galaxy exposes several distinct structures. Inwards from the nuclear ring, there are regions with positive and negative velocity differences of amplitudes $\sim$20--30\kms, but these regions are rather amorphous and show no point symmetry around the nucleus. Outside the nuclear ring, there is a coherent structure exhibiting positive velocity differences, with amplitudes of $\sim$10\kms which roughly coincides with the coherent structure with positive velocities in the residual velocity map. Within the region associated with overshot gas, and west of it, the velocity differences are elevated, exhibiting amplitudes of $\sim$15\kms or more; this observation supports the presence of complex dynamics there. 

Furthermore, the BPT map shows the emission of LINER type in the nucleus. The nuclear ring and the regions with clumpy \ha and \nii emission are predominantly star-forming. Within the remaining regions outside the nuclear ring, there is predominant LINER emission.

NGC\,4321 reveals strong coherent continuous structures in the residual velocity map, \textcolor{black}{extending over several kiloparsecs (see also features outlined in Fig.\,\ref{fig:Sample_extended_structures_perturbed_galaxies_residualvel}) and exhibiting large deprojected residual velocity amplitudes up to $\sim$55\kms, which are consistent with the concept of extended shocks in gas.}

\subsection{NGC\,4380}
\label{sec:NGC4380_notes}
The diagnostic images and maps of NGC\,4380 are presented in Fig.\,\ref{fig:NGC4380}. As the large scale and colour image show, NGC\,4380 is an unbarred galaxy with a bright core encircled by flocculent SF and dusty spiral arms. The MUSE FOV encompasses the core and a portion of the dusty spirals around it. As presented in the unsharp mask image, the prominent spiralling dust features around the core are revealed.

\renewcommand{\thefigure}{\thesubsection}
\begin{figure*}
    \centering
    \includegraphics[width=1\textwidth]{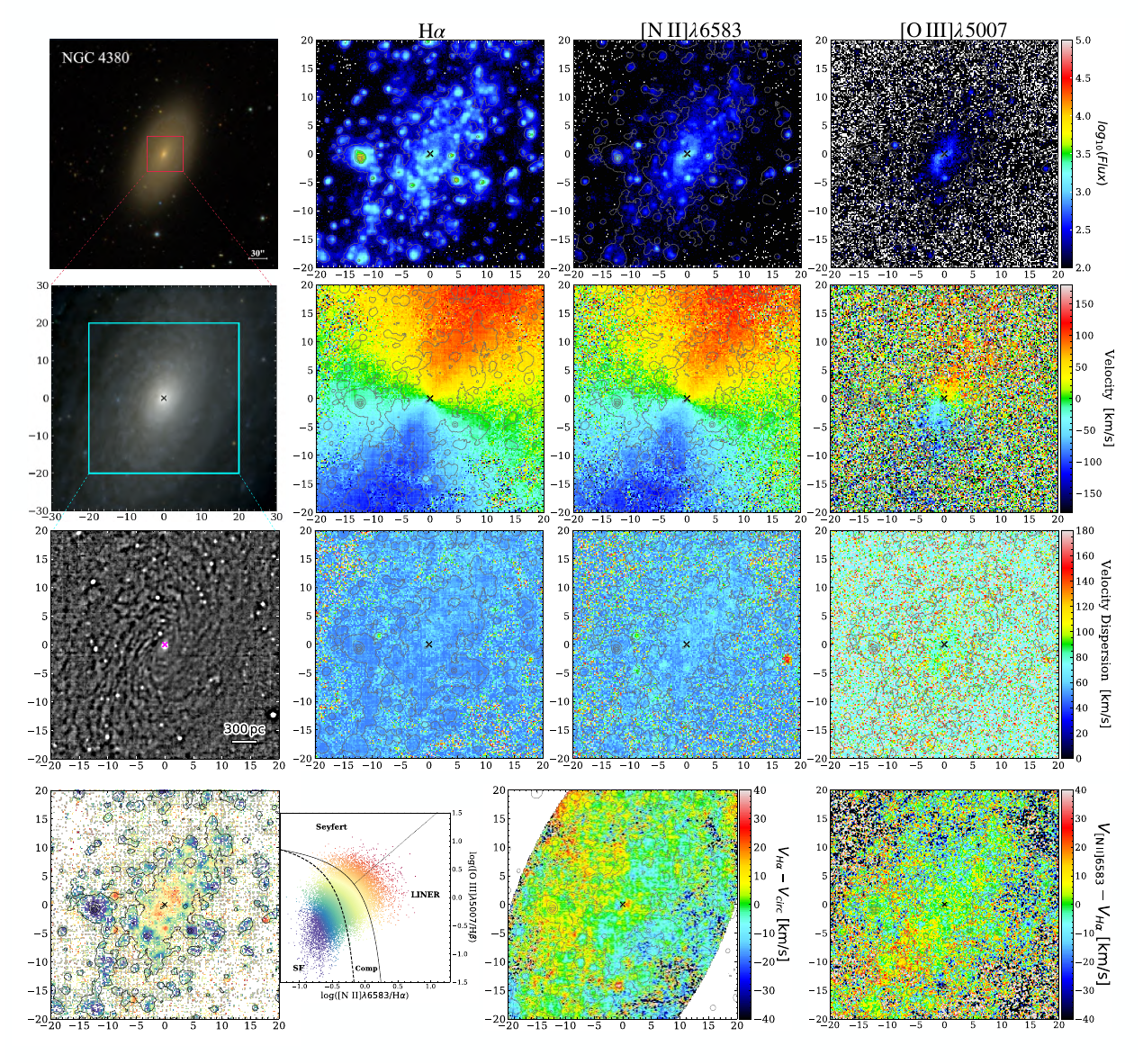}
    \caption{Images and diagnostics maps of NGC\,4380. Panels show the same as in Fig.\,\ref{fig:NGC4303}}
    \label{fig:NGC4380}
\end{figure*}

As shown in the flux maps, emission across all line groups is weak, particularly \oiii. \ha and \nii emission is clumpy. The velocity fields of \ha and \nii are mostly unperturbed. The zero velocity line remains straight. The velocity dispersion of \ha and \nii remains low, around $\sim$50--60\kms throughout the entire FOV. 

The PA of the LON=161\degr and \textit{i}=59\degr estimated by fitting the velocity field with \texttt{Kinemetry} agrees well with the parameters obtained from large-scale photometry (PA of the LON=157\degr and \textit{i}=59\degr). The residual velocity map of the galaxy exposes faint and noisy $m$=1 harmonic terms; after finely adjusting the PA of the LON in 1\degr increments, we settled on using 159\degr since this value raises the weakest artefacts. This demonstrates that once the velocity field is undisturbed, \texttt{Kinemetry} can estimate the disk orientation parameters consistent, within couple of degrees, with those obtained from large-scale photometric fits. The deprojected residual velocity amplitudes throughout the faint structures seen in the map are small, 20\kms or below, indicating an absence of a shock nature. Similarly, the velocity difference map is overall noisy and exposes no prominent structures.

The BPT map of the galaxy shows prominent SF interspersed with LINER emission spots in the spiral arms, where clumpy \ha and \nii emission is seen. Moreover, within $\sim$3--4\arcsec~radius, ionisation is dominated by LINER emission. While the central LINER emission can be associated with contribution from the core, the LINER emission along the spiral arms can be attributed to hot old low-mass stars or potentially to supernovae.

NGC4380 has relatively unperturbed kinematic fields and does not present signatures of extended shocks. This quiescence possibly can be attributed to the unbarred morphology of the galaxy, similar to our findings for NGC\,4237. 

\subsection{NGC\,4457}
\label{sec:ngc4457_notes}

\begin{figure*}
    \centering
    \includegraphics[width=1\textwidth]{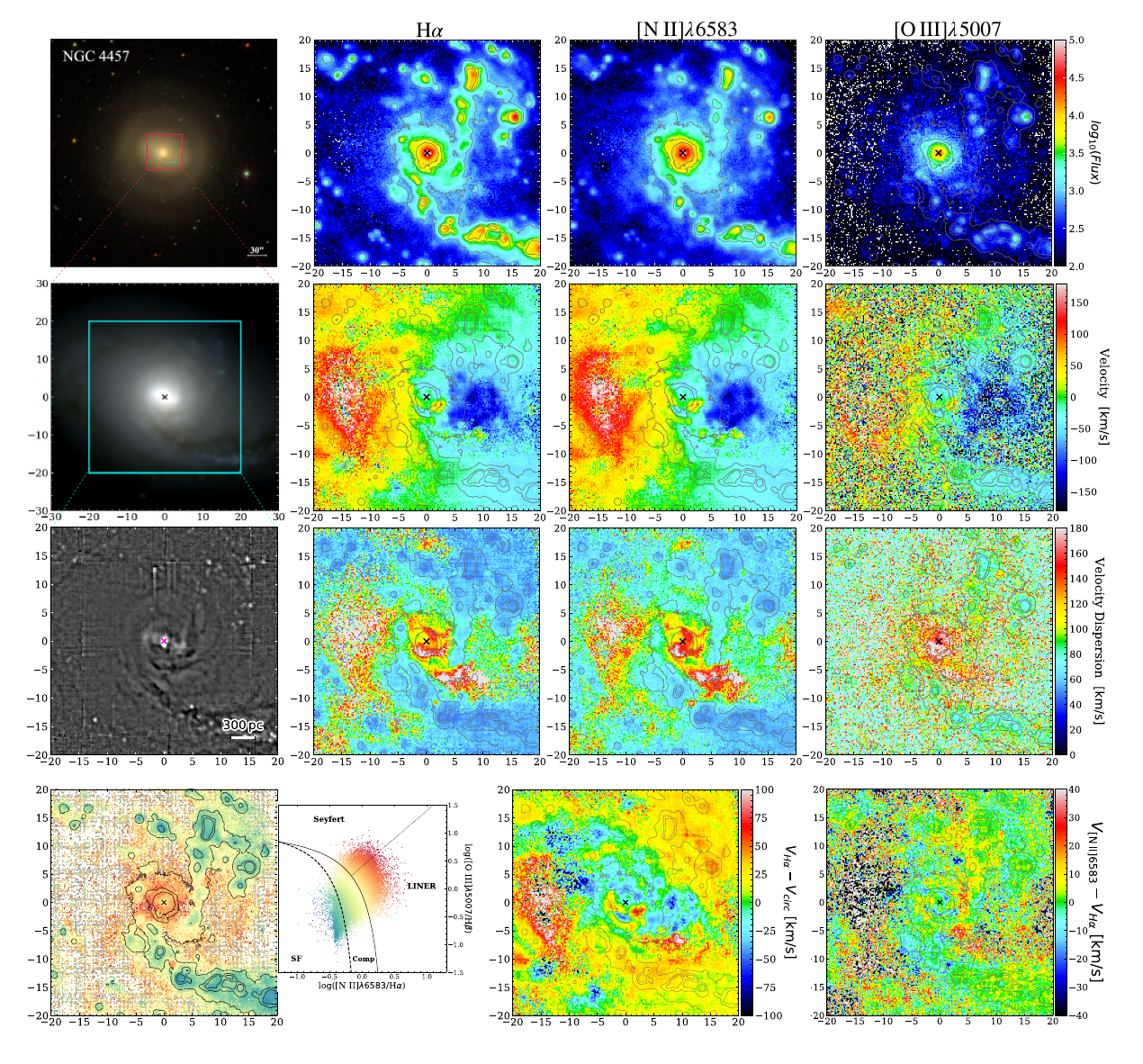}
    \caption{Images and diagnostics maps of NGC\,4457. Panels show the same as in Fig.\,\ref{fig:NGC4303}}
    \label{fig:NGC4457}
\end{figure*}

NGC\,4457 has been well-studied in many former works \citep{Vollmer_2013,Nemmen_2014,Cortes_15,Querejeta_2021,Stuber_21,Pan_2022,Munoz_22,Roberts_23}. The images and diagnostic maps of the galaxy are presented in Fig\,\ref{fig:NGC4457}. As shown in the large-scale and colour images, the galaxy has a bright core, which is classified as LINER type \citep{Terashima_2000,Munoz_22}, and a bar of $\sim$2--3\arcsec~radius. Moreover, based on its brightness profile, NGC\,4457 is classified as having a lensed morphology \citep{Querejeta_2021}; the lens is visible within the $\sim$30\arcsec~radius. The MUSE FOV covers almost the entire lens, the bar and the core. In colour images, two dust lanes are seen, resembling standard dust lanes along bars but instead in an oval with major axis $\sim$80\degr, measured from North to the East. A string of young stars is seen downstream the dust lanes. The formation of these stars is likely triggered by the increased gas density downstream of the extended shocks, unlike in standard bar lanes where strong shear usually inhibits SF. In the unsharp mask image, the dust lane that extends from the southwest end of the frame that curves inwards at east-southeast is more pronounced than its counterpart. The most prominent region of the counterpart dust lane is seen extending north to south within the inner 5\arcsec~radii. 

As shown in the flux maps, the emission for all line groups is concentrated within the $\sim$3--4\arcsec~radius. A string of \ha and \nii emission is seen along the dust lane to the southwest. Several \ha and \nii clumps, aligned with the north to south direction at $\sim$5\arcsec~radius, are present downstream from the counterpart dust lane. Examining the velocity maps of \ha and \nii, we notice disturbances and velocities inconsistent with the rotating disk. Within $\sim$4\arcsec~radius, the zero-velocity line has strong indents, and positive velocities are present to the southwest of the centre where negative velocities would be expected for a rotating disk. A string of regions with \ha and \nii velocities positive or close to zero extends into the region where velocities are expected to be blueshifted, at $\Delta\delta \sim -7$, starting at $\Delta\alpha \sim$ 0\arcsec~and stretching up to $\Delta\alpha \sim$ 10\arcsec. 

The velocity dispersion of \ha and \nii is high in the core and around it; within $\sim$4\arcsec~radius values exceed 150\kms. Similarly, high velocity dispersion is recorded to the east from the centre, up to a 15\arcsec~distance. Additionally, in a region extending from $\sim$6\arcsec~to $\sim$15\arcsec~of the centre, the velocity dispersion values are as high as the values seen in the core. This region roughly coincides with the abovementioned string of regions with anomalous positive velocities. In regions with clumpy \ha and \nii, the velocity dispersion of \ha and \nii remain around $\sim$60\kms. The velocity and velocity dispersion maps of \oiii are rather noisy due to poor signal, yet the values suggest the same trends as the values seen in \ha and \nii velocity and velocity dispersion maps.

The PA of the LON=80\degr and \textit{i}=20$^\circ$ estimated using \texttt{Kinemetry} agree well with the parameters obtained from large-scale fits within CBS (see Table\,\ref{tab:Disk_orientation_parameters}). Consequently, we adopt these parameters for the final disk model of the galaxy. The residual velocity map of the galaxy reveals several regions within $\sim$4\arcsec~radius with velocity residuals of amplitudes $\sim$30--50\kms. Four of these regions exhibit negative velocities. Three of them are towards the northeast to northwest and one towards the south. The other two regions exhibit positive velocities, one to the east and one to the west. There are no signs of wrongly fitted \textit{i} or the PA of the LON in the residual velocity map, indicating that the structures seen in the residual velocity map, particularly within $\sim$4\arcsec, are not due to the modelling of the disk and are likely caused by the AGN outflow cone piercing the disk and raising complex signatures in the kinematics \citep{Fischer_2013}.

In the region to the southwest, where elevated velocity dispersions and anomalous velocities are seen, the velocity residuals exhibit coherent positive values with amplitudes exceeding 50\kms. Moreover, high positive residual velocities are partially coinciding with the emission lane to the southwest. This structure potentially continues towards the east of the frame where values reach above 50\kms. In the same region to the east, the velocity dispersion of both \ha and \nii is high. Although several coherent structures can still be noticed, which can be associated with the inflow in the bar, the residual velocities throughout the emission lane are noisy and patchy. The corresponding deprojected residual velocity amplitudes across the extent of these structures are up to 120--125\kms. Due to complex mechanisms within the innermost regions of the galaxy, none of the structures that is relatively extended can be traced to the nucleus.

The velocity difference map of the galaxy is rather noisy. Inside the north to south running dust lane, at $\sim$4--5\arcsec, velocity differences are increased with amplitudes around 15\kms. 

Moreover, the BPT map exhibits LINER-type emission within the $\sim$4\arcsec~radius. This supports the idea that the AGN might be contributing to disturbances of gas in the core region. In regions of strong \ha emission, there is composite emission with a contribution from SF.

Despite potential signatures of extended shocks within the bar region, the presence of strong local SF and AGN activity within the innermost regions of the galaxy prevents us from tracing these signatures to the innermost regions. Therefore, we do not classify this galaxy as showing signs of coherent extended shocks, and it is grouped under \textit{other perturbations} category.

\subsection{NGC\,4643}

\begin{figure*}
    \centering
   \includegraphics[width=1\textwidth]{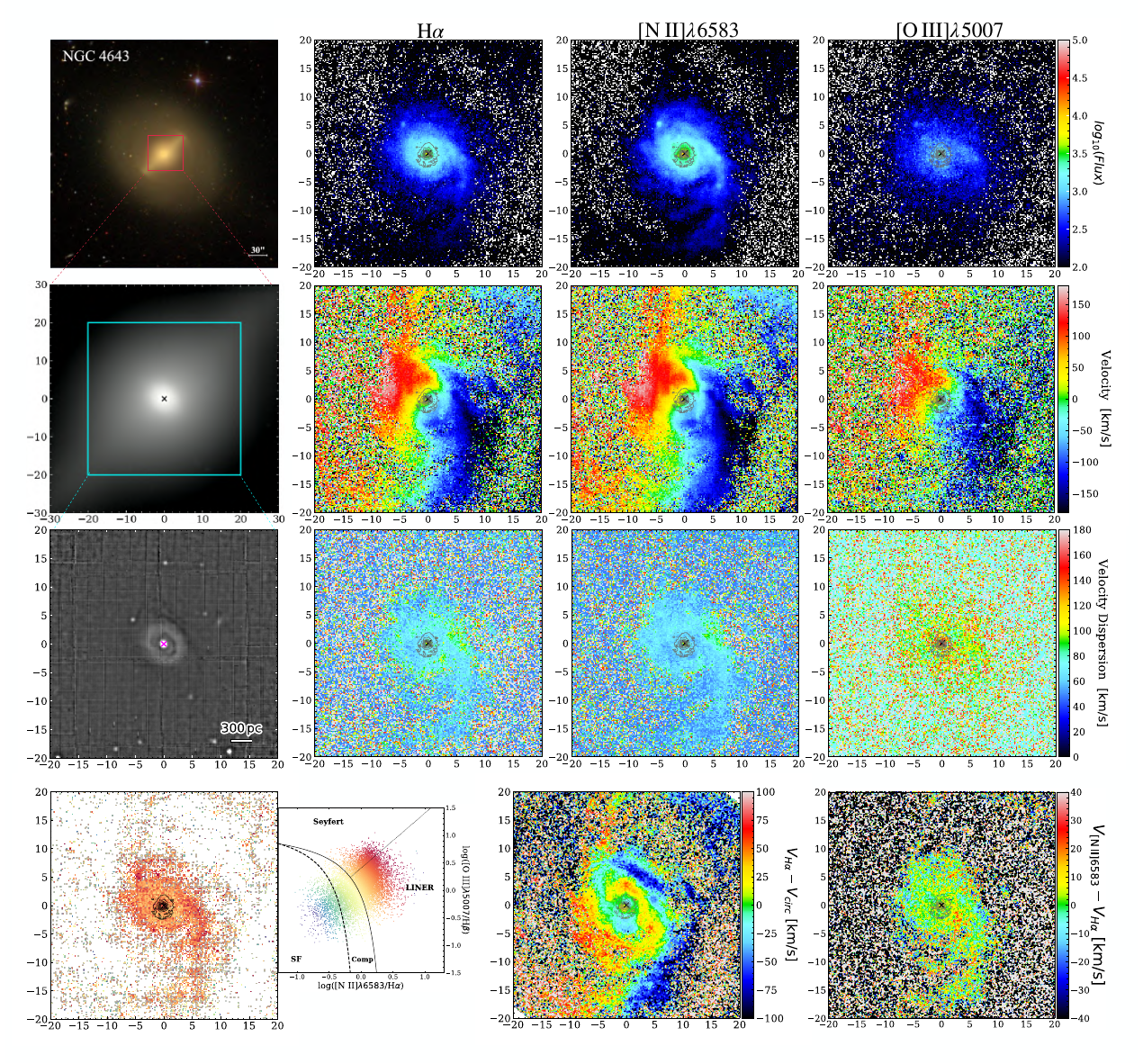}
    \caption{Images and diagnostics maps of NGC\,4643. Panels show the same as in Fig.\,\ref{fig:NGC4303}}
    \label{fig:NGC4643}
\end{figure*}

NGC\,4643 is one of the galaxies for which CBS provides a detailed morphological, photometric, and stellar kinematic study \citep{Erwin_21}.
We present the images and diagnostic maps of the galaxy in Fig.\,\ref{fig:NGC4643}. As shown in the large scale image, NGC\,4643 has a bright core and a strong bar surrounded by a ring. The MUSE FOV covers the core and almost the entire bar region. The unsharp mask image of the galaxy highlights a faint nuclear ring with $\sim$3--4\arcsec~radius. However, this ring appearance is a side effect of Galaxy's nuclear disk having a broken-exponential radial profile; more details can be found in \citet{Erwin_21}. Therefore, we refer to this structure as a nuclear disk instead of nuclear ring.

As shown in the flux maps, compared to other galaxies we studied in this work, NGC\,4643 has the weakest emission across all line groups. The strongest emission is confined within $\sim$4--5\arcsec~radius, and there is no sign of the nuclear ring in emission. Weak emission lanes are seen extending from the western side of the nuclear disk to the south, and a weak one from the eastern side to the north. These may be the inner parts of the bar lanes curving onto the nuclear disk. As can be seen in the velocity maps of all line groups, the zero-velocity line has a wiggly shape; within the central $\sim$1--2\arcsec~there is an intrusion of positive velocities to the south and negative to the north. Further to the north, intrusions of negative-positive velocities repeat interchangeably forming indents in the zero-velocity line.

The \ha and \nii velocity dispersion maps reveal high velocity dispersion values in the core, with values reaching $\sim$100\kms. Within the nuclear disk, and along the bar lanes, the values decrease and remain around 70--60\kms. The velocity and velocity dispersion of \oiii are not reliably estimated due to poor signal, the maps are rather noisy but still show a similar trend to \ha and \nii fields.

The PA of the LON=58\degr and \textit{i}=43\degr estimated by using \texttt{Kinemetry} differs by 5\degr from the parameters estimated from large-scale photometric fits in \citet{Erwin_08} (PA of the LON=53\degr and \textit{i}=38\degr). This difference can be attributed to the kinematics within central regions being dominated by the strong bar, which can influence the parameter estimations in that region. Once we model the rotating disk with parameters adopted from large-scale fits, the resultant residual velocity map is dominated by the $m$=1 harmonic terms, which are the artefacts of wrongly fitted PA of the LON. The artefacts are minimised when incorporating the parameters obtained from \texttt{Kinemetry}, yet subtracting the disk with parameters taken from large scale photometric fits does not affect the residual maps significantly. Therefore, we chose to adopt both parameters from the \texttt{Kinemetry}, while constructing the final disk model.

The residual velocity map of the galaxy exposes two strong coherent continuous structures. One exhibits positive velocity residuals, while the other exhibits negative velocity residuals, both with amplitudes of approximately 40--50\kms. The structure with negative velocities extends from the south edge of the frame, completing almost a full turn around the nuclear disk as it is winding inwards to connect the disk. This structure exhibits deprojected residual velocity amplitudes up to 110\kms. The counterpart to this structure is seen in positive velocities, exhibiting deprojected residual velocity amplitudes up to 70\kms. These two structures provide strong evidence of extended shocks in gas. Additionally, $\sim$0.5--1\arcsec~to the north and southeast of the centre, two regions with enhanced residual velocities are seen. These regions exhibit similar amplitudes to those found around the nuclear disk. One of these structures exhibits negative velocities, while the other exhibits positive velocities, suggesting a likely inflow to the core.

The velocity difference map of the galaxy is noisy and does not exhibit any strong structures. The BPT map is dominated by LINER emission.

NGC\,4643 exhibits weak emission across all line groups and has relatively less disturbed velocity and velocity dispersion fields compared to other sample galaxies we studied in this work. \textcolor{black}{Nevertheless, as explained in paragraphs above and stated in Table \ref{tab:galaxies_shock_classes}, we identify coherent kinematic features extending over several kiloparsecs and exhibiting large deprojected residual velocity amplitudes of up to $\sim$70--110\kms. Their coherent morphology, large spatial extent, and high residual velocities support their association with extended shocks.}

\subsection{NGC\,4941}
\label{sec:ngc4941_notes}

\begin{figure*}
    \centering
   \includegraphics[width=1\textwidth]{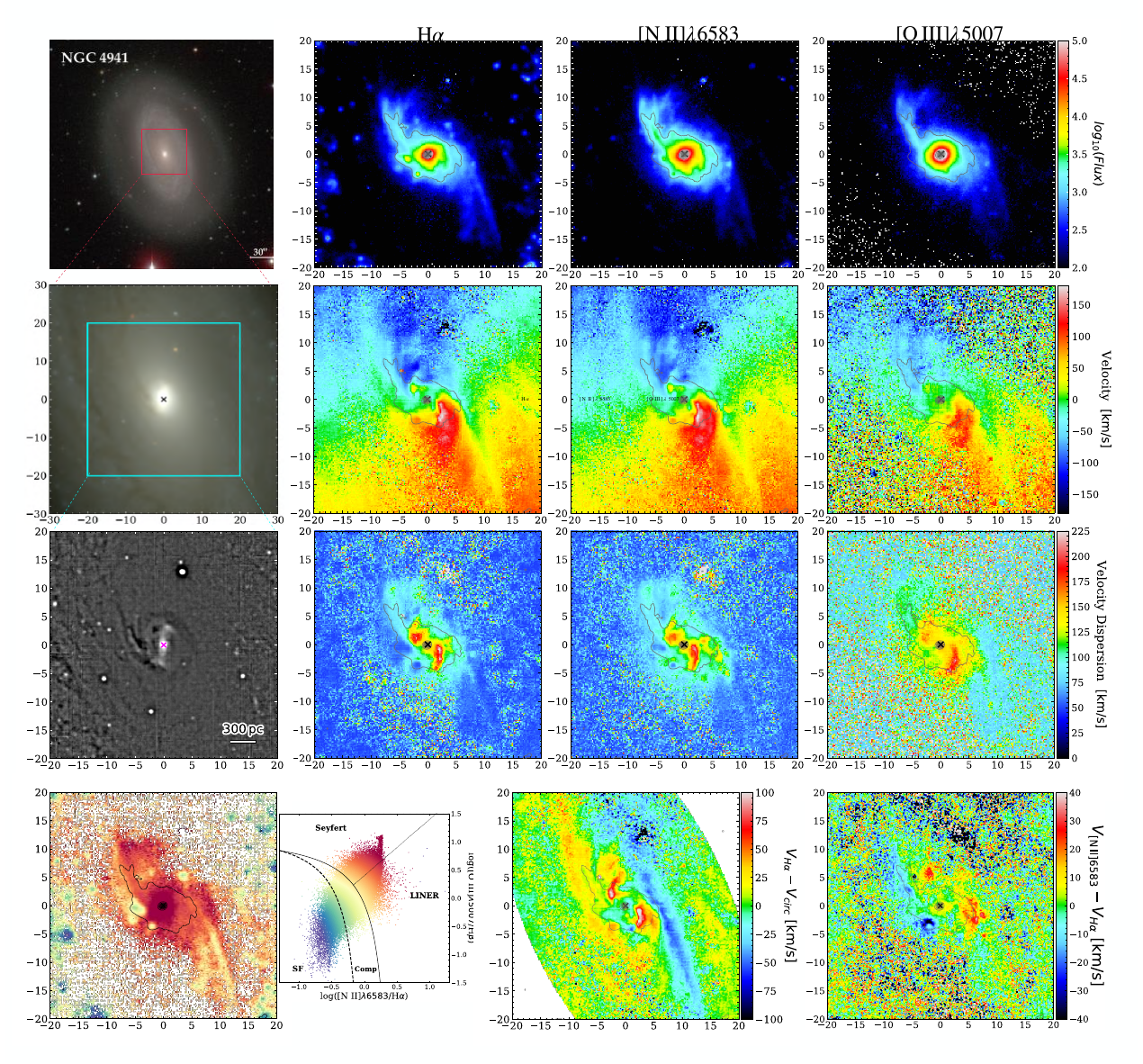}
    \caption{Images and diagnostics maps of NGC\,4941. Panels show the same as in Fig.\,\ref{fig:NGC4303}}
    \label{fig:NGC4941}
\end{figure*}

The images and diagnostic maps of NGC\,4941 are shown in Fig.\,\ref{fig:NGC4941}. NGC\,4941 has a bright, dust-obscured \citep{Jana_22} Seyfert 2 type \citep{Vernon-Cetty_2006,Asmus_2012} core, and an inner bar \citep{Greusard_2000} within $\sim$5--6\arcsec~radius. This bar can be more clearly seen in our unsharp mask image of the galaxy. Additionally, the existence of a weak boxy outer bar with a radius of $\sim$20\arcsec, which is surrounded by a dusty star-forming ring, is also proposed in this galaxy, both from visual and kinematic studies \citep{Eskridge_02, Erwin_04, Stuber_23}. The existence of the outer bar remains a subject of an ongoing debate \citep{deVaucouleurs_1991,Greusard_2000}. As shown in the large scale and colour image of the galaxy, the MUSE FOV covers the core, the inner bar, and a portion of the proposed outer bar. In the unsharp mask image, several dust lanes are seen along the proposed outer bar
. One such dust lane, extending inwards from the northeast, appears broken into segments and connects to the inner bar at the southeast, and another, as a much fainter counterpart to the former, connects to the inner bar at the northwest. 

As shown in the flux maps, the strongest emission across all line groups is confined within the $\sim$2\arcsec~radius for \ha and \nii, and $\sim$3--4\arcsec~radius for \oiii, but it is not extended along the inner bar. Beyond this region, two weak emission lanes are seen extending towards the northeast and southwest, stronger in the northeast
. Both lanes appear to fork into two components $\sim$8-10\arcsec~from the centre. In the northeast lane, \nii flux dominates over \ha, and in the southwest lane the opposite is true. These emission lanes mostly align with the dust lanes along the proposed outer bar.

The velocity maps reveal pronounced signs of strong non-circular motions for all emission line groups. Approximately 3\arcsec~to the northeast of the centre, there is a region with kinematics inconsistent with a rotating disk. Notably, the zero velocity line exhibits prominent indents at a radius of $\sim$8--10\arcsec. The velocity dispersion maps of all emission line groups, reveal very peculiar structures. In two regions $\sim$4--5\arcsec~towards northeast and southwest from the centre the velocity dispersion values exceed 170\kms particularly in \ha. These structures have a bi-symmetric nature, indicative of strong outflows from the AGN, which is in line with the north structure having kinematics inconsisitent with a rotating disc. Along the emission lane to the northeast, both \ha and \nii velocity dispersions are notably high, reaching values around 100\kms. On the other hand, in the outer regions of the southwest lane, the velocity dispersion drops to $\sim$50\kms, while in its inner region \nii velocity dispersions are around 130\kms. While \oiii velocity dispersions exhibit some noise, a clear trend mirrors what is seen in \ha and \nii velocity dispersion. In the remaining regions, the velocity dispersion remains around $\sim$50--70\kms.

While the PA of the LON=25\degr and \textit{i}=60\degr estimated from \texttt{Kinemetry}, slightly differs from parameters estimated from large-scale fits (e.g., \citealp{Gutierrez_11}: PA of the LON=21\degr, \textit{i}=48\degr), the differences can be attributed to the outer bar affecting the kinematic fits within the innermost regions. Since, we do not notice signs of artefacts from incorrectly fitted geometry in the residual velocity map, we use the parameters derived from \texttt{Kinemetry} fits. As seen in the residual velocity map, two strong coherent continuous structures with amplitudes of $\sim$40\kms are seen within the proposed outer bar region highlighted with the dashed lines in Fig.\,\ref{fig:Sample_extended_structures_perturbed_galaxies_residualvel}. One of these structures has positive velocities, and extends from the north edge of the frame to the south, east of the centre, while the other with negative values extends from the south edge of the frame to the north, west from the centre. These two structures manifest as straight lines parallel to the LON and follow the dust lanes, exhibiting deprojected velocity amplitudes up to 40 to 60 \kms. Therefore, we attribute these structures to straight shocks, indicative of inflow along the bar, providing compelling evidence supporting the existence of the outer bar. At $\sim$2--5\arcsec~from the centre towards the northeast and southwest, there are two regions with high positive velocity residuals, with amplitudes exceeding $\sim$50\kms. These structures coincide with the enhanced velocity dispersion seen across all emission line groups and further indicate its association with the AGN outflows. Although the two lane structures appear to wind towards inner regions, the structures associated with the AGN prevent us from tracing them further in.

The velocity difference map is noisy but there are regions of coherent velocity differences of magnitude $\sim$10--20\kms, in alignment with the straight shocks observed along the bar in the residual velocity map. The region to the south exhibits positive and the one to the north exhibits negative velocity differences, providing further support for the disturbance of gas and the inflow along the bar. Moreover, in the regions associated with the AGN outflows, there are positive velocity differences with amplitudes around $\sim$15--30\kms. 

The BPT map of the galaxy reveals strong Seyfert emission within the core due to irradiation from the AGN. Beyond this central region, ionisation is mostly dominated by LINER emission, except within the emission lanes, where there is also contribution from composite emission and very weak SF.

The kinematic fields of NGC\,4941 reveal pronounced kinematic structures elongated \textcolor{black}{along the outer bar, spanning across several kiloparsecs, with deprojected residual velocity magnitudes reaching up to $\sim$60\kms. These features are associated with extended shocks along the bar but} due to the strong AGN outflows, we cannot trace them to the innermost regions. \textcolor{black}{Therefore, based on the grouping explained in Sect.\,\ref{sec:grouping-shock-class}, this galaxy is classified under \textit{extended shocks and outflows} category.}

\subsection{NGC\,5248}
\label{sec:notes_5248}
\begin{figure*}
    \centering
    \includegraphics[width=1\textwidth]{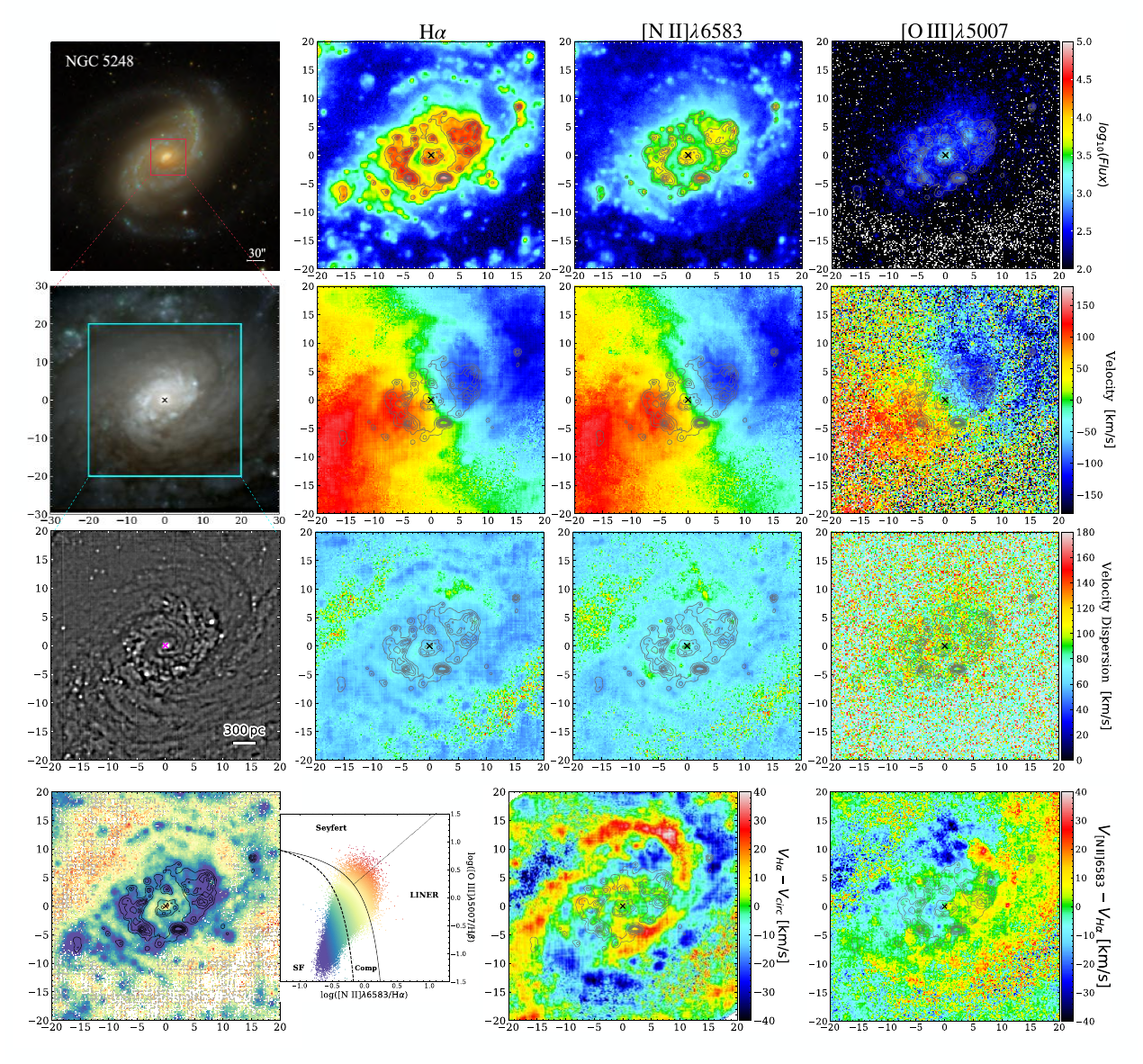}
    \caption{Images and diagnostics maps of NGC\,5248. Panels show the same as in Fig.\,\ref{fig:NGC4303}}
    \label{fig:NGC5248}
\end{figure*}

Images and the diagnostic maps of the NGC\,5248 are presented in Fig.\,\ref{fig:NGC5248}. As shown in the large scale and colour image, the galaxy has a bright core surrounded by an inner nuclear ring with $\sim$2\arcsec~radius and an outer nuclear ring with $\sim$7\arcsec~radius. It also has a weak bar with $\sim$20--25\arcsec~radius, surrounded by spiral arms. The MUSE FOV covers the core, the inner and outer nuclear rings, the bar and a portion of the spiral arms. The unsharp mask image of the galaxy reveals prominent dust structures throughout the entire FOV, many extending as spirals from the outer regions towards the outer nuclear ring and then towards the innermost regions.

As shown in flux maps, the strongest \ha and \nii emission is confined within the inner and outer nuclear rings and there is a \nii emission peak in the nucleus; between the two rings the emission decreases. There is clumpy \ha and \nii emission outside the rings, some aligned with the dusty spirals seen in the colour image. The \oiii emission across the entire FOV is weak but it has a peak at the nucleus. The \ha and \nii velocities are relatively unperturbed within the nuclear rings. However, outwards from the outer nuclear ring, the zero velocity line shows a strong indent at $\Delta\delta$ roughly 13\arcsec--14\arcsec~and $-5$\arcsec~indicating streaming motion along the spiral arms.

The velocity dispersion of \ha and \nii is low, with values $\sim$60--70 within both rings. Between the two rings, the \nii velocity dispersion shows local peaks with values around $\sim$90--100\kms in several regions to the east and southeast of the centre. Moreover, the velocity dispersion of \ha and \nii is around $\sim$60\kms where clumpy \ha and \nii emission are seen outside the rings. However, outside the rings, the velocity dispersion shows an increase in two arms upstream from the spiral dust lanes, possibly indicating some shocks there.

The disk orientation parameters, the PA of the LON=114\degr and \textit{i}=44\degr, obtained from \texttt{Kinemetry} agree well with parameters estimated from the large-scale fits \citep{Haan_08,Haan_09}, as shown in Table\,\ref{tab:Disk_orientation_parameters}. Thus, we employ these parameters for the final disk model. The residual velocity map of the galaxy exposes two coherent continuous spiral structures -- one with positive velocities and the other with negative velocities. Both of these structures are winding from the outer edge of the FOV in clockwise direction towards the innermost regions. These two spiral structures appear as counterparts of one another and are consistent with extended shocks in gas propagating to the innermost regions. Throughout the extent of these coherent structures, the deprojected residual velocity amplitudes are around 50 to 60\kms. These structures reach the outer nuclear ring, even though further in they are more patchy; there is an indication that the red structure is crossing and entering the inner ring, down to $\sim$1\arcsec.

The velocity difference map shows enhancements with amplitudes of $\sim$10--20\kms. Especially inwards from the outer ring and extending to the innermost spaxel, the enhancements in velocity differences remain prominent, maintaining the characteristic positive and negative pattern. The strongest amplitudes are within the inner $\sim$1\arcsec. These consistent signatures provide further support for the presence of extended shocks propagating down to the innermost regions.

The BPT map of the galaxy reveals LINER emission in the core and prominent SF within both the outer and inner nuclear rings. The region between the rings, where signatures of extended shocks are seen both in velocity residual and velocity difference maps, shows a contribution from composite emission, indicating the complex interaction of a high-energy mechanism there. Outside the rings, there is composite emission with a strong contribution from SF; the SF is particularly strong where clumpy \ha emission is seen. Additionally, outside the rings, there is a contribution from LINER emission in the regions where enhanced velocity dispersions are seen.

\textcolor{black}{As outlined in Fig.\,\ref{fig:Sample_extended_structures_perturbed_galaxies_residualvel}, NGC\,5248 exhibits compelling signatures of extended shocks, with coherent structures extending from large radii towards the innermost regions. These structures span several kiloparsecs and show large deprojected residual velocity magnitudes of up to $\sim$60\kms (see also Table\,\ref{tab:galaxies_shock_classes}), as identified through the different methodologies applied in this work. The extended shock signatures are particularly pronounced and show strong agreement with the expected line-of-sight velocity residual patterns presented by \citet{Davies_09}.}

\subsection{NGC\,7177}
\label{sec:ngc7177_notes}
\begin{figure*}
    \centering
    \includegraphics[width=1\textwidth]{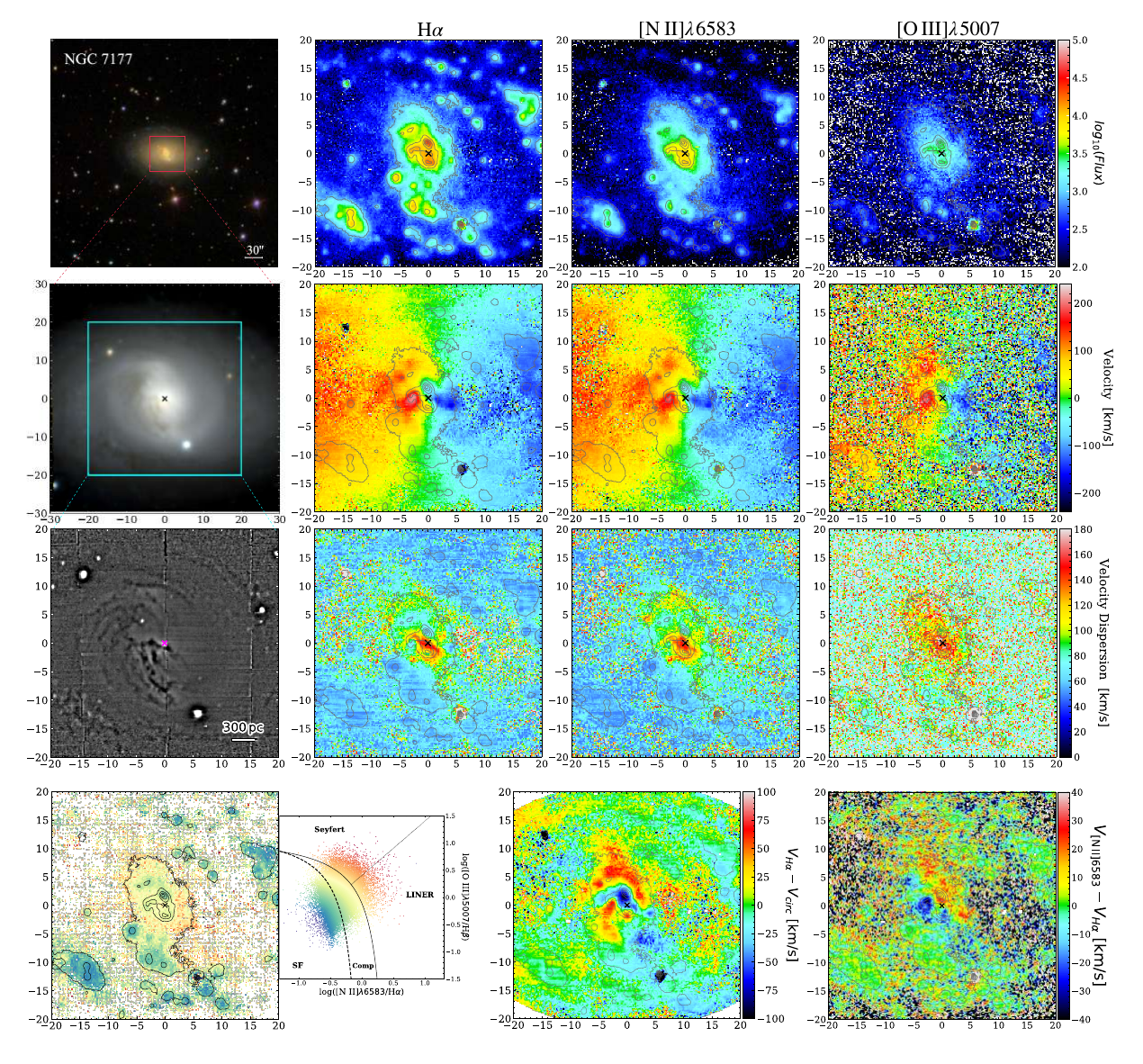}
    \caption{Images and diagnostics maps of NGC\,7177. Panels show the same as in Fig.\,\ref{fig:NGC4303}}
    \label{fig:NGC7177}
\end{figure*}

NGC\,7177 is a well-studied galaxy in many kinematic and photometric studies \citep{de_Jong_1996,Misselt_1999, Diaz_2000, Simien_02,Erwin_2005, Erwin_08,Silchenko_2010}. The images and diagnostic maps of the galaxy are presented in Fig.\,\ref{fig:NGC7177}. As shown in the large-scale and the colour image, NGC\,7177 has a bright dusty core, and a bar within $\sim$10\arcsec~radius \citep{Erwin_2005} surrounded by a ring. The MUSE FOV encompasses the core, the bar and the ring. The unsharp mask image of the galaxy reveals several dust lanes are seen within the bar region, more prominent in the East side of the galaxy.

As the flux maps show, the strongest emission across all line groups is confined within the inner $\sim$3\arcsec~radius with two clumps $\sim$2\arcsec~north and south of the nucleus, roughly along the major axis of the bar, seen in \ha emission. We also notice extended emission to the east, especially in \oiii. There is clumpy \ha and \nii emission indicative of local SF throughout the FOV.

Moreover, the velocity across all line groups shows non-circular motions. Within the $\sim$3--4\arcsec~radius, the zero velocity line has a wiggly shape, with indents at the locations of the two \ha clumps. Velocity amplitude at the emission extension to the east is $\sim$50\kms higher than on the other side of the nucleus. The velocity dispersion of \ha and \nii is high in the core and in an extended region $\sim$3\arcsec~towards the northeast and southwest of the centre. Coinciding with this region, there appears a dust lane almost perpendicular to the bar. Upon a careful examination of the spectra of spaxels with relatively high velocity dispersions, we confirm that there is no need for a secondary component in the fitting during the kinematic extraction, and the high velocity dispersion values are intrinsic. Where clumpy \ha and \nii emission is seen, throughout the FOV, the velocity dispersion of \ha and \nii is around $\sim$50--60\kms. There are velocity dispersion minima in the emission clumps $\sim$2\arcsec~south and north of the nucleus, and another minimum $\sim$1--2\arcsec~south of the east emission extent, spatially coincident with anomalously high LOS velocity. The velocity dispersions within the remaining regions in the bar are relatively high with values around $\sim$70--120\kms.

Fitting the velocity field with \texttt{Kinemetry}, we determined the optimal disk orientation parameters as PA of the LON$=$90\degr and \textit{i}$=$46\degr. These values deviate only slightly from the values obtained from large-scale photometric fits (e.g., \citealp{Erwin_08}: PA of the LON$=$83\degr and \textit{i}$=$48\degr). However, when using the large-scale parameters in the disk model, the residual maps exhibit slight signatures of $m$=1 harmonic terms, while using \texttt{Kinemetry}-deduced parameters minimises those artefacts. Therefore, we settle on using parameters estimated from \texttt{Kinemetry}.

The residual velocity map of the galaxy reveals several peculiar structures. Specifically, two relatively extended structures with positive residual velocities, exceeding 50\kms, coincide with the pattern of the dust lanes. One of these structures, forms an arch to the north of the centre with further extension to the north, while the other extends almost as a straight line over $\sim$7\arcsec~southeast of the centre. In between these two structures, a curved structure with negative velocities, of amplitudes exceeding 70\kms, reaches the nucleus from the north. These structures appear coherent, although they do not have point symmetry. The strong positive and negative velocity residuals are separated by high velocity dispersion region, which may be indicating two streams of gas colliding, as in on-axis inflow along the bars. Thus, the low velocity dispersion regions, with local minima at $\sim$1--2\arcsec~north and south of the nucleus, appear also to inflow towards the nucleus. 


Furthermore, the velocity difference map reveals two regions with consistent velocity differences, with amplitudes around $\sim$15--30\kms; one region with positive differences to the northwest of the centre, and the other region with negative differences to the east--southeast. Particularly, there are three peaks with maximum velocity differences, coinciding with the three minima in velocity dispersion maps.

The BPT map of the galaxy exhibits composite emission with a slight contribution from LINER emission in the core. This confirms that the contribution from the AGN to the overall gas disturbance in the galaxy is weak, indicating that the high velocity dispersions seen in the core, in an extended region $\sim$3\arcsec~towards the northeast and southwest of the centre, is not associated with the AGN and is caused by another physical mechanism, potentially due to collision of gas along on-axis inflow. Throughout the FOV there is composite emission with a contribution from star formation.

\textcolor{black}{The central regions of NGC\,7177 exhibit complex kinematic behaviour. The emission-line maps show high and relatively extended nuclear velocity dispersion, accompanied by coherent structures spanning roughly a kiloparsec. These structures exhibit large amplitudes in both the deprojected residual velocity map ($\sim$70\kms) and the velocity-difference map (up to $\sim$30\kms). Although the BPT diagnostics do not indicate a dominant contribution from AGN-driven outflows within the inner kiloparsec, the overall kinematic complexity is likely influenced by AGN activity. The observed kinematics are also broadly consistent with on-axis inflow along the bar \citep{Athanassoula_1992}. Given the presence of extended-shock signatures, together with indications that the inner regions are affected by AGN activity, we classify this galaxy in the \textit{extended shocks and outflow} category.}

\subsection{NGC\,7513}
\label{sec:NGC7513_notes}

\begin{figure*}
    \centering
    \includegraphics[width=1\textwidth]{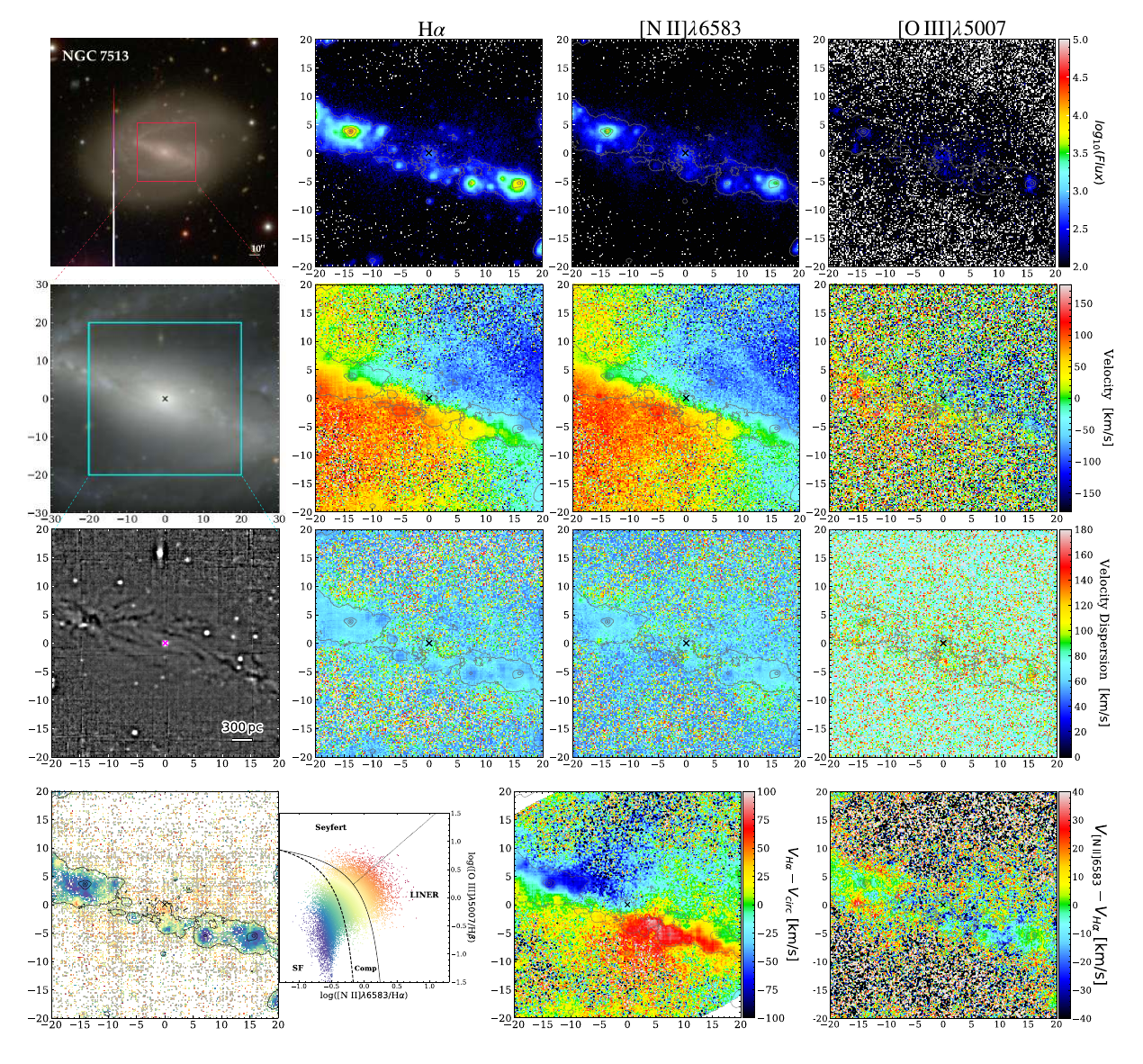}
    \caption{Images and diagnostics maps of NGC\,7513. Panels show the same as in Fig.\,\ref{fig:NGC4303}}
    \label{fig:NGC7513}
\end{figure*}

The images and diagnostic maps of NGC\,7513 are presented in Fig.\,\ref{fig:NGC7513}. As shown in large-scale image, the galaxy has a bright core, a strong bar \citep{Buta_09} within $\sim$30\arcsec~radius, and two dusty star-forming spiral arms connecting to the ends of the bar. The MUSE FOV shown in the colour image covers the core, almost entire bar region with some portion of the spiral arms. The unsharp mask image of the galaxy reveals straight dust lanes within 12\arcsec~along the bar; on the leading side of the bar, they pass the nucleus at $\sim$2\arcsec~distance and continue $\sim$5\arcsec~into the trailing sides. 

As shown in the flux maps, within $\sim$6\arcsec~radius, the emission across all line groups is weak, particularly for \oiii. Along the bar there is clumpy \ha and \nii emission, weak in the inner 12\arcsec~radius, significantly stronger outside of it, indicating local SF along the bar lanes. In velocity maps of \ha and \nii, the zero velocity line is bent by 90\degr at $\sim$10\arcsec~radius, roughly at the locations where line emission becomes strong. The velocity dispersion of \ha and \nii is generally low. The values are $\sim$70\kms in the core and $\sim$50\kms within the regions with clumpy emission along the bar. 

The disk inclination obtained by \texttt{Kinemetry} converged around 46\degr. However, the radial distribution of the PA of the LON ranged between 100--130\degr. Examining different residual velocity maps, after altering the fixed PA of the LON between the accepted range of 100-130\degr, we settled on using 105\degr as this value raised the least amount of artefacts. The final fixed parameters in the disk model, \textit{i} 46\degr and PA of the LON 105\degr agree well with the parameters estimated from large-scale fits.

The residual velocity map of galaxy exhibits two straight coherent features along the bar appearing as counterparts, on its trailing side; one exhibiting positive and one exhibiting negative velocities, \textcolor{black}{extending across few kiloparsecs}. To guide the reader, these extended features are outlined with dashed lines in the deprojected residual velocity map shown in Fig.\,\ref{fig:Sample_extended_structures_perturbed_galaxies_residualvel}. These features extend beyond the bend of the zero velocity line, continuing outside the frame, all the way to the emanation of the spiral arms from the bar. They do not reflect the bend in the zero velocity line, which indicates different gas kinematics in the outer regions of the bar, where strong line emission is present, and the inner regions.
\textcolor{black}{These features have deprojected residual velocity magnitudes around 80--100\kms, comparable to the structures seen in the other galaxies we studied in this work, and are similar to the extended features seen in the residual velocity map of NGC\,289 (see Sect.\,\ref{sec:NGC289_notes}). Although these features are on the trailing side of the bar, rather than the leading side as the typical extended shocks are, we still classify them as structures indicative of extended shocks due to their large coherent extent, magnitude and similarities shown in other galaxies.}

The velocity difference map of galaxy is noisy and shows no outstanding structures. Moreover the BPT map exhibits a combination of composite LINER type emission in the core and contribution from SF along the bar lanes, where clumpy emission is seen.

%% file: supplementary_appendix_figures.tex
\renewcommand{\thefigure}{\thesection 1}
\section{Supplementary maps and figures}
\label{app:suplementary_sample_plots}

In this section, we present supplementary figures and maps that support the analysis discussed in the main body of this article.

\begin{figure*}
    \centering
    \includegraphics[width=1\textwidth]{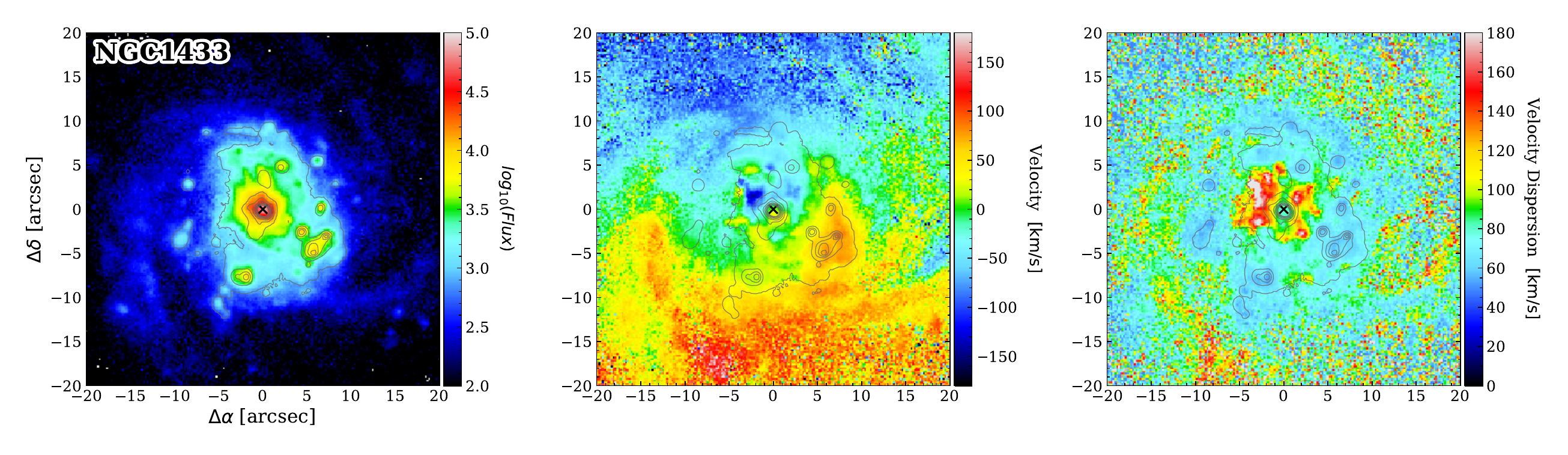}
    \includegraphics[width=1\textwidth]{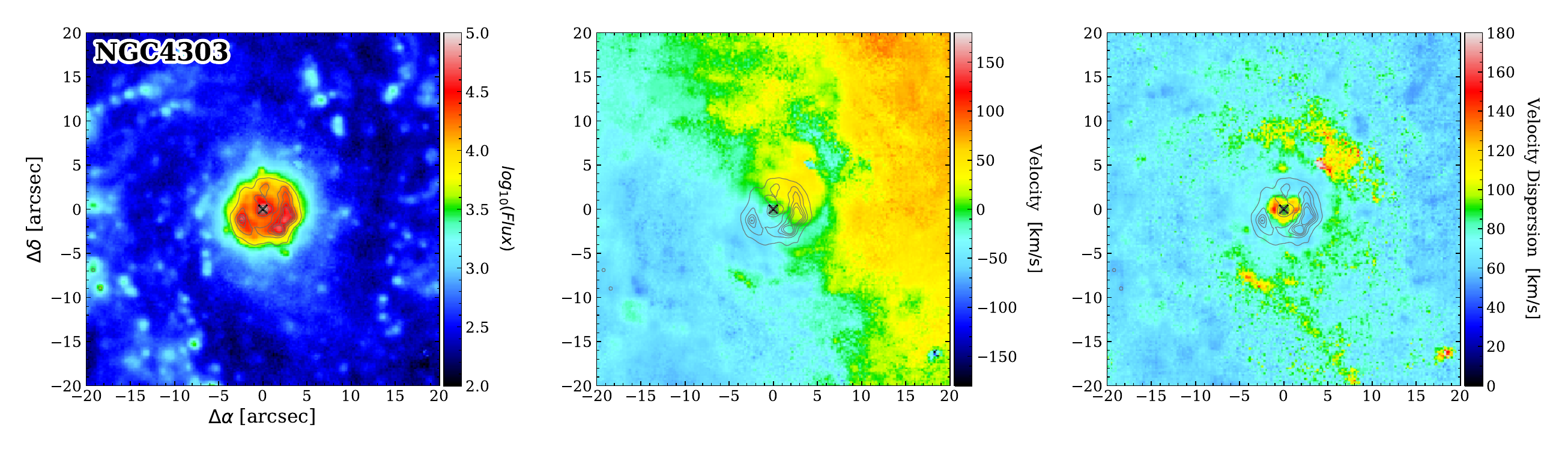}
    \includegraphics[width=1\textwidth]{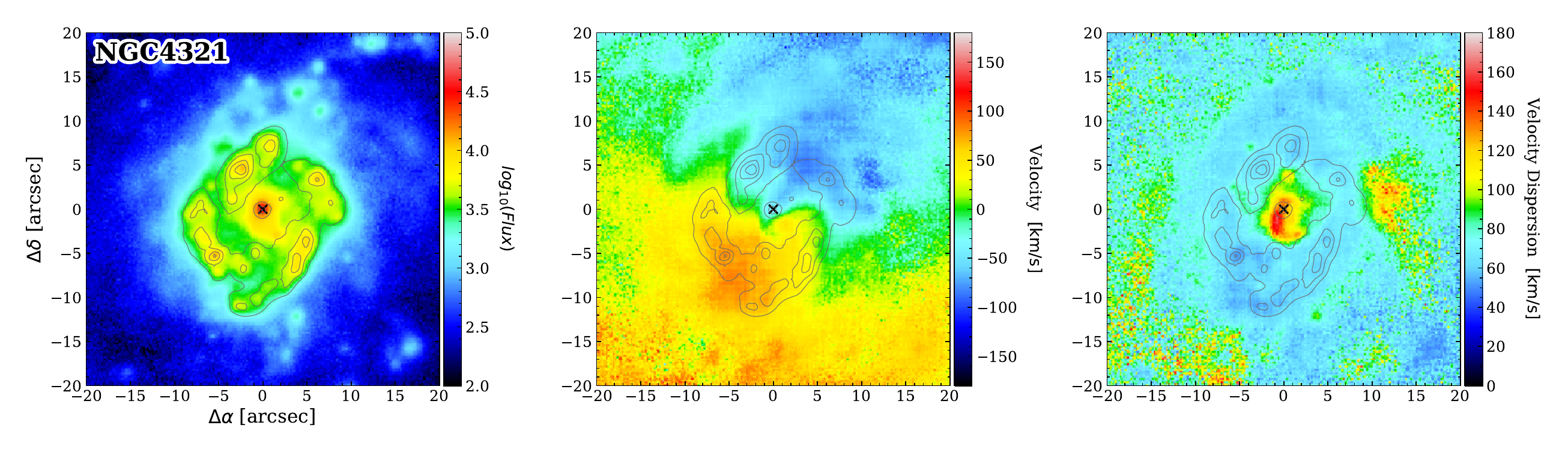}
    \caption{The \nii flux, velocity and velocity dispersion maps of NGC\,1433 (top), NGC\,4303 (middle) and NGC\,4321 obtained from single Gaussian fits. Gray contours represent the \ha emission. Centre of the galaxy is marked with an "x".}
    \label{fig:NGC1433_NGC4303_NGC4321_R1}
\end{figure*}

Fig.\,\ref{fig:NGC1433_NGC4303_NGC4321_R1} shows the flux, velocity, and velocity dispersion maps of NGC\,1433, NGC\,4303, and NGC\,4321, derived from single-Gaussian fitting.

Fig.\,\ref{fig:sample_vcircmaps} shows the circular velocity maps for the galaxies in our sample. These maps are derived by fitting a rotating disk model to the \ha velocity field of each galaxy using \texttt{Kinemetry}, with the fixed geometry, as described in Sects.\,\ref{met:residual_velocities} using orientation parameters (PA of LON and inclination) provided in Table \ref{tab:Disk_orientation_parameters}.

Fig.\,\ref{fig:deprojected_velocity_residuals} presents the deprojected residual velocity maps for galaxies in our sample that do not present extended structures, thus do not meet the criteria for extended shocks outlined in Sect. \ref{sec:met:criteria}. Although some structures are visible, consideration of the full set of diagnostic maps does not support their interpretation as extended shocks. These maps are included for completeness.

Fig.\,\ref{fig:AON_distributions} shows the distributions of \ha\ velocity dispersion and amplitude-over-noise (AON) for all spaxels within an elliptical aperture of semi-major axis 1\,kpc, corrected for the distance and inclination of each galaxy; as discussed in Sect.\,\ref{sec:average_perturbation_against_shock_amplitudes}, this figure forms the basis for estimating a robust average \ha\ velocity dispersion within the central kiloparsec of each system. Table\,\ref{tab:gas_disturbance_weigthed} presents the average \ha velocity dispersion values for each galaxy calculated using different weighting approaches: unweighed, flux and AON weighted.

\begin{table}
\centering
\caption{Comparison of the H$\alpha$ velocity dispersion for each galaxy with different weighting approach used within calculation. Columns show the Galaxy ID, unweighted mean; flux-weighted mean and AON-weighted mean.}
\label{tab:gas_disturbance_weigthed}
\resizebox{\columnwidth}{!}{%
\begin{tabular}{lccc}
\hline
{Galaxy ID} & {Unweighted} & {Flux} & {AON} \\
\hline
IC2051   & 77.6 & 72.8 & 73.8 \\
NGC289   & 72.9 & 73.8 & 72.0 \\
NGC0613  & 103.9 & 94.3 & 98.3 \\
NGC1097  & 83.0 & 76.2 & 78.4 \\
NGC1300  & 72.4 & 68.9 & 68.6 \\
NGC1433  & 76.1 & 76.6 & 73.1 \\
NGC1566  & 94.7 & 197.8 & 106.1 \\
NGC3351  & 85.8 & 70.1 & 76.2 \\
NGC3368  & 104.7 & 105.9 & 102.9 \\
NGC3489  & 70.3 & 71.0 & 69.3 \\
NGC3626  & 76.1 & 87.3 & 71.9 \\
NGC4237  & 54.2 & 53.9 & 54.0 \\
NGC4303  & 71.4 & 65.7 & 67.2 \\
NGC4321  & 67.9 & 65.1 & 66.0 \\
NGC4380  & 53.9 & 53.5 & 53.6 \\
NGC4457  & 87.1 & 102.5 & 85.7 \\
NGC4643  & 70.9 & 73.0 & 71.6 \\
NGC4941  & 89.4 & 134.2 & 101.2 \\
NGC5248  & 63.4 & 61.1 & 61.9 \\
NGC7177  & 81.8 & 89.0 & 83.1 \\
NGC7513  & 59.6 & 59.2 & 59.0 \\
\hline
\end{tabular}}
\end{table}

\renewcommand{\thefigure}{\thesection 2}
\begin{figure*}
    \begin{minipage}{1\textwidth}
    \includegraphics[width=0.97\textwidth]{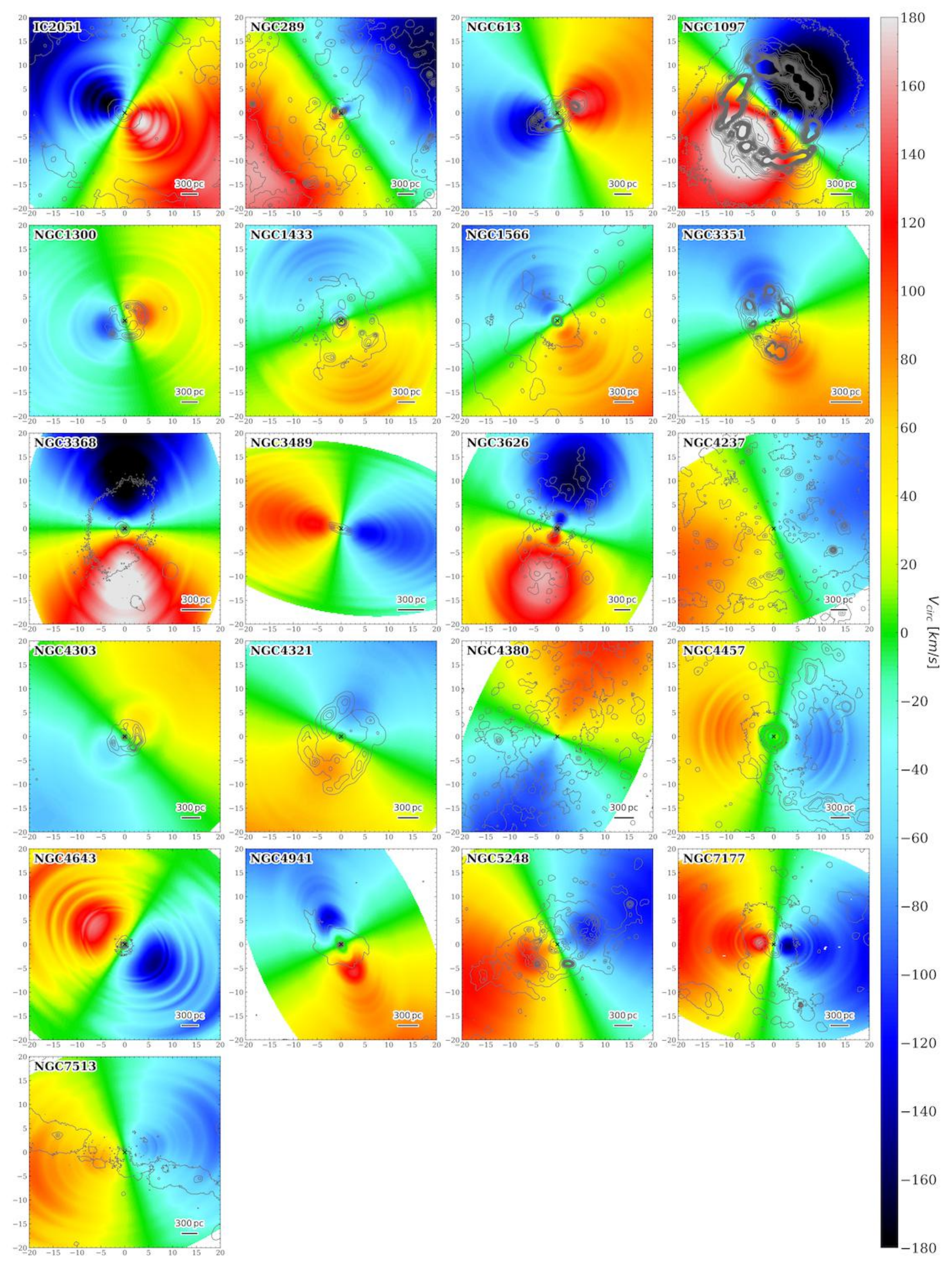}
    \end{minipage}
    
    \caption{Maps of the circular velocity of a rotating disc, constructed with \texttt{Kinemetry} using a fixed centre, PA and inclination, with the adopted parameters listed in Table\,\ref{tab:Disk_orientation_parameters}. The models are based on the \ha line-of-sight velocity field.}
    \label{fig:sample_vcircmaps}
\end{figure*}

\renewcommand{\thefigure}{\thesection 3}
\begin{figure*}
    \begin{minipage}{1\textwidth}
    \includegraphics[width=0.97\textwidth]{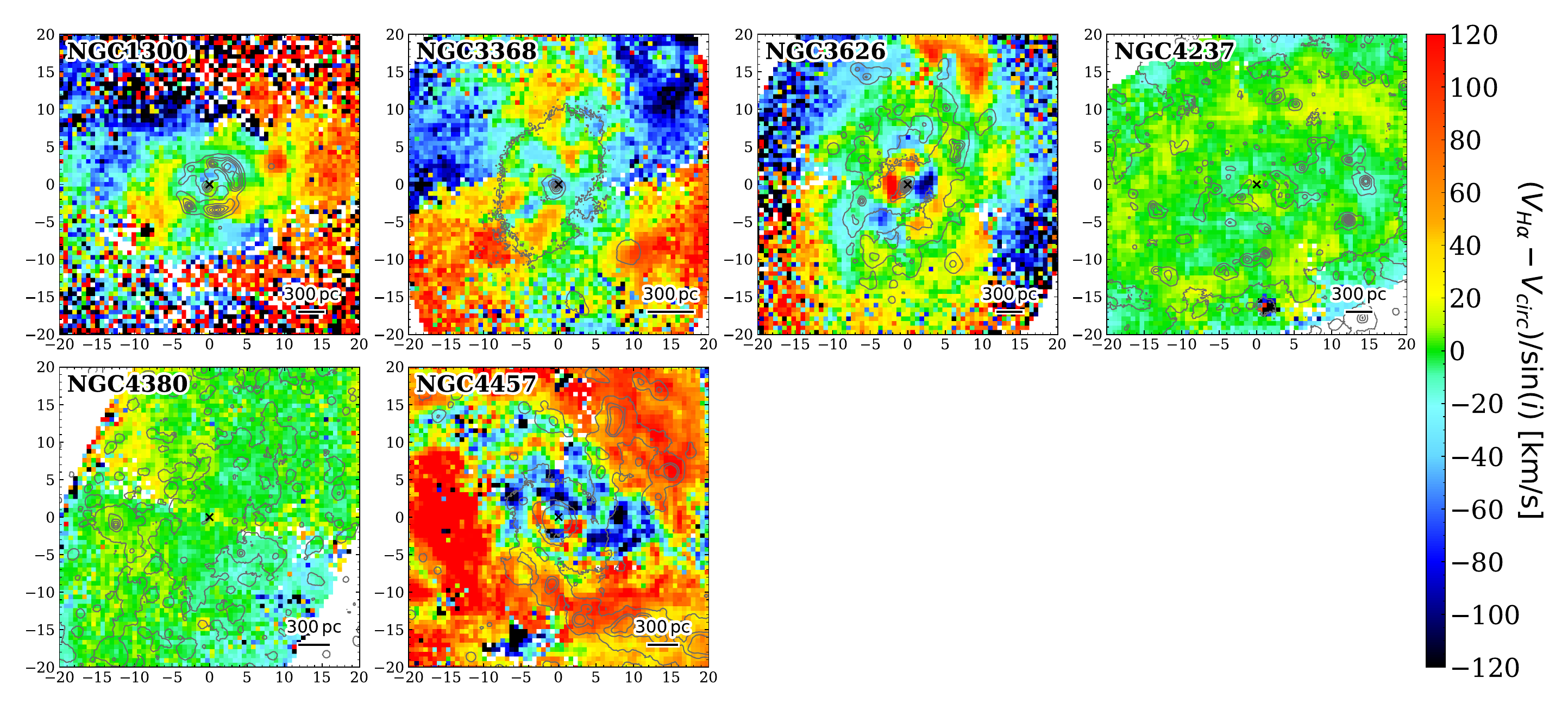}
    \end{minipage}
    
    \caption{Maps of deprojected residual velocity for the galaxy sample studied in this work in which no extended structures are identified. The value of each spaxel is the average over $3\times3$ spaxels' box centred on it. The grey contours represent the \ha emission, dashed lines outline the coherent structures we identify on the maps.}
    \label{fig:deprojected_velocity_residuals}
\end{figure*}

    

\renewcommand{\thefigure}{\thesection 4}
\begin{figure*}
    \begin{minipage}{1\textwidth}
    \includegraphics[width=0.95\textwidth]{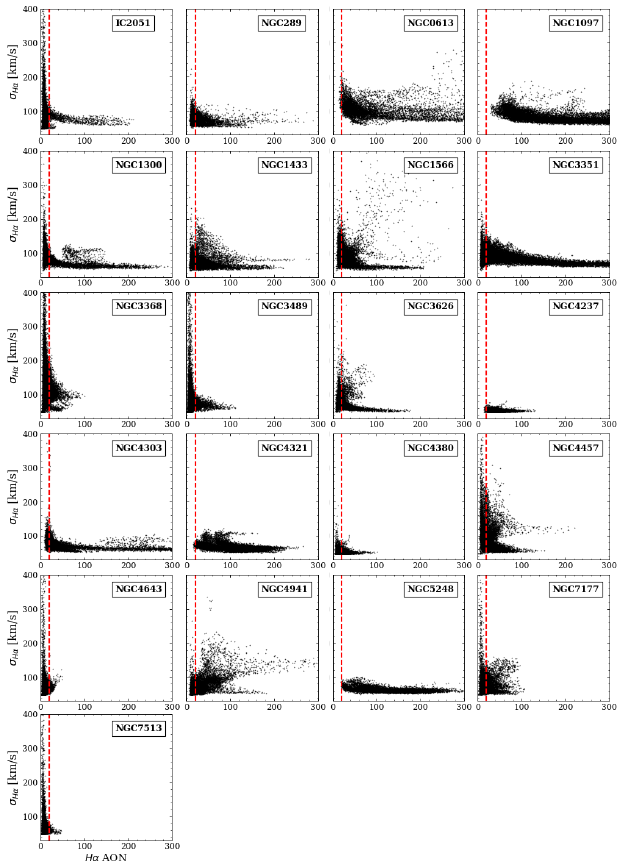}
    \end{minipage}
    \caption{The distribution of spaxels within the inner kpc on the plane defined by the \ha velocity dispersion and AON for each galaxy studied in this work. The red dashed line represents the AON=20 threshold.}
    \label{fig:AON_distributions}
\end{figure*}